\documentclass[12pt]{article}
\usepackage[T1]{fontenc}
\usepackage[dvips]{graphicx}
\graphicspath{{images/}}
\usepackage{verbatim}
\usepackage{amsmath,amsthm}
\usepackage{xspace}
\usepackage{pifont}
\usepackage{graphicx}
\usepackage{amssymb}
\usepackage{epic, eepic}
\usepackage{dsfont}
\usepackage{amssymb}
\usepackage{makeidx}
\usepackage{mathrsfs}
\usepackage{exscale}
\usepackage{color} 
\usepackage{overpic} 
\usepackage{bm}
\usepackage{bbm}
\usepackage{booktabs} 
\usepackage{color, colortbl}
\usepackage{subcaption}
\usepackage{float}
\usepackage[numbers]{natbib}
\usepackage[title]{appendix}

\RequirePackage[colorlinks,citecolor=blue,urlcolor=blue]{hyperref}

\definecolor{Gray}{gray}{0.9}

\usepackage{amsmath,afterpage}
\usepackage{epsf}
\usepackage{graphics,color}
\RequirePackage[colorlinks,citecolor=blue,urlcolor=blue]{hyperref}
\usepackage[export]{adjustbox}

\usepackage{tcolorbox}
\def\e{\mathbf{e}}

\def\0{\mathbf{0}}

\def\lam{\lambda}
\def\rr{\rightarrow}

\def \< {\langle}
\def \> {\rangle}

\def\ol{\overline}

\def\beqa{\begin{eqnarray}}
\def\eeqa{\end{eqnarray}}
\def\beqas{\begin{eqnarray*}}
\def\eeqas{\end{eqnarray*}}

\newtheorem{theorem}{Theorem}[section]
\newtheorem{lemma}[theorem]{Lemma}

\newtheorem{proposition}[theorem]{Proposition}

\newtheorem{corollary}[theorem]{Corollary}

\newtheorem{remark}[theorem]{Remark}

\newtheorem{definition}[theorem]{Definition}
\newtheorem{assumption}[theorem]{Assumption}
\numberwithin{equation}{section}
\newcommand{\hatd}[1]{{}}

\newcommand{\bd}{\begin{displaymath}}
\newcommand{\ed}{\end{displaymath}}
\newcommand{\be}{\begin{equation}}
\newcommand{\ee}{\end{equation}}
\newcommand{\bq}{\begin{eqnarray}}
\newcommand{\eq}{\end{eqnarray}}
\newcommand{\bn}{\begin{eqnarray*}}
\newcommand{\en}{\end{eqnarray*}}
\newcommand{\dl}{\delta}

\def\wt{\widetilde}

\def\one{\mathds{1}}

\def\E{{\mathbb{E}}}

\usepackage[normalem]{ulem}

\usepackage{mathtools}
\mathtoolsset{showonlyrefs}

\usepackage{authblk}

\title{Unwinding Toxic Flow with Partial Information}
\author[1]{Alexander Barzykin }
\author[2]{Robert Boyce \thanks{RB is supported by the EPSRC Centre for Doctoral Training in Mathematics of Random \mbox{Systems}: Analysis, Modelling and Simulation (EP/S023925/1).}}
\author[2]{Eyal Neuman }
\affil[1]{HSBC}
\affil[2]{Department of Mathematics, Imperial College London }

\begin{document}

\vspace{-0.5cm}
\maketitle

\begin{abstract}
We consider a central trading desk which aggregates the inflow of clients’
orders with unobserved toxicity, i.e. persistent adverse directionality. The desk chooses either to internalise the inflow or externalise it to the market in a cost-effective manner. In this model, externalising the order flow creates both price impact costs and an additional market feedback reaction for the inflow of trades. The desk’s objective is to maximise the daily trading P\&L subject to end-of-day inventory penalisation. We formulate this setting as a partially observable stochastic control problem and solve it in two steps. First, we derive the filtered dynamics of the inventory and toxicity, projected to the observed filtration, which turns the stochastic control problem into a fully observed problem. Then we use a variational approach in order to derive the unique optimal trading strategy. We illustrate our results for various scenarios in which the desk faces momentum and mean-reverting toxicity.
\end{abstract} 

\begin{description}
{\small \item[Mathematics Subject Classification (2010):]  91G10, 49N10, 49N90, 93E20, 93E11, 60G35 
\item[JEL Classification:] C61, C73, G11, G24, G32 
\item[Keywords:] central risk book, market making, optimal liquidation, price impact, partially observable stochastic control  
}
\end{description}


\section{Introduction}
Information asymmetry between buyers and sellers has long played a central role in the modelling of financial markets. In particular, models of informed traders have been at the core of financial research ever since the pioneering work of \citet{GLOSTEN198571,Kyle_85,EASLEY198769}. Toxic order flow is trading activity that adversely selects market makers by exploiting an informational advantage, causing them to provide liquidity at a loss. This phenomenon has a profound effect on market making and on high-frequency trading strategies. The statistical properties of toxic flows and indicators that predict them, such as volume imbalance and trade intensity, have received a considerable amount of attention in \cite{Cartea_26,Easley:2012aa,Easley:2011aa,Easley:2011ab} among others. 

A related problem is how to unwind stochastic client flow while controlling inventory and execution costs. Market makers and brokers can offset opposing client
trades (known as internalisation) and unwind residual risk in the market (known as externalisation).
\citet{Butz:2019aa} proposed a simple model for internalisation using queuing theory, in which the desk can skew their prices in order to control inventory risk. \citet{Barzykin:2023aa} proposed a stochastic optimal control model which combines internalisation and externalisation. In their model, the dealer sets quotes to attract flow while they simultaneously hedge in a separate liquidity pool. The common situation of an informed trader with information about the trend of an asset price and a broker who trades at a loss was studied in \citet{cartea_25}. However, in their setting the broker takes advantage of knowing the parameters of the toxic order flow which he uses as a signal in order to decide whether to internalise or externalise, depending on the state in the market.

In pursuit of minimising firm-wide trading costs due to price impact, many institutions have centralised trading activity, including into desks commonly known as central risk books (CRBs), which net opposite orders of clients. In a recent work \citet*{nutz2023unwinding} studied the problem faced by a CRB which optimally unwinds an order flow arriving from clients by using both internalisation and externalisation, while taking into account transient impact induced by the outflow. In their model Nutz et al. assume that the order flow is a diffusion process with parameters which are known to the desk. The sign of the drift coefficient of the inflow determines the type of toxicity, which could be mean-reverting or momentum driven. Moreover, in \cite{nutz2023unwinding} the externalisation does not affect the inflow created by informed traders. From the mathematical perspective, the model of Nutz et al. differs from standard optimal execution problems such as  \cite{cartea2016orderflow,lehalle2019signals,neuman2022optimalsignaltrading} insofar as in their model the inventory to be executed is random and determined by the inflow. In \citet{bergault2025hedge} the authors studied a similar problem as \cite{nutz2023unwinding} but with permanent price impact rather than transient price impact and with strictly convex instantaneous price impact costs.

The main objective of this paper is to combine the two streams of literature, i.e. the statistical estimation of toxicity and the optimal unwinding in the presence of toxicity, into a unified tractable framework. The data analysis in Section \ref{sec-data} of institutional FX flow implies that the inflow's drift parameter, which represents toxicity, can change on a daily basis from being mean reverting to having momentum (i.e. positive trend). While the trajectories of the inflow are observed by the desk, the drift and the noise of the inflow are unobservable and are subject to inference throughout the trading day. This turns the liquidation problem faced by the desk into a partially observed stochastic control problem. Moreover, as suggested by \citet{brunnermeier2005predatory}, if the desk externalises their accumulated inventory, other traders in the market can interact with this flow introducing feedback. 
The desk is also allowed to exploit price predicting signals in the unwind strategy.   
We will capture all three aforementioned phenomena, which are not included in the model proposed by \citet{nutz2023unwinding}. As a result, the optimal unwinding problem therefore becomes a partially observable stochastic control problem.

In order to tackle the partially observable stochastic control problem faced by the desk, we use methods from filtering theory. We recall that filtering concerns the estimation of an unobserved stochastic process, given noisy observations of the process. Linear filtering dates back to the pioneering work of \citet{kalman1960filter,kalman1961inear} while in the non-linear case we refer to the seminal contribution of Zakai, Kushner and Stratonovich (see Chapter 3 of \cite{bain2009filtering} and \cite{crisan2009signaldependentnoise}). The area of partially observable stochastic control considers optimal control problems with incomplete information, i.e. when some of the state variables are unobserved by the desk. In our setting the unobserved quantities are the drift (toxicity) and the noise of the order flow. In contrast to the full information case, in this class of problems the control is required to be adapted to a smaller filtration generated by the observable states, which in turn is influenced by the control (see Section \ref{subsec:posc note} for further details). Recently, \citet{sun2023lqposc} introduced a novel approach for solving a class of quadratic optimal control problems with partially observable linear dynamics of the states, by utilising the Fujisaki-Kallianpur-Kunita method from stochastic filtering theory \cite{Fujisaki:1972aa}. Moreover, \citet{wonham1968sepearation} showed that the known effect of the control can be removed from the observation without changing its filtration.

Partial information in optimal execution has been studied by \citet{colaneri2020partial} and \citet{chevalier2024incomplete}, who use the Zakai and Kushner-Stratonovich approaches, respectively, to infer hidden Markov regimes. In the former, the regime governs price jump dynamics, while in the latter it governs market order arrival intensities, which in turn determine price dynamics. \citet{dammann2023incomplete} instead filter an unknown constant price drift taking two possible values.

In order to optimally unwind a stochastic order flow without observing its toxicity, we first derive the dynamics of the state processes in terms of the projection of the toxicity to the observed filtration (see  Theorem \ref{thm:filtering processes}). This step is inspired by the methodology of \citet{sun2023lqposc}. Having obtained the dynamics of the filtering processes, we use a variational approach in order to derive the optimal unwind strategy, which minimises the desk's cost functional (see Theorem \ref{thm:optimal control}).  In Section \ref{sec:illustration} we perform a detailed numerical analysis and provide empirical evidence for the model’s implications. Our study provides the following new insights: 
\begin{itemize} 
\item[\textbf{(i)}] 
Using institutional FX flow, we show that the directional character of aggregate client flow can differ markedly across trading days; Figure \ref{fig:gbpusd_cumulative_intraday_flow.pdf} illustrates a near-martingale day and a day with pronounced positive directionality.
\item[\textbf{(ii)}] We show that the feedback of externalisation has a prominent effect on performance, even for small values of the feedback parameter (see Table \ref{tab:naive_b_tests}). We also observe that in cases where the feedback is of large magnitude, the desk can exploit this to its own advantage and control the inflow and the transient price impact in order to get better trading performance (see discussion in Section \ref{sec-naive}). 
\item[\textbf{(iii)}] 
We study the performance gap between the partially observable problem and the full information case and find the regret to be of order less than $0.1\%$ in all tested scenarios (see Table \ref{table:strategy performance}).
\end{itemize}

\paragraph{Mathematical contribution.} In addition to our financial observations we comment on the mathematical contribution of this work with respect to \citet{sun2023lqposc} and \citet{nutz2023unwinding}.
\begin{itemize} 
\item[\textbf{(i)}] Our model extends the model of \cite{nutz2023unwinding} to the case where the drift of the inflow is unknown to the desk (see \eqref{def:Z}). We also include the feedback of the desk's trades on the inflow process (see \eqref{def:Z} and \eqref{def:theta}) which was not taken into account in \cite{nutz2023unwinding}, or in earlier papers. These two major changes in the model introduce a partially observable stochastic control problem which is solved by introducing a new two-step approach as described below. Moreover, in contrast to \cite{nutz2023unwinding} where only martingale prices were considered, our fundamental price also includes a general finite variation drift (see $A$ in \eqref{ass:P}), often referred to as \emph{alpha}. Alphas serve as short-term price predictors, and they are an important ingredient in portfolio choice and execution strategies (see \citep{A-N-T24,GARLEANU16,lehalle2019signals,neuman2022optimalsignaltrading,webster2023handbook} among others). As a result, the optimal execution rate includes a new term which incorporates the signal's expected future values. Moreover, our notion of toxicity is truly adverse to the desk's performance because it is anticorrelated to the average price movements arising from the alpha. This affects P\&L, as seen in Section \ref{sec:toxicity-pnl}. We demonstrate the effects of these components in Section \ref{sec:illustration}.

\item[\textbf{(ii)}] Our filtering argument departs from the proof of \citet{sun2023lqposc} at the martingale representation step, after removing the known effect of the control from the observation as in \citet{wonham1968sepearation}. In their setting, the partial filtration is generated by the observation process, whereas in our setting the desk also observes the independent martingale and finite variation components of the fundamental price. Consequently, the desk's partial filtration contains information that is not generated by the order flow observation, so the Fujisaki, Kallianpur and Kunita theorem cannot be applied directly. We overcome this by first reducing the filtering problem to the information generated by the order flow observation and only then applying the theorem. In the second step of the derivation of the optimal strategy, we solve the stochastic control problem using a variational method (see Theorem \ref{thm:optimal control}), which is orthogonal to the method of verification via a linear-quadratic ansatz, which was used in \cite{sun2023lqposc}. 
\end{itemize} 

\textbf{Organisation of the paper:} In Section \ref{sec:model}, we introduce the model for stochastic order flow and other market processes and define the desk's optimal control problem. In Section \ref{sec:main-results}, we present our main results, including the filtering process satisfied by the conditional expectation of the drift of the inflow, as well as the optimal unwind strategy for the desk. In Section \ref{sec:illustration} we provide empirical evidence and present simulations of the numerical implementation of the model. Section \ref{sec:conclusion} provides some concluding remarks. Appendix \ref{sec:filtering proof} provides the proof of the filtering result, Theorem \ref{thm:filtering processes}, while Appendix  \ref{sec:control proof} gives the proof of the optimal control result, Theorem \ref{thm:optimal control}. Finally, Appendix \ref{sec:full representation of control} comprehensively details the representation of the optimal control and Appendix \ref{sec:pf-full} gives the proof of the full information result.

\section{Model setup}\label{sec:model}
We consider a central trading desk of an investment bank or a large trading firm
which aggregates clients’ orders. The desk continuously chooses an externalisation
rate, while residual inventory is warehoused and may be offset by subsequent client
flow.
In order to describe this setup mathematically, we adopt the main features of the model from Section 2.1 of \citet{nutz2023unwinding}.

Let $T>0$ denote a finite deterministic time horizon and let $(\Omega, \mathcal{F},(\mathcal {F}_{t})_{t\in[0,T]},\mathbb{P})$ be a filtered probability space satisfying the usual conditions of right continuity and completeness. 
We consider a risky asset which follows a semimartingale price process $P=(P_t)_{0 \leq t\leq T}$ whose canonical decomposition $P = p + \ol M + A$ into a (local) martingale $\ol M=(\ol M_t)_{0 \leq t\leq T}$ and a predictable finite-variation process $A=(A_t)_{0 \leq t\leq T}$ satisfies
\begin{equation} \label{ass:P} 
E \left[ \langle  \ol M  \rangle_T \right] + E\left[\left( \int_0^T |dA_s| \right)^2 \right] < \infty.
\end{equation}
We assume that the inflow (or order flow) of buy/sell orders of the risky asset arriving to the desk satisfies the following dynamics, 
\be \label{def:Z} 
dZ^{q}_t = \theta^{q}_t dt + \sigma dW^Z_t, \quad 0  \leq t\leq T, \quad  Z^{q}_0=0,
\ee
where $W^Z = (W^Z_t)_{t\geq 0}$ is a Brownian motion and $\sigma$ is a positive constant. The trend $\theta^{q}= (\theta^{q}_t)_{t \geq 0}$ is an adapted stochastic process, unknown to the desk, which follows
\begin{equation} \label{def:theta}
    d\theta^{q}_{t}= \left(a_{t}\theta^{q}_{t} + b_{t}q_{t}\right)dt + c_{t}dA_{t} + d_{t}dW^{\theta}_{t}, \quad 0\leq t \leq T, \quad  \theta_{0}\in\mathbb{R}, 
\end{equation}
where $a=(a_{s})_{s\in[0,T]}, b=(b_{s})_{s\in[0,T]}, c=(c_{s})_{s\in[0,T]}$ and $d=(d_{s})_{s\in[0,T]}$ are measurable deterministic functions, which are bounded on $[0,T]$. Additionally, $b$ must be differentiable and it must hold that 
$\sup_{0\leq t\leq T}|b'_t|<\infty$. 
Here $W^\theta=  (W^\theta_t)_{t\geq 0}$ is a Brownian motion not depending on $W^Z$ and $q=(q_s)_{s\in[0,T]}$ denotes the desk's trading rate, to be specified later. We assume that $\overline{M}$, $A$, $W^Z$, and $W^\theta$ are mutually independent. The parameters of $\theta$ are assumed to be known from historical data of previous trades. Throughout this work, superscript $q$ indicates that a process is controlled by the desk's trading rate, which will be defined later. The sign of the function $a_t$ determines whether the toxicity $\theta^{q}$ exhibits momentum (when $a_t>0$) or reversion (when $a_t<0$).
The function $b_t$ represents the feedback effect on the inflow of the desk's unwind trades. This function is typically chosen to be negative in order to model the behaviour of predatory traders. For example, when the desk unwinds inventory (i.e. $q_t<0$), predatory traders can make profits by first buying the asset in the market, and then closing their position by placing sell orders, which may increase the desk's inventory. We refer to \citet{brunnermeier2005predatory} for a detailed description of this phenomenon.
The function $c$ in \eqref{def:theta} represents the tendency of toxic flow to correlate with the price signal in a way which is adverse to the desk. Finally, the non-negative function $d$ represents the volatility of the toxicity $\theta^{q}$. 

Note that the inflow model specified in \eqref{def:Z} and \eqref{def:theta} is a slight variant of the inflow dynamics in Section 2.1 of \cite{nutz2023unwinding}, which is compatible with a partially observable stochastic control setting and also provides additional flexibility from the modelling perspective. The process $Z^{q}$ in \eqref{def:Z} represents the cumulative aggregated inflow the desk faces. A positive increment of $Z^{q}$ corresponds to ask orders placed by one of the clients. 
In Section 2.1 of \cite{nutz2023unwinding} the inflow dynamics are assumed known to the desk. This can be restrictive in practice, since the data in Section \ref{sec-data} show substantial day-to-day variation in the directional character of client flow.
Moreover, as suggested by \cite{brunnermeier2005predatory}, if a large trader needs to sell due to risk management or other considerations, other traders may react and also sell, and subsequently buy back the asset. This phenomenon, which was not captured in \cite{nutz2023unwinding}, leads to price overshooting and to adverse execution revenues for the desk. We address both of these important issues in the dynamics of the inflow in \eqref{def:Z} and \eqref{def:theta}. 
In order to handle the uncertainty of the parameters of the inflow and the feedback effect of externalisation, we introduce the notion of the partial filtration. 
 
\begin{definition}[Partial filtration]\label{part-f} 
Let $\theta^0$ and $Z^0$ denote the solutions of \eqref{def:theta} and \eqref{def:Z}, respectively, when $q\equiv0$. We define the desk's partial filtration $\mathcal{Y}^0=\{\mathcal{Y}^0_t\}_{t\in[0,T]}$ as the usual augmentation of the natural filtration
\begin{equation}
\label{def:reference-filtration}
    \mathcal{Y}^0_t
    :=\sigma\left(\ol M_s,A_s,Z^0_s:0\leq s\leq t\right),\quad 0\leq t\leq T.
\end{equation}
Proposition \ref{prop:fixed-filtration} below shows that this filtration is generated equivalently by the desk's actual observations and its own trading rate. In particular, the definition does not assume that the desk directly observes the counterfactual inflow $Z^0$, and the toxicity $\theta^q$ remains unobserved. This fixed reference filtration construction adapts the idea in Section 4 of \citet{wonham1968sepearation}, where the component induced by the control is subtracted from the observation to obtain an observation independent of the control that generates the same filtration.
\end{definition} 

The desk's goal is to optimally decide between unwinding or internalising the order flow by controlling its own trading rate $q$ which is selected from the following class of admissible strategies: 
\begin{equation}
\label{def:admissset}
\mathcal A :=\left\{ q \, : \, q \textrm{ is }\{\mathcal{Y}^{0}_{t}\}_{t\in[0,T]}-\textrm{ prog measurable s.t. }  \mathbb E\Big[  \int_0^T q_t^2 dt  \Big] <\infty \right\}.
\end{equation}

For $q\in\mathcal A$, define the control induced shifts
\begin{equation}
\label{eq:control-shifts}
\begin{split}
    \Delta\theta^q_t
    &:=\int_0^t b_s\exp\left(\int_s^t a_rdr\right)q_sds,\\
    \Delta Z^q_t
    &:=\int_0^t\Delta\theta^q_sds,\quad 0\leq t\leq T.
\end{split}
\end{equation}
We further define the adjusted observed inflow
\begin{equation}
\label{def:adjusted-inflow}
\begin{split}
    \overline{Z}^{q}_{t}
    &:=Z^{q}_{t}-\Delta Z^q_t,\quad 0\leq t\leq T.
\end{split}
\end{equation}
We also denote by $\mathcal G^q=\{\mathcal G^q_t\}_{t\in[0,T]}$ the usual augmentation of the natural filtration generated by everything observable, including the desk's own control $(\ol M,A,Z^q,q)$, so that
\begin{equation}
\label{def:observable-filtration}
\begin{split}
    \mathcal G^q_t
    &:=\sigma\left(\ol M_s,A_s,Z^q_s,q_s:0\leq s\leq t\right),\quad 0\leq t\leq T.
\end{split}
\end{equation}

\begin{proposition}
\label{prop:fixed-filtration}
For every $q\in\mathcal A$ we can write 
\begin{equation}
\begin{split}
    \theta^q=\theta^0+\Delta\theta^q,
    \qquad
    Z^q=Z^0+\Delta Z^q,
    \qquad
    \overline{Z}^q=Z^0.
\end{split}
\end{equation} 
Moreover, $\mathcal G^q=\mathcal Y^0$, $Z^q$ is $\mathcal Y^0$-adapted, and $\mathcal A$ is a vector space.
\end{proposition}
We define the desk's cumulative outflow trades as follows, 
\begin{equation} \label{def:X}
Q^{q}_{t} = \int_{0}^{t}q_{s}ds, \quad 0\leq t \leq T. 
\end{equation}
We assume that the desk's trading activity causes price impact on the risky asset's execution price in the sense that their orders are filled at prices
\be \label{def:S}
S^{q}_{t} = P_{t} +  Y^q_t +\frac{1}{2}\epsilon q_{t}, \quad 0 \leq t \leq T, 
\ee
where $\epsilon>0$ and the transient price impact $Y^q$ is formulated according to the Obizhaeva and Wang model \cite{obizhaeva2013dynamics}, 
\begin{equation} \label{def:Y}
    Y^{q}_{t} = \int_{0}^{t}e^{-\beta(t-s)}\lambda q_{s}ds, \quad 0 \leq t \leq T, 
\end{equation}
with constants $\lam,\beta>0$. The term $\frac{1}{2}\epsilon q_{t}$ represents the instantaneous price impact as in \cite{AlmgrenChriss2} and it is also used in \cite{nutz2023unwinding} as a proxy for the trading costs associated with the bid-ask spread.

The desk's inventory during the trading period is given by the sum of the outflow and the inflow, where a negative sign of outflow implies execution, 
\begin{equation} \label{inv} 
    X^{q}_{t} = x + Q^{q}_{t} + Z^{q}_{t}, \quad 0\leq t\leq T. 
\end{equation}
It follows from \eqref{def:admissset}, \eqref{def:X}, and Proposition \ref{prop:fixed-filtration} that $Q^q$ and $X^q$ are $\mathcal Y^0$-adapted.
Here $x\in\mathbb{R}$ denotes the desk’s initial inventory, so $X_{0}^{q}=x$.
The expected execution costs of $q$, conditional on $(P_0,X_0^q)=(p,x)$, are given by
\begin{equation} \label{def:objective}
\begin{split}
    C(q;p,x) &= \mathbb{E}_{p,x}\left[\int_{0}^{T}S^{q}_{t}dQ_{t} - X^{q}_{T}P_{T} + \alpha (X^{q}_{T})^{2}\right]  \\
    &= \mathbb{E}_{p,x}\left[\int_{0}^{T}\left(P_{t}q_{t} + Y^{q}_{t}q_{t} + \frac{1}{2}\epsilon q^{2}_{t}\right)dt - X^{q}_{T}P_{T} + \alpha (X^{q}_{T})^{2}\right].
\end{split}
\end{equation}
The first term on the right-hand side of~\eqref{def:objective} represents the desk's terminal costs from trading; that is, the final cash position including the trading costs which are induced by the spread and the transient price impact as prescribed in~\eqref{def:S}. The term $X^{q}_{T}P_{T}$ represents the end of the day book value of the desk's position.
The term $\alpha (X^{q}_{T})^{2}$ with a constant $\alpha > 0$ imposes a cost for any discrepancies between the inflow and the outflow at the end of the day, hence it provides an incentive for the desk to unwind the order flow.  We therefore wish to find a trading rate $q \in \mathcal{A}$ such that 
\begin{equation} \label{def:value} 
    V(p,x) = \inf_{q \in \mathcal{A}}  C(q;p,x). 
\end{equation}
\textbf{The desk's observables:}  We assume that the model parameters $\beta, \lambda, \epsilon, \sigma, \alpha, \theta_0$ (which are constants) and $a, b, c, d$ (which are deterministic functions) in \eqref{def:Z}, \eqref{def:theta}, \eqref{def:S} and \eqref{def:Y} are known to the desk as they are estimated before the trading period. 
Moreover, the order flow $Z^{q}$ in \eqref{def:Z}, the execution price $S$ in \eqref{def:S} and the price signal $A$ are observed by the desk. Since the trading rate $q$ is also an observable, this means that the transient price impact $Y^{q}$ and the local martingale component $\overline{M}$ of the price are observables as well (see \eqref{ass:P} and \eqref{def:S}). However, realisations of the order flow's drift $\theta^{q}_{t}$ in \eqref{def:theta}  along with the Brownian motions $W^{Z}$ and $W^{\theta}$ are unobserved by the desk.

 \begin{remark}
This notion of partial information is the focus of our work and provides the main difference from the model studied in \cite{nutz2023unwinding}, in which the inflow $Z$ is uncontrolled by the desk and it takes the form,
$
    dZ_{t} = -\theta_{t}Z_{t}dt + \sigma_{t}dW_{t}
$
where $\theta$ and $\sigma$ are known deterministic functions. In this work we assume that the toxicity $\theta$ is unknown and needs to be inferred by the desk in a cost-effective manner, while taking into account the feedback effect of the desk's executed trades on the inflow (see \eqref{def:Z} and \eqref{def:theta}). 
\end{remark}

\subsection{A note on partially observable stochastic control} \label{subsec:posc note}

One of the main challenges in the theory of partially observable stochastic control is that the control must be progressively measurable with respect to the observable filtration, but since the observation process depends on the control, the stochastic control problem is ill-posed. We refer to Section 9.1 of \cite{bensoussan2018estimation} for an explanation of this so-called `chicken and egg' problem. In the case of linear-quadratic partially observable stochastic control problems, this can be resolved by introducing an additional limitation on the admissibility of the control, also explained in Section 9.1 of \cite{bensoussan2018estimation}. In our case, the control is additionally required to be adapted to the partial filtration $\mathcal Y^0$ in Definition \ref{part-f}, which is generated by $(\ol M,A)$ and the process $Z^0_t := \int_{0}^{t} \theta^0_sds + \sigma W^{Z}_{t}$, where $Z^0$ and $\theta^0$ denote the inflow and toxicity processes in \eqref{def:Z} and \eqref{def:theta} when $q\equiv 0$, i.e. in the case of no outflow trades. This does not assume that the desk directly observes the counterfactual inflow $Z^0$. Indeed, the adjusted inflow $\overline Z^q$ in \eqref{def:adjusted-inflow} is constructed from the observed inflow and the desk's own past trades, and Proposition \ref{prop:fixed-filtration} shows that $\overline Z^q=Z^0$. Hence $\mathcal Y^0$ contains only information which the desk can construct in real time, while the toxicity remains unobserved.

In our case after imposing this admissibility condition, we tackle the problem by using an idea introduced in \cite{sun2023lqposc}, whereby we first fix a control $q\in\mathcal{A}$, then derive the so-called \textit{filtering processes}. Proposition \ref{prop:fixed-filtration} also shows that $\mathcal A$ is a vector space, so the perturbations and convex combinations used in the variational argument remain admissible. By deriving the dynamics of the filtering processes (see Section \ref{sec-filt-res}) we manage to transform the partially observable stochastic control problem \eqref{def:value} into a fully observable problem and to derive its unique solution in Section \ref{sec-res-cont}.

In the following we provide additional details about the method. Let $(L_t)_{t\geq0}$ be a square-integrable stochastic process on $(\Omega, \mathcal{F},(\mathcal {F}_{t})_{t\in[0,T]},\mathbb{P} )$.  Recall that the conditional expectation of $L_t$ with respect to the partial sigma-algebra $\mathcal{Y}^{0}_{t}$ is the orthogonal projection onto the Hilbert space of square-integrable $\mathcal{Y}^{0}_{t}$-measurable random variables. We denote this projection $\hat{L}_{t}=\mathbb{E}\left[L_{t}|\mathcal{Y}^{0}_{t}\right]$. Since the inventory $X^{q}$ in \eqref{inv} is $\mathcal Y^0$-adapted by \eqref{def:X}, \eqref{inv}, and Proposition \ref{prop:fixed-filtration}, we can write its projection as $X^{q}_t=\hat{X}^{q}_t$, and apply this to \eqref{def:objective} to get
\begin{equation} \label{eq:cost Xhat}
    \mathcal{C}(q;p,x) = \mathbb{E}_{p,x} 
     \left[\int_{0}^{T}\left(P_{t}q_{t}+Y^{q}_{t}q_{t}+\frac{1}{2}\epsilon q^{2}_{t}\right)dt - P_{T}\hat{X}^{q}_{T}+\alpha (\hat{X}^{q}_{T})^{2}\right].
\end{equation}
Our goal in the upcoming section is to derive an SDE satisfied by $\hat{X}^{q}$, in terms of the projection of $\theta^q$ in \eqref{def:theta} to $\mathcal{Y}^{0}_{t}$ (i.e. $\hat{\theta}^{q}$). Having obtained the dynamics of the filtering process $\hat{X}^{q}$ in Corollary \ref{thm-filt}, we use a variational approach in order to derive the optimal unwind strategy $q^*$ in feedback form, which minimises the cost functional in Theorem \ref{thm:optimal control}.  

\section{Main Results}  \label{sec:main-results}
In this section we present our main results. We start with the results related to the filtering process in Section \ref{sec-filt-res}. Using these results, we derive the unique optimal strategy for the partially observable stochastic control problem \eqref{def:value} in Section \ref{sec-res-cont}. 

\subsection{The Filtering Processes} \label{sec-filt-res}
We define the joint inventory-toxicity process, 
\be \label{x-vec} 
 \mathbb{X}^{q}_{t} = 
       ( \theta^{q}_{t}, 
       X^{q}_{t})^{\top}, \quad t \geq 0. 
\ee
From  \eqref{def:Z}, \eqref{def:theta} and \eqref{inv} it follows that $ \mathbb{X}^{q}$ satisfies, 
\begin{equation} \label{eq:X SDE}
   d \mathbb{X}^{q}_{t}    = \left(A_{t}\mathbb{X}^{q}_{t} + B_{t}q_{t}\right)dt + C_{t}dA_{t} + D_{t}dW^{\theta}_{t} + E_{t}dW^{Z}_{t},
\end{equation}
where
\begin{equation} \label{eq:Y SDE}
\begin{split}
    &\mathbf{A}_{t} = 
    \begin{pmatrix}
        a_{t} & 0 \\
        1 & 0 & 
    \end{pmatrix}
    , \qquad
    \mathbf{B}_{t} = 
    \begin{pmatrix}
        b_{t} \\
        1 
    \end{pmatrix}
    , \qquad
    \mathbf{C}_{t} = 
    \begin{pmatrix}
        c_{t} \\
        0 
    \end{pmatrix}
    , \\ \qquad
    &\mathbf{D}_{t} = 
    \begin{pmatrix}
        d_{t} \\
        0
    \end{pmatrix},
    \qquad
    \mathbf{E}_{t} = 
    \begin{pmatrix}
        0 \\
        \sigma
    \end{pmatrix}
    .
\end{split}
\end{equation}
Recall that the partial filtration $\mathcal{Y}^0$ was defined in Definition \ref{part-f}. We further define 
\begin{equation}  \label{filt-x} 
    \hat{\mathbb{X}}^{q}_{t} = \mathbb{E}\left[\mathbb{X}^{q}_{t}|\mathcal{Y}^{0}_{t}\right]
    \qquad \text{and} \qquad 
    \wt{\mathbb{X}}^{q}_{t} = 
     \mathbb{X}^{q}_{t} - \mathbb{E}\left[\mathbb{X}^{q}_{t}|\mathcal{Y}^{0}_{t}\right], \quad t \geq 0,
\end{equation}
where we often use the following notation for the projected coordinates of  $\hat{\mathbb{X}}^{q}$: 
\be \label{x-th-hat} 
\hat {X}^{q}_t= \mathbb{E}\left[ {X}^{q}_{t}|\mathcal{Y}^{0}_{t}\right], \quad 
\hat {\theta}^{q}_t = \mathbb{E}\left[\theta^{q}_{t} |\mathcal{Y}^{0}_{t}\right], \quad t\geq 0.  
\ee
Since $\Delta\theta^q$ in \eqref{eq:control-shifts} is $\mathcal Y^0$-adapted,
\begin{equation}
\label{eq:filter-shift}
\begin{split}
    \hat\theta^q_t
    &=\hat\theta^0_t+\Delta\theta^q_t,\\
    \hat X^q_t
    &=X^q_t,\\
    \wt{\mathbb X}^q_t
    &=(\theta^0_t-\hat\theta^0_t,0)^\top,\quad 0\leq t\leq T.
\end{split}
\end{equation}
The inflow $Z^{q}$ in \eqref{def:Z}, which is often referred to as the `observation process', can then be written as, 
\begin{equation}\label{z-trans}
    dZ^{q}_{t}
    =  
    \e_1^{\top}
    \mathbb{X}^{q}_{t}dt + 
    \sigma
    dW^{Z}_{t},
\end{equation}
where 
\be \label{e1}
\e_1 = (1,0)^\top.
\ee
We also define the following covariance function,
\begin{equation} \label{cov} 
    \Sigma_{t} = \mathbb{E}\left[(\tilde{\mathbb{X}}^{q}_{t})(\tilde{\mathbb{X}}^{q}_{t})^{\top}\right], \quad t\geq 0,
\end{equation}
By \eqref{eq:filter-shift}, $\Sigma$ is independent of $q$
and we define the innovation process, 
\begin{equation} \label{inov-proc} 
    V_{t} := Z^{q}_{t} - \int_{0}^{t}\e_1^{\top}\hat{\mathbb{X}}^{q}_{s}ds
    =Z^0_t-\int_0^t\hat\theta^0_sds, \quad t\geq 0.
\end{equation}
The second equality follows from \eqref{eq:control-shifts} and \eqref{eq:filter-shift}. In particular, the innovation process $V$ is the same for every $q\in\mathcal A$.
\begin{theorem} \label{thm:filtering processes}
For any $q\in \mathcal A$ the process $\hat{\mathbb{X}}^{q}$ satisfies the following equation, 
\begin{equation} \label{eq:Xarrowhat SDE}
    d\hat{\mathbb{X}}^{q}_{t} = \left(\Sigma_{t}\e_1+\sigma \mathbf{E}_{t}\right)\sigma^{-2}dV_{t} + \left(\mathbf{A}_{t}\hat{\mathbb{X}}^{q}_{t}+\mathbf{B}_{t} q_{t}\right)dt +\mathbf{C}_{t}dA_{t},
\end{equation}
where $\Sigma$ in \eqref{cov} is given by the unique positive semi-definite solution to the Riccati equation, 
\begin{equation} \label{eq:Sigma ODE}
    \frac{d\Sigma_{t}}{dt} = \left(\mathbf{A}_{t}-\sigma^{-1} \mathbf{E}_{t} \e_1^\top \right)\Sigma_{t} + \Sigma_{t}\left(\mathbf{A}_{t}-\sigma^{-1} \mathbf{E}_{t} \e_1^\top\right)^{\top} - \sigma^{-2}\Sigma_{t} \e_1\e_1^\top\Sigma_{t} + \mathbf{D}_{t}\mathbf{D}_{t}^{\top},
\end{equation}
for $t> 0$ with $\Sigma_{0}=0$. 
\end{theorem}
The following corollary provides a convenient form for handling $ \hat{X}^{q}$ and $\hat{\theta}^{q}$ separately.  
\begin{corollary} \label{thm-filt} 
Let  $(\kappa^{(1)}_{t}, \kappa^{(2)}_{t})^\top = \sigma^{-2}\Sigma_{t}\e_1$. For any $q\in \mathcal A$, the process $\hat{X}^{q}$ satisfies,
    \begin{equation} \label{eq:Xhat SDE}
        \hat{X}^q_{t} =    x + \int_{0}^{t} \left(\kappa^{(2)}_{s} + 1\right)dV_{s} + \int_{0}^{t} \left(\hat{\theta}^{q}_{s} + q_{s} \right)ds, \quad t \geq 0, 
    \end{equation}
    and $ \hat{\theta}^q$ satisfies, 
   \begin{equation} \label{eq:thetahat}     \begin{split}
        \hat{\theta}^q_{t} 
        =& \theta_{0}\exp\left(\int_{0}^{t}a_{r}dr\right) + \int_{0}^{t}\exp\left(\int_{s}^{t}a_{r}dr\right)\kappa^{(1)}_{s}dV_{s} \\
        &+ \int_{0}^{t}\exp\left(\int_{s}^{t}a_{r}dr\right)b_{s}q_{s}ds +\int_{0}^{t}\exp\left(\int_{s}^{t}a_{r}dr\right)c_{s}dA_{s}, \quad t \geq 0. 
    \end{split}
    \end{equation}
\end{corollary}
The proofs of Theorem \ref{thm:filtering processes} and Corollary \ref{thm-filt} are given in Appendix \ref{sec:filtering proof}. 

\subsection{Optimal Control with Partial Information} \label{sec-res-cont} 

Thanks to Theorem \ref{thm:filtering processes} we can transform the partially observable stochastic control problem \eqref{def:value} to a fully observable stochastic control problem \eqref{eq:cost Xhat}, by including the filtering processes $\hat{X}^{q}$ and $\hat{\theta}^{q}$ which are adapted to the partial filtration $\mathcal Y^0$. Next we use a variational approach in order to solve \eqref{def:value}. 

Before we state our main results, we introduce some further definitions and notation. Recall that the functions $a_t,b_t$ were defined in \eqref{def:theta}. We define the deterministic function $r_{t}$ as follows, 
\begin{equation}
    r_t
    =
    b'_t
    \int_t^T
    \exp\left(\int_t^s a_r\,dr\right)ds
    -b_t,\quad 0\leq t \leq T. 
\end{equation} 
Moreover, we define the following matrix valued process $(L_{t})_{t\in[0,T]}$, which is based on the model parameters and functions introduced in Section \ref{sec:model},
\begin{equation} \label{def:L}
    \mathbf{L}_{t} := 
    \begin{pmatrix}
        0 & 1 & 0 & 1 & 0 & 0 & 0 & 0 \\
        0 & a_{t} & 0 & b_{t} & 0 & 0 & 0 & 0 \\
        0 & 0 & -\beta & \lambda & 0 & 0 & 0 & 0 \\
        0 & 0 & \frac{\beta}{\epsilon} & 0 & -\frac{\beta}{\epsilon} & \frac{a_{t}}{\epsilon} & -\frac{r_{t}}{\epsilon} & 0 \\
        0 & 0 & 0 & -\lambda & \beta & 0 & 0 & 0 \\
        0 & 0 & 0 & 0 & 0 & -a_{t} & r_{t} & 0 \\
        0 & 0 & 0 & 0 & 0 & 0 & 0 & 0 \\
        0 & 0 & 0 & 0 & 0 & 0 & 0 & 0  
    \end{pmatrix}. 
\end{equation}
Then consider the matrix valued process $(\Phi_{t})_{t\in[0,T]}$, given by the solution to the system, 
\begin{equation} \label{def:Phi}
    d\Phi_{t} = \mathbf{L}_{t}\Phi_{t}dt, \quad 0\leq t \leq T, \quad \Phi_{0}=I.
\end{equation}

\begin{proposition}\label{prop:Phi}
The system \eqref{def:Phi} has a unique absolutely continuous
solution $\Phi$ on $[0,T]$, satisfying \eqref{def:Phi} almost everywhere. Moreover, $\Phi_t$ is invertible for every
$t\in[0,T]$, and
\begin{equation}\label{eq:Phi_bound}
\begin{split}
    \sup_{0\leq t\leq T}\|\Phi_t^{-1}\|<\infty.
\end{split}
\end{equation}
\end{proposition}
The proof of Proposition~\ref{prop:Phi} is given in
Appendix~\ref{sec:control proof}.

We define 
\begin{equation} \label{s-mat} 
    S(t) := \Phi_{T}\Phi^{-1}_{t}, \quad  0\leq t \leq T. 
\end{equation}
We further define the following vector functions: 
 \begin{align}
    G(t) &:=
    \begin{pmatrix}
        \frac{2\alpha}{\epsilon} & 0 & \frac{1}{\epsilon} & 1 & 0 & 0 & 0 & 0 
    \end{pmatrix}
    S(t),  \label{eq:first terminal} \\
    H(t) &:= 
    \begin{pmatrix}
        0 & 0 & 0 & 0 & 1 & 0 & 0 & 0
    \end{pmatrix}
    S(t) ,  \label{eq:second terminal} \\ 
    I(t) &:= 
    \begin{pmatrix}
        0 & 0 & 0 & 0 & 0 & 1 & 0 & 0
    \end{pmatrix}
    S(t)  ,  \label{eq:third terminal}\\
    J(t) &:= 
    \begin{pmatrix}
        -2\alpha & 0 & 0 & 0 & 0 & 0 & 1 & 1
    \end{pmatrix}
    S(t).   \label{eq:fourth terminal}
\end{align}
We impose some further technical 
assumption on the elements of these vectors.
\begin{assumption} \label{assum:nonzeros}   
    We assume that $a_{t}, b_{t}, \epsilon, \beta, \lambda $ and $\alpha$ are chosen such that: 
    \begin{itemize} 
\item[\textbf{(i)}]
    \begin{equation} \label{eq:lower bound assumptions}  
    \begin{split}
        &\underset{0 \leq t \leq T}{\inf}G_{4}(t)>0, \qquad \underset{0 \leq t \leq T}{\inf}H_{5}(t)>0, \\
        &\underset{0 \leq t \leq T}{\inf}I_{6}(t)>0, \qquad \underset{0 \leq t \leq T}{\inf}J_{7}(t)>0,
    \end{split}
    \end{equation}
\item[\textbf{(ii)}]
    \begin{equation} \label{eq:I5,I6 assumptions}  
       \inf_t|I_6(t)H_5(t)-I_5(t)H_6(t)| >0,
    \end{equation}
\end{itemize} 
\end{assumption}
We will also need the following assumption on the functions $(h^{q},i^{q}, j^{q}, \tilde{h}^{R}, \tilde{i}^{R})$ which contain products, divisions and additions of the fundamental  functions $(G,H,I,J)$. Since their expressions are lengthy, they appear in Appendix \ref{sec:full representation of control}. 
\begin{assumption} \label{ass2}
We assume that
   \begin{equation} 
       \inf_{t\in [0,T]} \left(\left| 1+\frac{G_{5}(t)}{G_{4}(t)}h^{q}(t) + \frac{G_{6}(t)}{G_{4}(t)}i^{q}(t) + \frac{G_{7}(t)}{G_{4}(t)}j^{q}(t) \right| \wedge \left|1 + \frac{J_{5}(t)}{J_{7}(t)}\tilde{h}^{R}(t) + \frac{J_{6}(t)}{J_{7}(t)}\tilde{i}^{R}(t) \right| \right)>0,
    \end{equation}
\end{assumption} 
Note that Assumptions \ref{assum:nonzeros} and \ref{ass2} can be easily verified numerically. Indeed they were satisfied in all of our numerical experiments in Section  \ref{sec:illustration}. 

In the following theorem we derive the unique minimiser $q^*$ of the cost functional \eqref{def:objective} in terms of the filtered toxicity and inventory $\hat \theta^q$ and $\hat X^q$ from Corollary \ref{thm-filt}, along with the observed price distortion $Y$, the signal $A$ and the price process $P$. For the sake of readability, the expressions for the deterministic functions $g^{X}, g^{\theta}, g^{Y}, g^{P}$ and $g^{A}$ that appear in the statement of Theorem \ref{thm:optimal control} are given in Appendix \ref{sec:full representation of control}, in terms of the functions $G, H, I$ and $J$. 

\begin{theorem} \label{thm:optimal control}
Assume that Assumptions \ref{assum:nonzeros} and \ref{ass2} hold. Then, there exists a unique minimiser $q^*\in \mathcal A$ of the cost functional \eqref{def:objective}. Moreover $q^*$ satisfies, 
   \begin{equation} \label{eq:optimal q}
     q^{*}_{t} = g^{X}(t)\hat{X}^{q*}_{t} + g^{\theta}(t)\hat{\theta}^{q*}_{t} + g^{Y}(t)Y^{q*}_{t} + g^{P}(t)P_{t} + \mathbb{E}\left[\int_{t}^{T}g^{A}(s,t)dA_{s}\Big\vert\mathcal{Y}^{0}_{t}\right], 
    \end{equation}
    for $0\leq t \leq T$ where $g^{X}, g^{\theta}, g^{Y}, g^{P}$ and $g^{A}$ are deterministic functions defined in \eqref{def:g}. 
\end{theorem}
The proof of Theorem \ref{thm:optimal control} is given in Appendix  \ref{sec:control proof}.
As in many practical cases, the signal $A$ has the form of an integrated Ornstein-Uhlenbeck process. Some prominent examples of such signals are the limit order book imbalance signal, studied in \cite{lehalle2019signals}, or pairs trading signals, where two assets with similar features are traded together. The difference between the weighted returns of these assets can be treated as a trading signal and approximated by an Ornstein-Uhlenbeck process (see \cite{avellaneda2010statarb}). In these cases and others, it is assumed that the signal $A$ in \eqref{ass:P} satisfies, 
\begin{equation} \label{spec-a} 
    A_{t} = \int_{0}^{t}U_{s}ds, \quad t \geq 0, 
\end{equation}
where $U$ is an Ornstein–Uhlenbeck process, 
\begin{equation} \label{spec-u}
 dU_{t}=-\kappa U_{t}dt + \ell dW^{U}_{t},  \quad t>0, \quad U_{0}=\nu.
\end{equation}
Here $\kappa,\ell>0$ and $\nu \in \mathbb R$ are constants and $W^{U}$ is a Brownian motion independent of $W^Z, W^{\theta}$ and $\overline{M}$.

The following corollary, which derives the optimal trading rate when $A$ is given by \eqref{spec-a}, follows directly from Theorem \ref{thm:optimal control} by an application of integration by parts. 
\begin{corollary} \label{corol-ou}
Let $A$ be as in \eqref{spec-a}. Then under the same assumptions as in Theorem \ref{thm:optimal control} the unique minimiser $q^*\in \mathcal A$ of the cost functional \eqref{def:objective} satisfies, 
 \begin{equation}
    q^{*}_{t} = g^{X}(t)\hat{X}^{q*}_{t} + g^{\theta}(t)\hat{\theta}^{q*}_{t} + g^{Y}(t)Y^{q*}_{t} + g^{P}(t)P_{t} + g^{U}(t)U_{t}, \quad 0\leq t \leq T, 
\end{equation}
where
\begin{equation}
    g^{U}(t) = \int_{t}^{T}g^{A}(s,t)e^{-\kappa(s-t)}ds, \quad 0\leq t \leq T.
\end{equation}
\end{corollary}
 
\begin{remark} 
Theorem \ref{thm:optimal control} extends Theorem 2.14 in \cite{nutz2023unwinding} in the following directions: 

(i) We extend a variant of the model presented in \cite{nutz2023unwinding} to the case where the drift of the inflow $Z^q$ in  \eqref{def:Z}  is unknown to the desk, as this is often the case in the markets. In our framework the statistical properties of the toxicity $\theta^q$ in \eqref{def:theta}, such as mean and volatility are estimated from historical data. In order to solve this partially observable stochastic control problem \eqref{def:value},  we derive the dynamics of the projected toxicity and inventory in Corollary \ref{thm-filt}. We then substitute them into the original stochastic control problem as described in \eqref{eq:cost Xhat}, in order to derive an optimal solution to \eqref{def:value} on the partial filtration $\mathcal Y^0$.  

(ii) In our model we include the feedback of the desk's trades on the inflow process (see \eqref{def:Z} and \eqref{def:theta}). This feedback is another important feature in realistic trading models which was not taken into account in \cite{nutz2023unwinding} and is not well studied in the literature on order execution. From the technical perspective, the feedback makes the raw observation depend on the trading rate. Proposition \ref{prop:fixed-filtration} shows that the partial filtration $\mathcal Y^0$ coincides with the filtration generated by $(\ol M,A,Z^q,q)$. The proof then departs from \citet{sun2023lqposc} at the martingale representation step. Since $\mathcal Y^0$ also contains the independent price information $\ol M,A$, it is not generated by the order flow observation alone, and the Fujisaki, Kallianpur and Kunita theorem cannot be applied directly. We therefore first isolate the smaller filtration containing only the order flow information and apply their argument there.

(iii) In contrast to \cite{nutz2023unwinding} where only martingale prices are considered, our fundamental price $P$ also includes a general finite variation drift (see $A$ in \eqref{ass:P}), which is often called an alpha. Alphas, which are short term price predictors, are an important ingredient in portfolio choice and execution strategies, as argued in \citep{A-N-T24,GARLEANU16,lehalle2019signals,neuman2022optimalsignaltrading,webster2023handbook} among others. From the trading perspective, the addition of alpha signals to the model yields a new term in the optimal trading rate \eqref{eq:optimal q} which includes the signal's expected future values. We demonstrate the effects of this component on the model in Section \ref{sec:illustration}. 
\end{remark} 

\section{Numerical Implementation } 
\label{sec:illustration}

This section is structured as follows: In Section \ref{sec-data} we provide empirical evidence for the order flow model in \eqref{def:Z} and \eqref{def:theta}. Section \ref{sec:numerial results} details a numerical implementation of the optimal trading strategy from Corollary \ref{corol-ou} and compares its performance to the full information case. In Section \ref{sec-risk} we illustrate how externalisation helps to mitigate inventory risk. 
Section \ref{sec-naive} is dedicated to the study of the performance gap between a naive strategy that doesn't take into account the feedback effect of its own trades, and the optimal strategy from Corollary \ref{corol-ou}. Finally, Section \ref{sec:toxicity-pnl} investigates the effect on P\&L of the adversity of toxic flow, i.e. the parameter $c$.

\subsection{Empirical Evidence} 
\label{sec-data}

\begin{figure}[H]
     \centering
\begin{overpic}[scale=0.4]{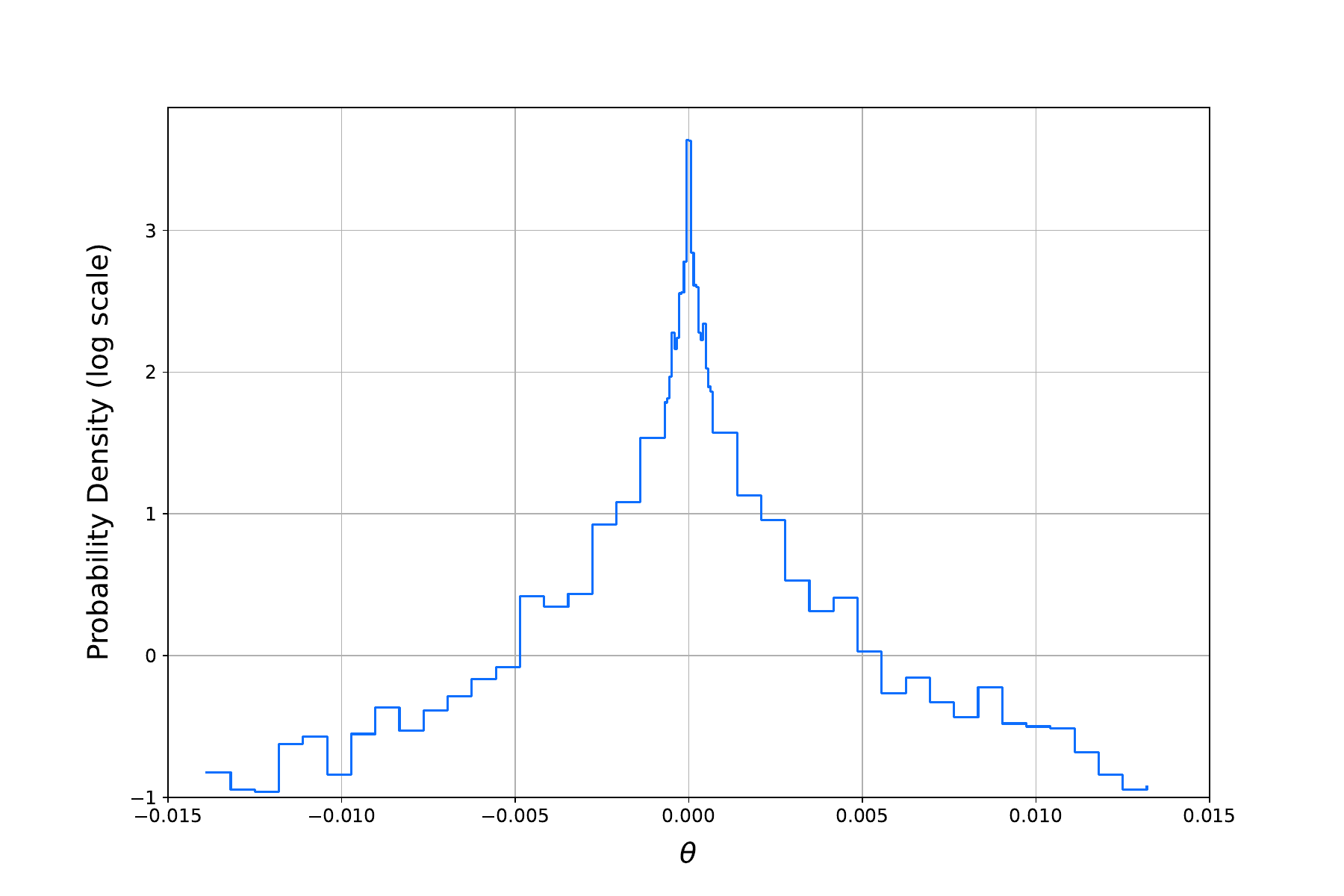}
\put(48,2.5){\colorbox{white}{\parbox{0.02\linewidth}{%
    ${ \scriptstyle
      \Delta Z_t}
    $}}}
\end{overpic}
         \caption{Histogram of one-minute increments of the GBPUSD flow from September 2023 to June 2024.} \label{fig:gbpusd_flow_pdf.pdf}
\end{figure}

We consider a subset of FX spot and outright trade flow in a single currency pair, GBPUSD, from September 2023 to June 2024 as experienced by
HSBC eFX desk.\footnote{This sample is sufficiently diverse to provide realistic results but by no means complete to fully represent HSBC FX market making franchise.} The time bins of the data series are of 1 minute duration where high frequency flow is summed up within each bin. Only  the flow with low pricing sensitivity was included (e.g. retail flow) in order to mimic the CRB setting.
Figure~\ref{fig:gbpusd_flow_pdf.pdf} shows the histogram of one-minute flow increments. Traded volume is normalised by average daily volume (ADV) and time is measured in days. The increment distribution is approximately symmetric with a tiny bias (the mean of $\theta$ is 0.02) and visibly heavy tails. The estimated one-day normalised flow volatility is $\hat \sigma = 0.052$.

\begin{figure}[H]
     \centering
     \includegraphics[scale=0.4]{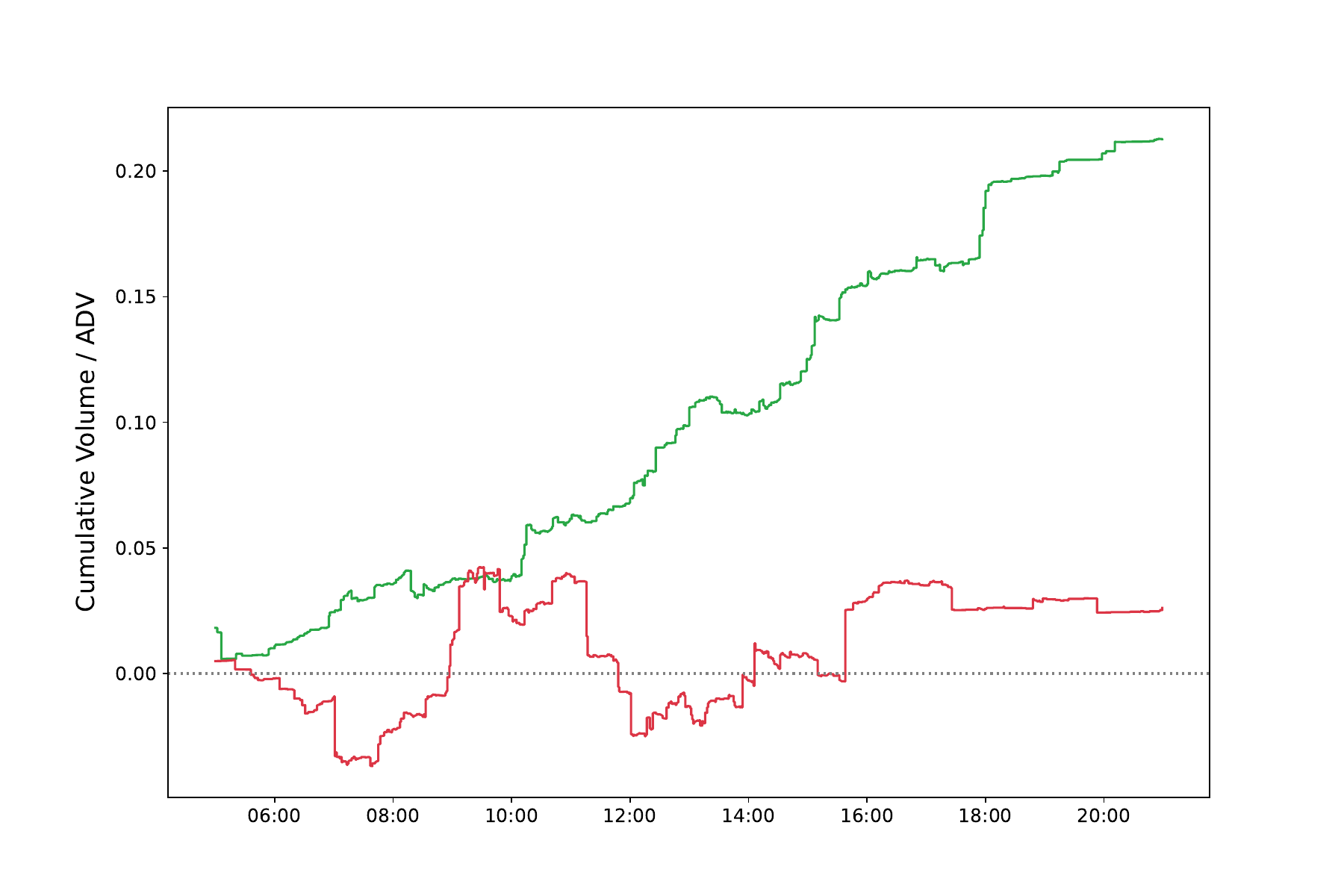}
     \caption{Two realisations of GBPUSD order flow: a near-martingale day (red) and a day with pronounced positive directionality (green).}
    \label{fig:gbpusd_cumulative_intraday_flow.pdf}
\end{figure}

\begin{figure}[H]
     \centering
    \includegraphics[scale=0.4]{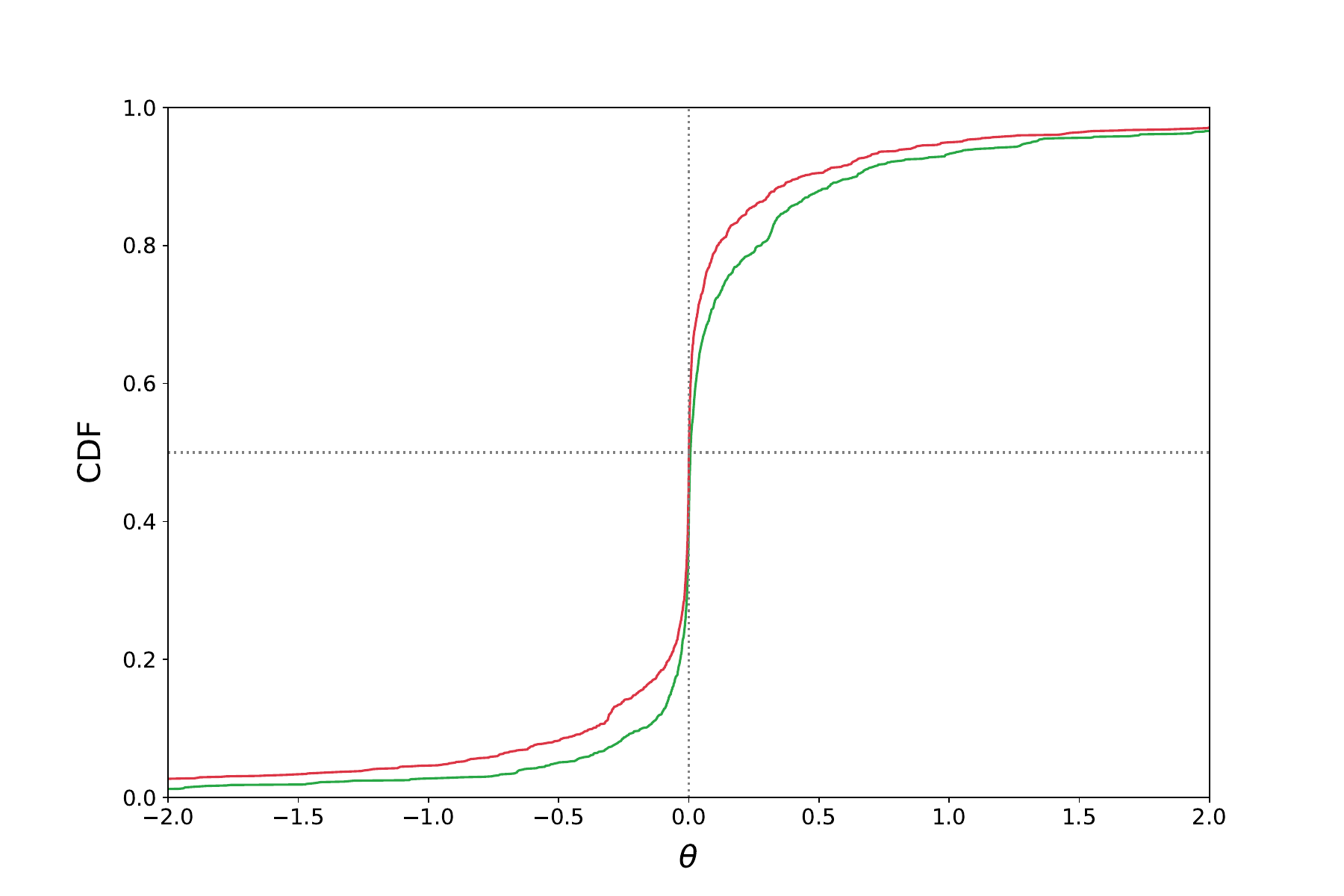}
    \label{fig:gbpusd_intraday_flow.pdf}
    \caption{CDF of $\hat \theta$ for the same days as in Figure \ref{fig:gbpusd_cumulative_intraday_flow.pdf}.}
    \label{fig:gbpusd_intraday_flow_cdf}
\end{figure}

In Figure \ref{fig:gbpusd_cumulative_intraday_flow.pdf} we plot the order flow for two particular days. 
The estimated flow
directionality differs substantially across these cases, with $\hat{\theta} = 0$ (red) and $0.2$ (green), corresponding to a near-martingale day and a day with pronounced positive directionality, respectively.
In Figure \ref{fig:gbpusd_intraday_flow_cdf} we plot the cumulative distribution function (CDF) of $\hat \theta$ for the same days. The lower branch of the curve corresponds to negative theta and the upper branch to positive theta.  

We also compare the desk's hedging (externalisation) flow with discrete-time changes in the one-minute client-flow increments and find a correlation coefficient of $-0.25$. Because hedging is endogenous and residual pricing sensitivity cannot be fully removed, we interpret this as reduced-form evidence consistent with a negative feedback coefficient $b$ in \eqref{def:theta}, rather than as a direct estimator of $b$.

\subsection{Numerical results} 
\label{sec:numerial results}

In this section, we investigate numerical simulations of the trading period under various parameter choices. In the first scenario we consider the case where the toxicity $\theta^{q}$ in \eqref{def:theta}  exhibits reversion, which makes filtering more challenging, due to the low \emph{drift to noise ratio}. In the second scenario, the toxicity exhibits momentum. This means the toxicity follows a more predictable pattern and thus filtering leads to better estimates. The third scenario concerns a shorter trading period and it also includes a price predictive signal (alpha).
Finally, we consider the full-information benchmark in which the desk observes the true inflow drift.

\paragraph{Scenario 1: mean-reverting toxicity.}
The first scenario represents our benchmark. The desk has one trading day ($T=1$), and starts the day with no initial inventory. The initial values for $X,\theta,P,$ are as shown in Table \ref{table:initial_conditions}. By definition, $Z_0=Y_0=0$.
\begin{table}[H]
\centering
\begin{tabular}{c c c c} 
 \hline
 $x$ & $\theta_{0}$ & $p$  \\ [0.2ex] 
 \hline 
 0 & 0.1 & 1.34  \\ 
 \hline
\end{tabular}
\caption{Initial conditions used for all scenarios}
\label{table:initial_conditions}
\end{table}
We model the local martingale component $\overline{M}$ in \eqref{ass:P} of the unaffected price $P$ as a Brownian motion
\begin{equation} \label{spe-M} 
    \overline{M}_{t} = \sigma_{M}W^{M}_{t}.
\end{equation}
We choose $\sigma_{M}=6.3\cdot10^{-3}$, which corresponds to 10\% annualised. We set the signal to be $A=0$ for now. The volatility of inflow in \eqref{def:Z} is taken to be $\sigma=0.1$, which is conservatively rounded up following the analysis in \ref{sec-data}. The coefficients in the dynamics of $\theta^{q}$ are then shown in Table \ref{table:coefficients}. The price impact parameters in \eqref{def:S} and \eqref{def:Y} are chosen to represent those of a liquid market and are also given in Table \ref{table:coefficients}. 
They are consistent with values reported by \cite{nutz2023unwinding} and internal desk estimations.
\begin{table}[H]
\centering
\begin{tabular}{c c c c c c c} 
 \hline
 $a$ & $b$ & $c$ & $d$ & $\epsilon$ & $\beta$ & $\lambda$ \\ [0.2ex] 
 \hline 
 -0.4 & -0.25 & -10 & 0.01  & 1.0 & 10 & 0.1 \\ 
 \hline
\end{tabular}
\caption{Coefficients in $\theta^{q}$ and price impact parameters}
\label{table:coefficients}
\end{table}
The parameter $a$ represents momentum in the drift of the inflow, so that for $a<0$ on average toxicity will decrease from its positive initial value as the day progresses. The parameter $b$ represents the market's response to the outflow trades. As discussed in Section \ref{sec:model}, this is chosen to be negative to model the behaviour of predatory traders.  Finally, the penalisation parameter in \eqref{def:objective} is set fairly high to $\alpha=100$ to strongly encourage the desk to unwind all of their inventory at the end of the trading period. 

\begin{figure}[H]
     \centering
     \includegraphics[scale=0.5]{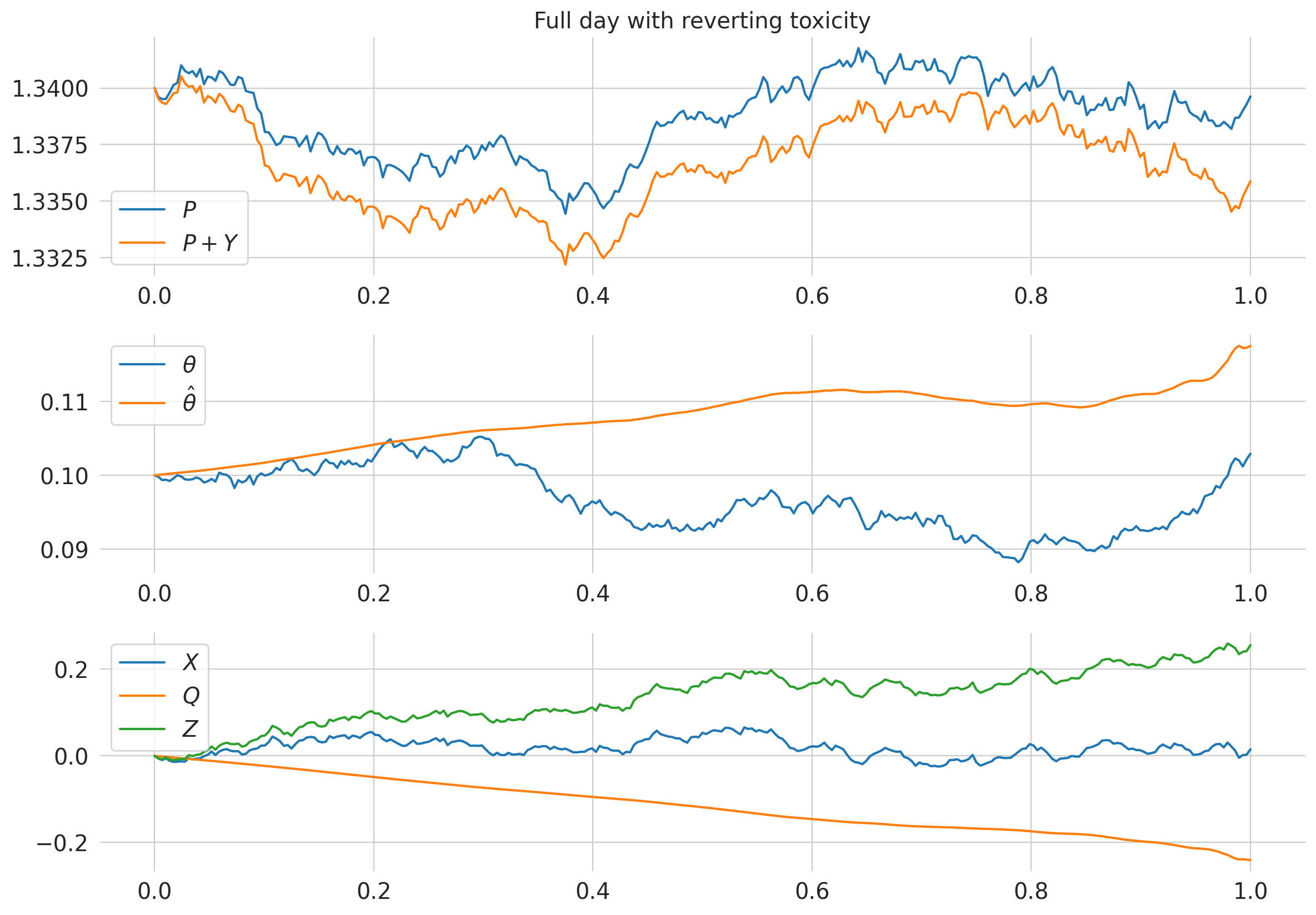}
     \caption{A simulation of the model's state variables at optimality in the case of mean-reverting toxicity. The trading P\&L in this simulation is 0.3042.}
    \label{fig:reversion}
\end{figure}

We simulate the model outputs, including the optimal trading rate by using Corollary \ref{corol-ou}. We notice that, due to the positive $\theta^{q}$ throughout the simulation, the cumulative inflow $Z^{q}$ consistently rises, with fluctuations driven by the noise term $W^{Z}$. Over the day, the desk progressively sells this inflow, leading cumulative $Q^{q}$ to decline by the end of the day. The result is that the desk manages to maintain inventory $X^{q}$ close to zero across the entire day. All of these processes are shown on the lower panel of Figure \ref{fig:reversion}.

The desk also continuously considers transient price impact $Y^{q}$. Towards the end of the day, the desk decides that avoiding terminal penalty is more cost-effective than managing transient price impact, and thus the magnitude of $Y^{q}$  rises rapidly. This is shown in the upper panel, via the difference between the price with transient price impact, and the unaffected price $P$.

This scenario is a marginally more challenging environment for filtering than Scenario 2, where toxicity exhibits momentum. Without a clear trend for the toxicity $\theta^{q}$, the true value is harder to estimate. Nonetheless, the filter $\hat \theta^{q}$ still performs well, as can be seen on the middle panel of Figure \ref{fig:reversion}. Due to calibration of the weights on the innovation process done in Appendix \ref{sec:filtering proof} (and presented in Theorem \ref{thm:filtering processes}), the estimate is kept close in magnitude to the true value. The terminal inventory $X^{q}_{T}$ is therefore still very close to zero as seen on the third panel of Figure \ref{fig:reversion}. 

In what follows, we refer to trading P\&L of a single realisation of the model,  which is defined as the negative of the trading cost (without the risk aversion term)
\begin{equation} \label{def:trading p&l}
    P\&L_\text{trading}^{q} = P_{T}X^{q}_{T} - \int_{0}^{T}S_{t}dQ^{q}_{t}.
\end{equation}

\begin{figure}[H]
     \centering
     \includegraphics[scale=0.5]{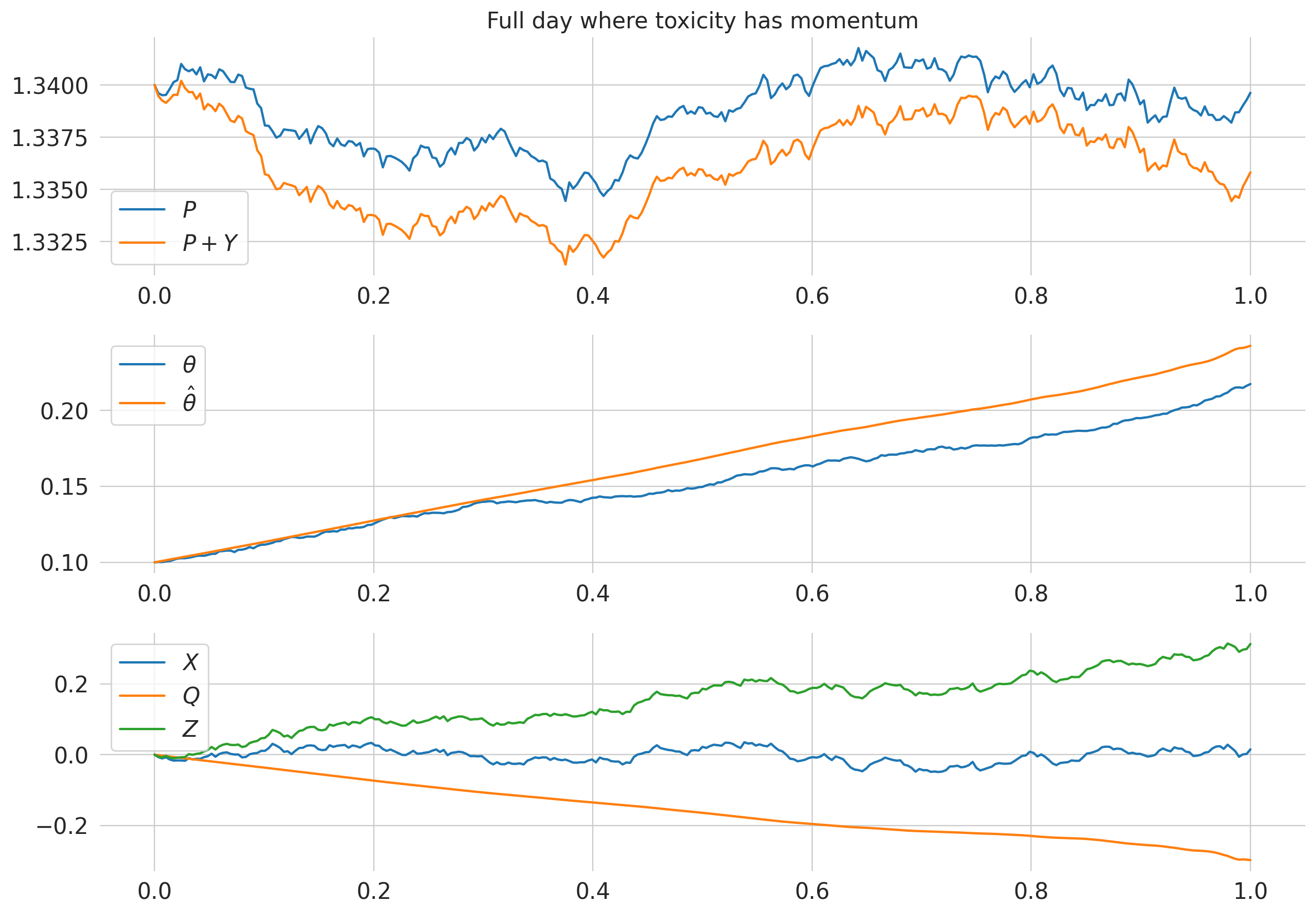}
     \caption{A simulation of the model where toxicity exhibits momentum. The trading P\&L in this simulation is 0.3643.}
    \label{fig:momentum}
\end{figure}

\paragraph{Scenario 2: momentum toxicity.}
In the second scenario, all parameters from the previous scenario are unchanged except the value of $a$ which is now set to $0.4$. 
The toxicity $\theta^{q}$ now has substantial momentum, and increases in magnitude over the day more consistently than in the reversion case. This pattern is much easier to filter for, due to the higher drift to noise ratio of $\theta^{q}$ as seen on the middle panel of Figure \ref{fig:momentum}.
The cumulative inflow $Z^{q}$ is now larger than the value in the previous scenario, which gives the desk more inventory to sell. At first this may seem a good outcome, but this can sometimes result in more price impact and instantaneous costs being paid.

\paragraph{Scenario 3: incorporating alphas.}
In contrast to the previous two scenarios considered, we now look at a desk that only trades for a fraction of the whole day ($T=0.04$). The most notable change is that we assume the desk has access to a finite-variation price predicting signal over this window. The signal $A$ is expressed as the integral of an Ornstein-Uhlenbeck process, as described in \eqref{spec-a} and \eqref{spec-u}. The parameters appearing in \eqref{spec-u} are $\kappa=1$, $\ell=1$ and $\nu=0$. All other parameters are identical to those of Scenario 1. Due to the relatively small volume traded, the transient price impact $Y^{q}$ is much smaller than in the case with toxicity reversion over an entire day. 

\begin{figure}[H]
     \centering
     \includegraphics[scale=0.5]{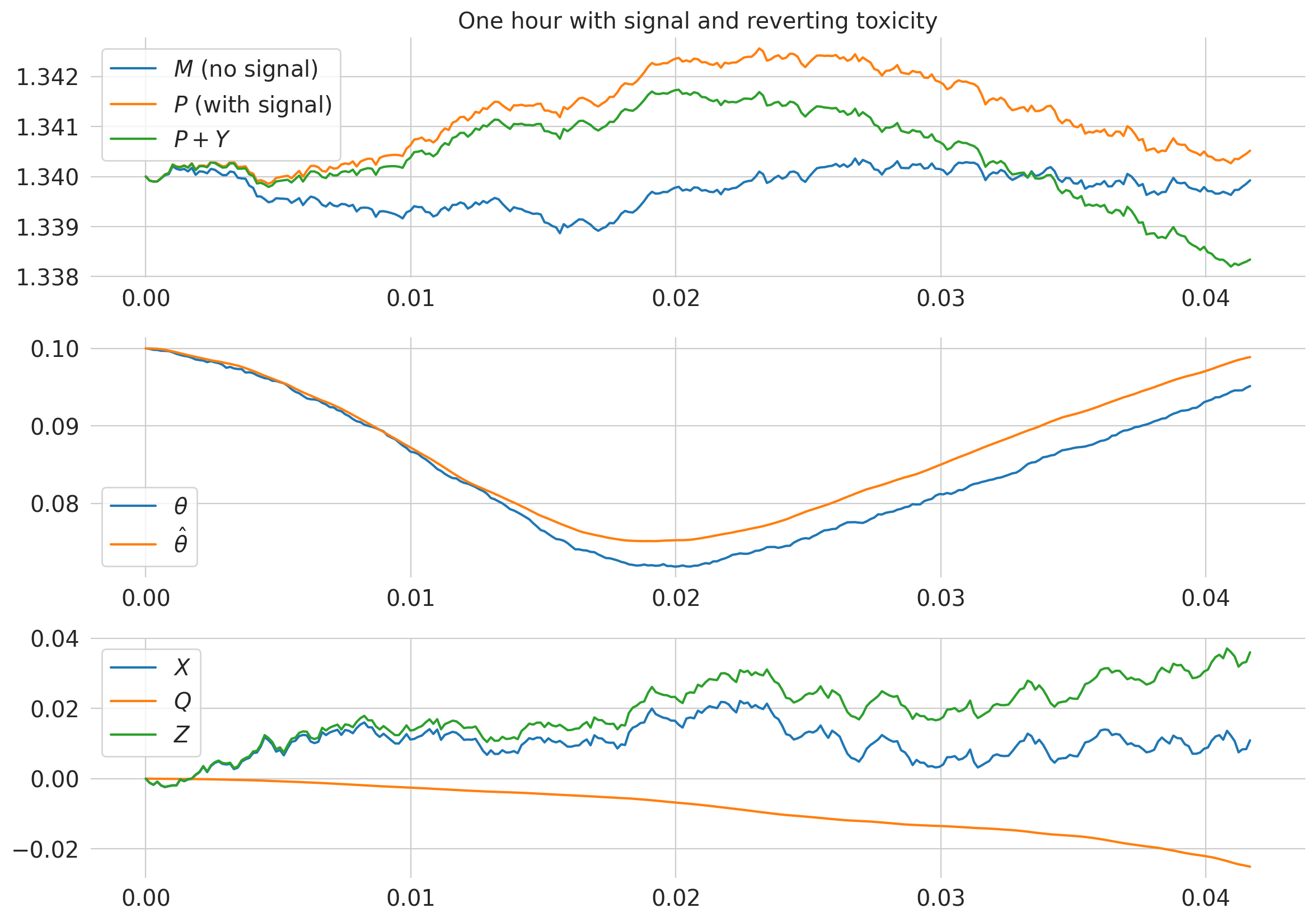}
     \caption{A simulation of the model over a short time horizon with a signal. The trading P\&L of this simulation is 0.0373. }
    \label{fig:short}
\end{figure}

\paragraph{Comparison to the full information case.}

We compare the partially observable model to the case where the desk has access to the true value of the toxicity $\theta^{q}$ and does not need to filter. Specifically, the problem is unchanged from the partially observable stochastic control problem described in Section \ref{sec:model}, except for the class of admissible controls, which is now given by
\be \label{def:admissset full info} 
\mathcal A :=\left\{ q \, : \, q \textrm{ is }\{\mathcal{F}_{t}\}_{t\in[0,T]}-\textrm{ prog measurable s.t. }  \mathbb E\Big[  \int_0^T q_t^2 dt  \Big] <\infty \right\}.
\ee
In this case, there is no need to obtain filtering processes. The optimal control is of the same form. 
This naturally leads us to the corresponding result to Theorem \ref{thm:optimal control}.
\begin{theorem} \label{thm:optimal control full information}
    Let assumptions analogous to \ref{assum:nonzeros} and \ref{ass2} be satisfied. Then, the optimal control $q^{*}$ in the full information case is given by
    \begin{equation} \label{eq:optimal q1}
     q^{*}_{t} = g^{X}(t)X^{q*}_{t} + g^{\theta}(t)\theta^{q*}_{t} + g^{Y}(t)Y^{q*}_{t} + g^{P}(t)P_{t} + \mathbb{E}\left[\int_{t}^{T}g^{A}(s,t)dA_{s}\Big\vert\mathcal{F}_{t}\right],
    \end{equation}
    where $g^{X}, g^{\theta}, g^{Y}, g^{P}$ and $g^{A}$ are the same deterministic functions as in Theorem \ref{thm:optimal control}, and are given in Appendix \ref{sec:full representation of control}. 
\end{theorem}
 
We first plot the optimal model outputs which follow from \eqref{eq:optimal q1}, using model parameters identical to those of Scenario 1. We see that, due to the strong performance of the filter in the partially observable case shown in Scenario 1, the performance and trading decisions made are almost identical. 

\begin{figure}[H]
     \centering
     \includegraphics[width=14cm]{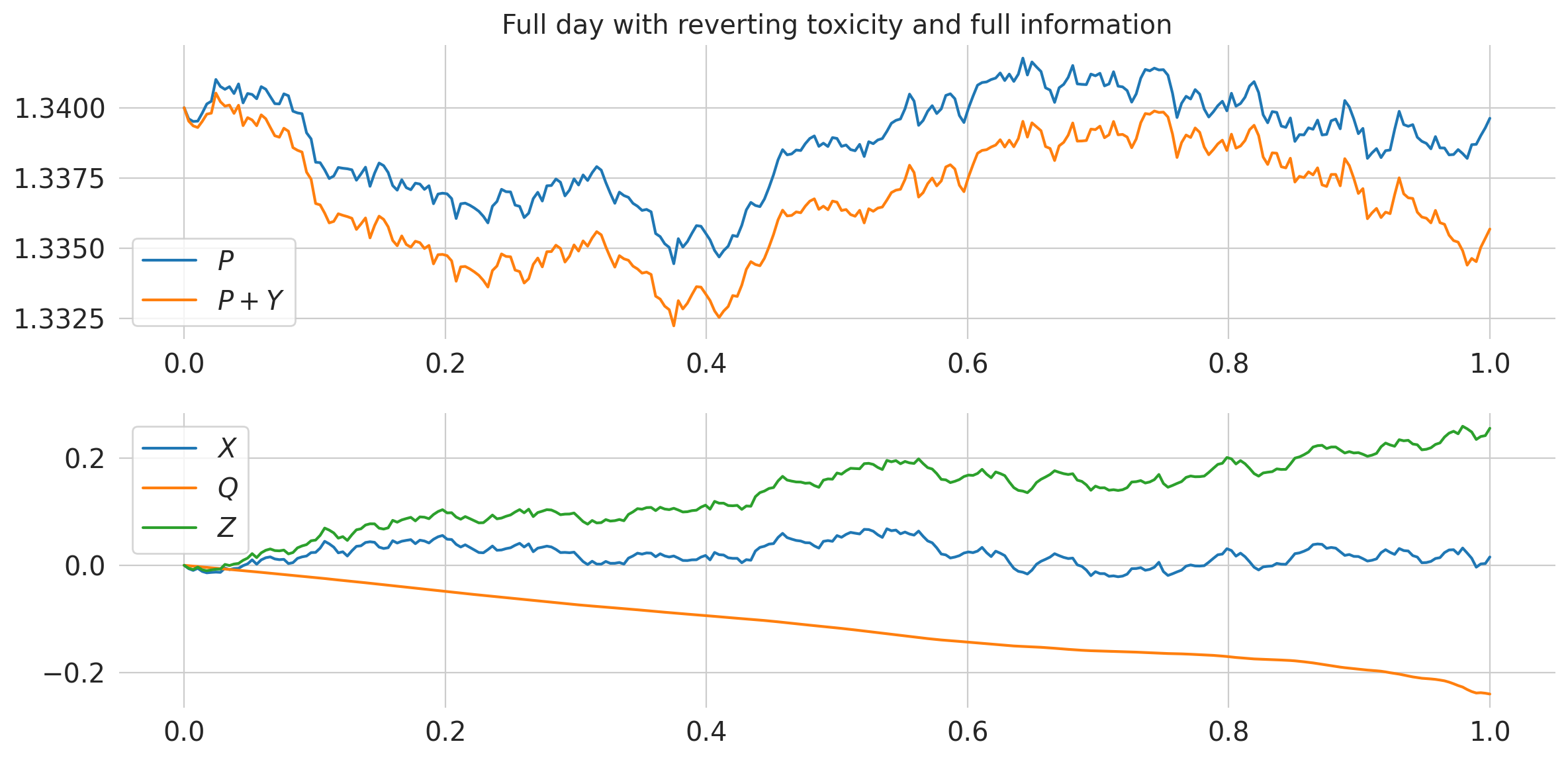}
     \caption{A simulation of the model where toxicity reverts and the desk has access to the true value of $\theta$. The trading P\&L in this simulation is 0.3030, which is very close to the trading P\&L for the simulation shown in Figure \ref{fig:reversion}.}
    \label{fig:full info reversion}
\end{figure}

The performance of the optimal strategy under each of the Scenarios 1-3 is shown in Table \ref{table:strategy performance}. The Monte Carlo estimate of the cost functional, with 10,000 simulations used for each, is shown on the right-hand side. One takeaway is that the case where toxicity $\theta^{q}$ exhibits momentum (and subsequently the cumulative inflow $Z^{q}$ is of larger magnitude) has a lower cost functional than the case where toxicity reverts to zero. The other main takeaway from Table \ref{table:strategy performance} is that the performance of the desk who does not know the true value of $\theta$ is very close to that of the desk with full information.
The estimated costs agree to four decimal places in every scenario. This is a consequence of the strong performance of filtering in our model.

\begin{table}[H]
\centering
\begin{tabular}{c || c | c} 
 \hline
 Scenario & Partial information & Full information \\  
 \hline 
 \hline
 Toxicity reverts
  & -0.09456 & -0.09459 \\ 
 \hline
 Toxicity has momentum
  & -0.16015 & -0.16020 \\ 
 \hline
 Short period with signal
  & 0.00582 & 0.00582 \\ 
 \hline
\end{tabular}
\caption{Monte Carlo estimates, to five decimal places, of the cost functional $\mathcal{C}$ in \eqref{def:objective} using 10,000 simulations for each scenario, where $\theta$ is unknown and known.}
\label{table:strategy performance}
\end{table}
\subsection{Externalisation for mitigating risk} \label{sec-risk}

\begin{figure}[H]
     \centering
     \includegraphics[width=14cm]{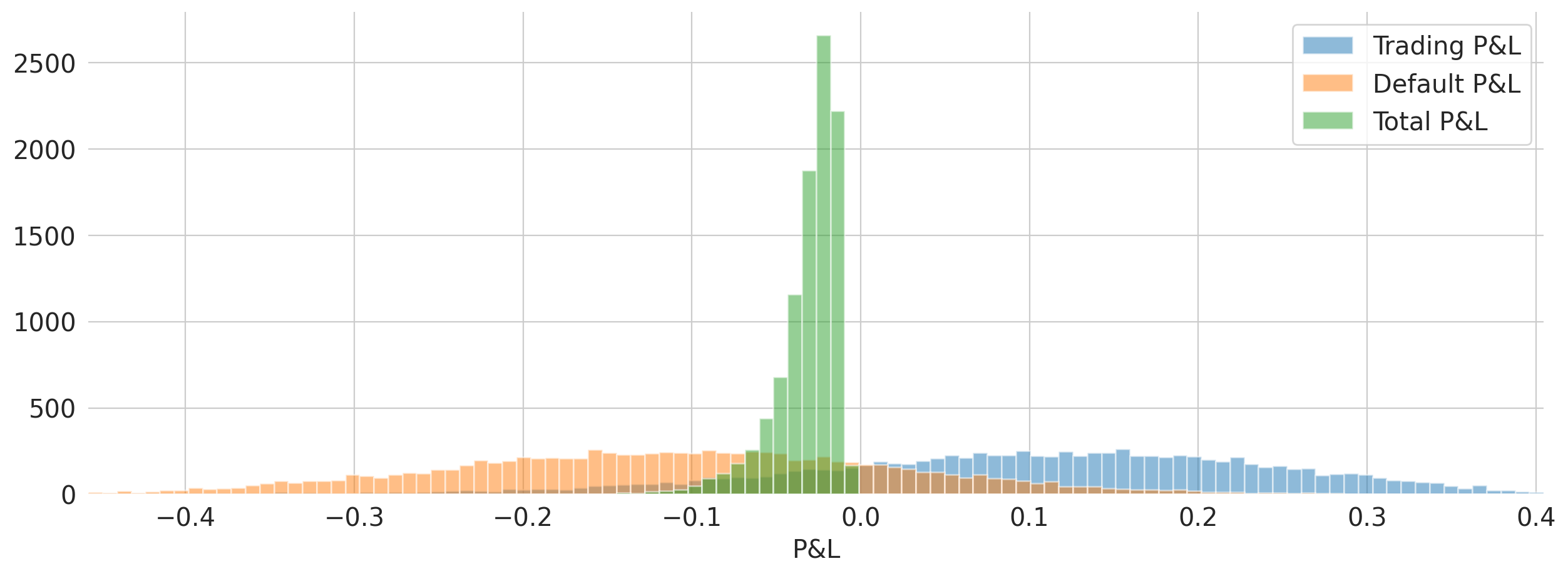}
     \caption{Probability distribution of total P\&L, trading P\&L and default P\&L without trading for the parameters in Scenario 1.}
    \label{fig:p&l hist}
\end{figure}

Stochastic inflow carries the following P\&L exposure for the book 
\begin{equation}  
    P\&L^{0} = -\int_{0}^{T} P_{t}\,dZ^{0}_{t},
\end{equation}
which is associated with the risk of holding inventory. 
The desk can reduce this risk by trading, at the expense of price impact costs.
For the strategy $q$, the corresponding total P\&L uses the actual controlled inflow $Z^q$ as follows
\begin{equation}  
    P\&L_\text{total}^{q} = P\&L_\text{trading}^{q} - \int_{0}^{T} \left(P_{t}+Y^{q}_{t}\right)\,dZ^{q}_{t}.
\end{equation}
The optimisation criterion in \eqref{def:objective} consists of expected trading costs together with a terminal inventory penalty. In our study, we have minimised the trading costs with an additional risk aversion penalty.
Figure \ref{fig:p&l hist} compares the distributions of P\&L components for the parameter choices of Scenario 1 obtained by Monte Carlo simulation (10,000 trajectories).
The total P\&L can be seen to have a significantly lower variance than P\&L without trading,
demonstrating the suitability of externalisation for managing the overall risk for the book.

\subsection{A naive desk and model misspecification} \label{sec-naive}
\begin{figure}[H]
     \centering
     \includegraphics[width=14cm]{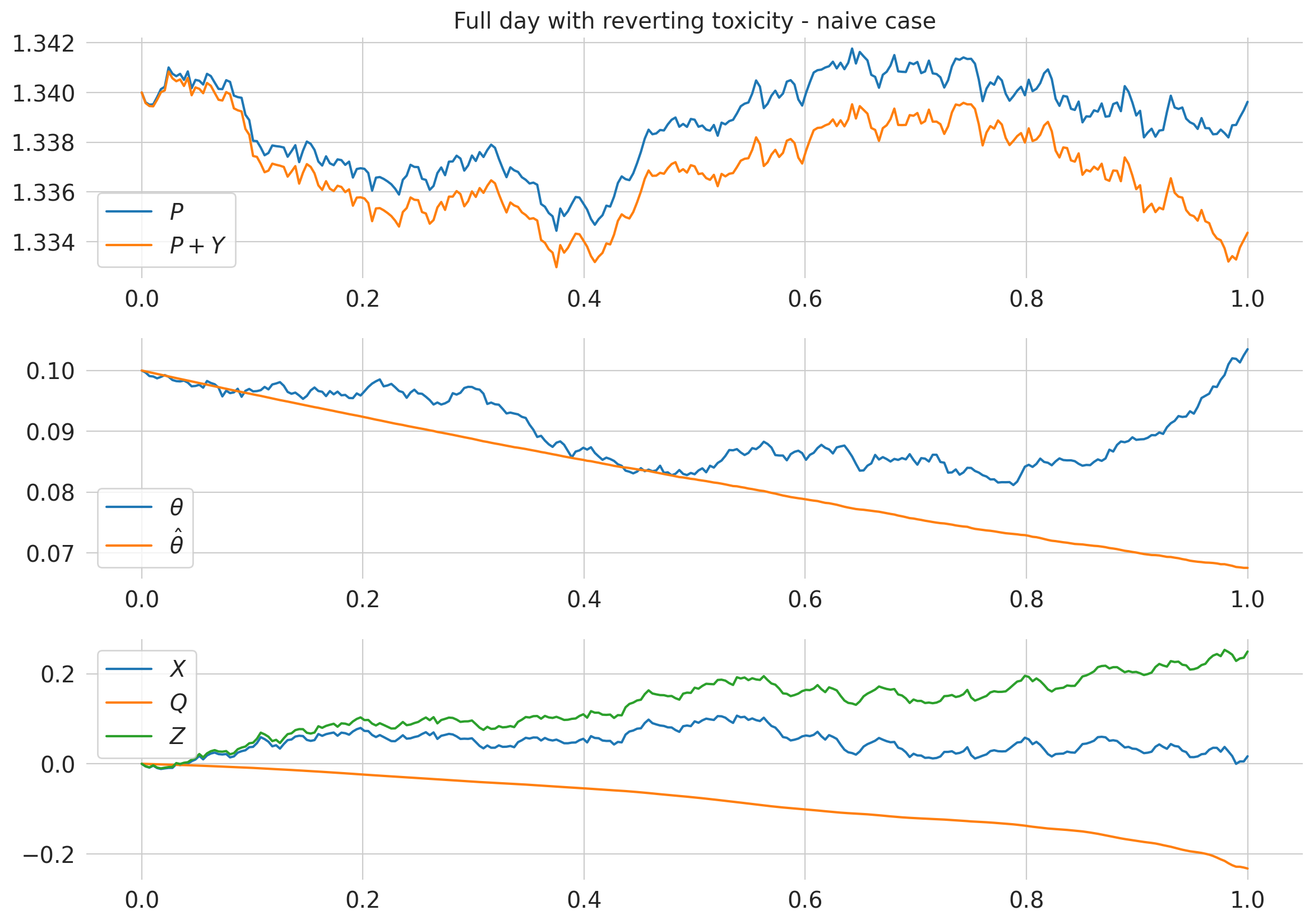}
     \caption{A simulation of the model for a naive desk who believes $b=0$ when in fact $b=-0.25$. The trading P\&L for this simulation is 0.2898, which is $5\%$ lower than in Scenario 1.}
    \label{fig:naive}
\end{figure}

Lastly, we consider a desk that believes $b=0$, while the true value is $b=-0.25$ in this case. The naive desk believes that there is no market response to outflow trades and thus no ability to influence inflow. This naive desk then filters and trades accordingly. All other parameters are the same as those of the case where toxicity $\theta^{q}$ reverts to zero, described in Scenario 1 of Section \ref{sec:numerial results}. This leads to worse estimates of $\theta^{q}$ than if they had used the correct value of $b$. The poor performance of the filter is clear in the second panel of Figure \ref{fig:naive}, especially when compared to the other scenarios in Section \ref{sec:numerial results}. The naive desk in this simulation achieves a trading P\&L of 0.2898, which is lower than in Scenario 1, which uses the same parameters but considers a desk that knows the correct value of $b$. This can partially be explained by the worse performance of the filter, especially at the end of the trading period, where fast unwinding leads to a change in $\theta$ which the naive desk is unaware of. 

We also compare the performance of the naive desk with one who knows the true value of $b$, for example from calibration prior to the start of the day. Table \ref{tab:naive_b_tests} shows the Monte Carlo estimate of the cost functional found using 2,000 simulations in both the case of the naive desk and the optimal desk. We show negative values of $b$ which represent the `predatory trading' effect discussed in Section \ref{sec:model}, but positive values of $b$ are also possible providing Assumptions \ref{assum:nonzeros} and \ref{ass2} are satisfied. 

We see that both the naive and optimal desk's performance is better when $b$ is of larger magnitude. The desk then has the ability to influence the inventory $X^{q}$ more strongly via $Z^{q}$, allowing for more flexibility when controlling transient price impact $Y^{q}$. One example is that the desk has been selling (as is usually the case when $\theta_{0}>0$) and thus transient price impact $Y^{q}$ has become negative. The desk can then buy, worsening the long inventory position, to offset the transient price impact. In the case where $b$ is far from zero, the increased inventory from buying will be partially offset by the market feedback, while the price impact will still be reduced in magnitude. Such a trading decision may lead to higher trading costs later if $b$ is close to zero, due to the need to unwind the acquired inventory. Note that even a naive desk may decide to make such a trade due to the necessity of controlling transient price impact and the possibility of inventory internalisation due to inflow. We also observe, as expected, that the optimal desk consistently achieves a slightly lower cost functional.

\begin{figure}[H]
     \centering
     \includegraphics[width=14cm]{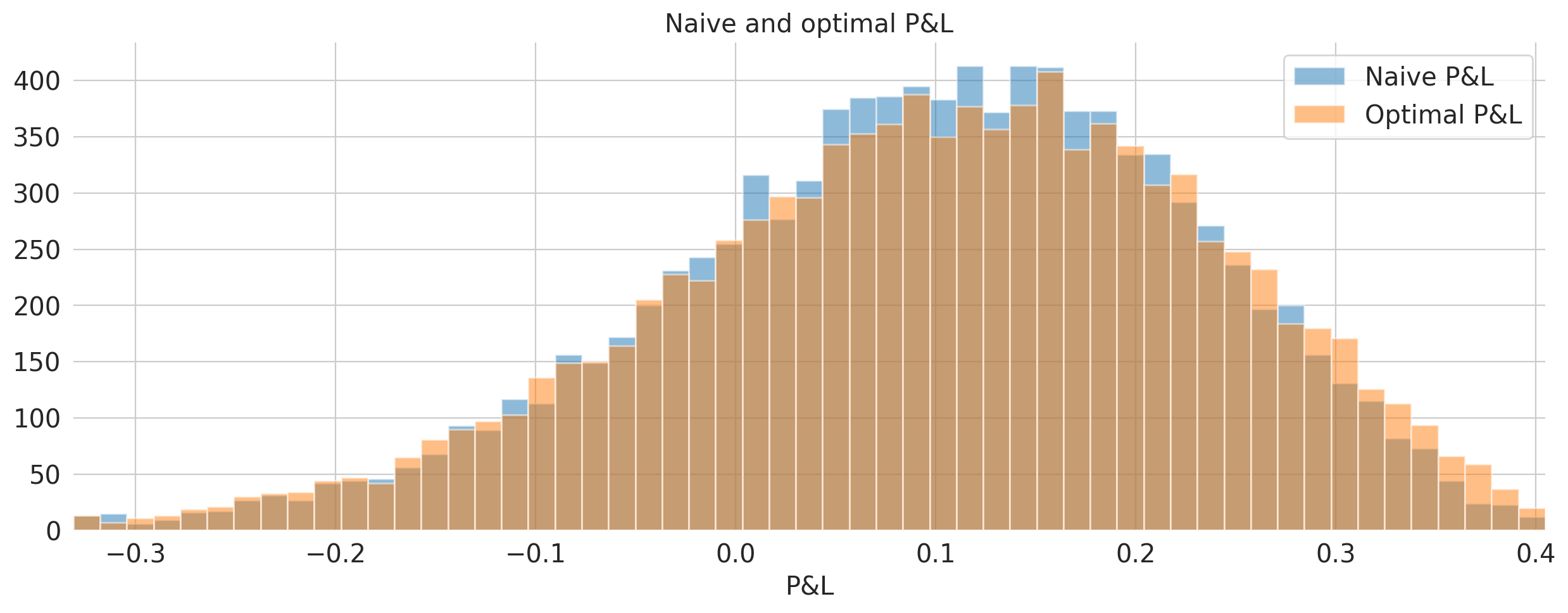}
     \caption{Histogram of trading P\&L for the naive desk (blue) and the optimal desk (orange) for the true value of $b=-0.25$.}
    \label{fig:naive hist}
\end{figure}

We also present in Figure \ref{fig:naive hist} a histogram of the trading P\&L for the case where $b=-0.25$, but a naive desk believes $b=0$. We see that while the optimal strategy slightly outperforms the naive strategy, this difference is modest. The value of knowing $b$ may therefore be limited, which suggests that the model is somewhat robust to parameter misspecification.

\begin{table}[H]
    \centering
    \begin{tabular}{c||c|c}
         True $b$ & Naive cost MC & Optimal cost MC\\
         \hline \hline
         -0.1 & -0.0834 & -0.0840 \\
         \hline
         -0.2 & -0.0876 & -0.0903 \\
         \hline
         -0.3 & -0.0918 & -0.0985 \\
         \hline
         -0.4 & -0.0961 & -0.1091   
         \end{tabular}
    \caption{The sensitivity of the model to the market's trading response, and the impact of misspecification. In each case, the naive desk believes $b=0$ and filters and unwinds accordingly, while the optimal desk filters and unwinds using the correct value of $b$.}
    \label{tab:naive_b_tests}
\end{table}

\subsection{Effect of toxicity on P\&L} \label{sec:toxicity-pnl}

To isolate the effect of toxicity on performance, we repeat the short horizon experiment with the signal from Scenario 3 while varying the constant sensitivity parameter $c$. Recall that $c$ determines how the toxicity responds to increments in the price signal $A$ in \eqref{def:theta}. Negative values of $c$ produce the adverse relationship described in Section 2, whereby movements in the signal induce changes in the conditional mean order flow that are unfavourable to the desk. Increasing the magnitude of $c$ therefore strengthens the adverse selection component of the incoming flow and makes the resulting inventory more costly to manage.

\begin{figure}[H]
     \centering
     \includegraphics[width=12cm]{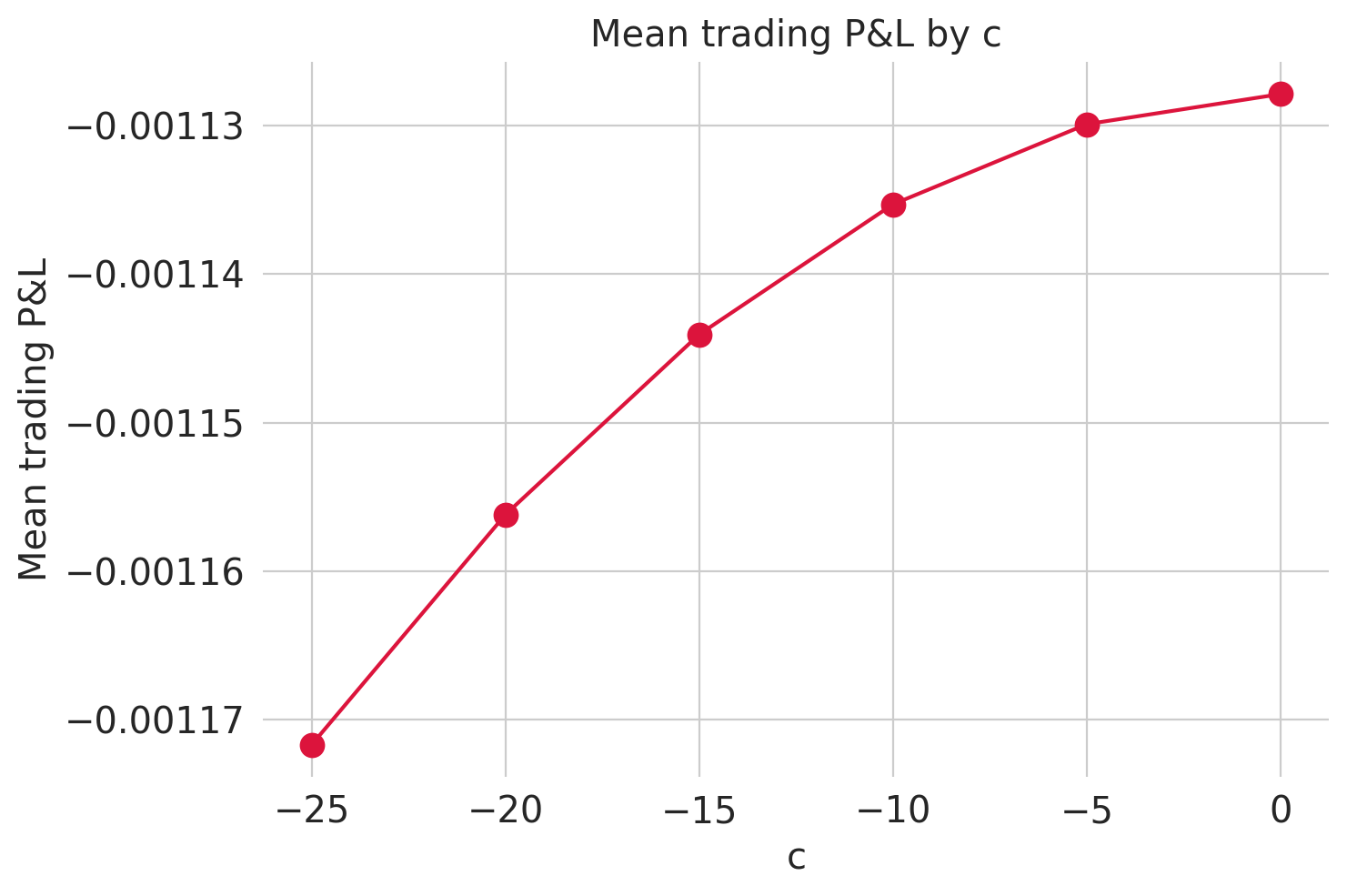}
     \caption{Mean trading P\&L as the constant sensitivity of toxicity to the signal $c$ varies.}
    \label{fig:vary c}
\end{figure}

This effect can be seen in Figure \ref{fig:vary c}, which shows that mean trading P\&L increases monotonically with $c$. The fact that mean P\&L remains negative when $c=0$ reflects the baseline costs of externalisation, including instantaneous and transient price impact, while the difference across values of $c$ isolates the additional cost generated by toxic flow.

\section{Conclusion}  \label{sec:conclusion} 
In this paper we have considered a model for a central trading desk which aggregates the inflow of clients' orders with unobserved toxicity. The desk then faces a continuous dilemma of whether to internalise their inventory and save on transaction costs and market impact but bear the risk of adverse selection,
or to externalise in a cost-effective manner and reduce risk. 
Institutional FX data illustrate substantial day-to-day variation in flow directionality and provide reduced-form evidence consistent with a negative externalisation feedback.

The desk’s typical objective is to maximise the risk-adjusted daily P\&L. We have formulated the aforementioned setting as a partially observable stochastic control problem and solved it in two steps. First, we derived the filtered dynamics of the inventory and toxicity, projected to the partial filtration, which turns the stochastic control problem into a fully observed problem. Then we used a variational approach in order to derive the unique optimal trading strategy. We have illustrated our results for various scenarios in which the desk faces momentum and mean-reverting toxicity.  We have shown that feedback of externalisation has a prominent effect on the P\&L, even for small values of the feedback parameter.  We then studied the P\&L performance gap between the partially observable problem and the full information case and found the regret to be of order $0.1\%$ for all tested market scenarios. Incorporating continuous assessment of flow toxicity into optimal control in real time offers a dynamic solution to the internalisation-externalisation dilemma.

\section*{Acknowledgments} 
We are grateful to the EPSRC Centre for Doctoral Training in Mathematics of Random \mbox{Systems}: Analysis, Modelling and Simulation for supporting this research. The views expressed are those of the authors and do not necessarily reflect the views or the practices at HSBC.

\begin{appendices}

\section{Proof of Theorem \ref{thm:filtering processes}} \label{sec:filtering proof}

The proof of Theorem \ref{thm:filtering processes} uses ideas from the proof of Theorem 3.1 of \cite{sun2023lqposc}. The main idea in this approach is to use the innovation process $V$ in \eqref{inov-proc} as the noise term of the filtering processes $\hat{\mathbb{X}}^{q}$ in \eqref{filt-x} 
 which is weighted by a time-varying deterministic process, which is a solution to a matrix differential equation as in \eqref{eq:Sigma ODE}. These weights can be thought of as analogous to the Kalman gain originally studied in \citet{kalman1961inear}. The weights are found using the Fujisaki-Kallianpur-Kunita theorem \cite{rogers2000diffusions} and ideas from stochastic analysis. 

We first prove that the adjustment in \eqref{def:adjusted-inflow} removes precisely the effect of the desk's trading rate from the observed inflow. This also establishes the vector space property of the admissible class used in Appendix \ref{sec:control proof}.

\begin{proof}[Proof of Proposition \ref{prop:fixed-filtration}]
Let $q\in\mathcal A$. Subtracting the equation for $\theta^0$ from \eqref{def:theta} gives
\begin{equation}
\label{eq:toxicity-difference}
\begin{split}
    d\left(\theta^q_t-\theta^0_t\right)
    &=\left(a_t\left(\theta^q_t-\theta^0_t\right)+b_tq_t\right)dt,\\
    \theta^q_0-\theta^0_0&=0.
\end{split}
\end{equation}
Using an integrating factor method in \eqref{eq:toxicity-difference}, we obtain
\begin{equation}
\label{eq:toxicity-difference-solution}
\begin{split}
    \theta^q_t-\theta^0_t
    =\int_0^t b_s\exp\left(\int_s^t a_rdr\right)q_sds
    =\Delta\theta^q_t,\quad 0\leq t\leq T.
\end{split}
\end{equation}
It follows from \eqref{def:Z} and \eqref{eq:toxicity-difference-solution} that
\begin{equation}
\label{eq:inflow-difference-solution}
\begin{split}
    Z^q_t-Z^0_t
    =\int_0^t\left(\theta^q_s-\theta^0_s\right)ds=\int_0^t\Delta\theta^{q}_{s}ds=\Delta Z^q_t,\quad 0\leq t\leq T.
\end{split}
\end{equation}
Combining \eqref{def:adjusted-inflow} and \eqref{eq:inflow-difference-solution} yields
\begin{equation}
\label{eq:adjusted-inflow-identity}
\begin{split}
    \overline Z^q_t
    &=Z^q_t-\Delta Z^q_t=Z^0_t,\quad 0\leq t\leq T.
\end{split}
\end{equation}
In particular, \eqref{eq:adjusted-inflow-identity} proves the identity $\overline Z^q=Z^0$. Since $q$ is $\mathcal Y^0$-progressively measurable, \eqref{eq:toxicity-difference-solution} shows that $\theta^q-\theta^0=\Delta\theta^q$ is $\mathcal Y^0$-adapted. It then follows from \eqref{eq:inflow-difference-solution} that $\Delta Z^q$, and hence $Z^q=Z^0+\Delta Z^q$, are $\mathcal Y^0$-adapted. Therefore $\mathcal G^q_t\subseteq\mathcal Y^0_t$. Conversely, \eqref{def:adjusted-inflow} and \eqref{eq:adjusted-inflow-identity} show that $Z^0=\overline Z^q$ is $\mathcal G^q$-adapted. Since $\ol M$ and $A$ are also contained in $\mathcal G^q$, Definition \ref{part-f} gives $\mathcal Y^0_t\subseteq\mathcal G^q_t$. Hence $\mathcal G^q=\mathcal Y^0$.

Finally, it remains to show that $\mathcal{A}$ is a vector space. If $q,u\in\mathcal A$ and $\alpha,\delta\in\mathbb R$, then $\alpha q+\delta u$ is $\mathcal Y^0$-progressively measurable and, by Young's inequality,
\begin{equation}
\label{eq:admissible-linear-combination}
\begin{split}
    \mathbb E\left[\int_0^T\left(\alpha q_t+\delta u_t\right)^2dt\right]
    &\leq 2\alpha^2\mathbb E\left[\int_0^Tq_t^2dt\right]+2\delta^2\mathbb E\left[\int_0^Tu_t^2dt\right]<\infty.
\end{split}
\end{equation}
Hence $\mathcal A$ is a vector space. Taking the appropriate values of $\alpha$ and $\delta$ gives both $q+\delta\eta\in\mathcal A$ and $\delta q+(1-\delta)u\in\mathcal A$. This concludes the result.
\end{proof}

We next consider an affine transformation of the innovation process and prove that it is a standard Brownian motion. Recall that the class of admissible unwind strategies $\mathcal A$ was defined in \eqref{def:admissset}. Recall that $V$ was defined in \eqref{inov-proc}.
\begin{lemma} \label{lemma:V tilde}
For any $q \in \mathcal A$ the process 
    \begin{equation} \label{tilde-v}
        \wt{V}_{t}:= \sigma^{-1}(V_{t}-V_{0}), \quad t\geq 0, 
    \end{equation}
    is a $\{\mathcal{Y}^{0}_{t}\}_{t\in[0,T]}$-Brownian motion.
\end{lemma}

\begin{proof} 
Let $q\in \mathcal A$. First note that from \eqref{inov-proc} it follows that $(V_t)_{ t\in [0,T]}$ is $(\mathcal {Y}^0_t)_{t\in[0,T]}$ adapted. 
From \eqref{filt-x}, \eqref{z-trans} and \eqref{inov-proc} it follows that, 
    \begin{equation} \label{eq:substituted V}
        V_{t} = \int_{0}^{t}\e_{1}^{\top}\wt{\mathbb{X}}^{q}_{s}ds + \sigma W^{Z}_{t}. 
    \end{equation}
Since $\wt{\mathbb{X}}^{q}\perp\mathcal{Y}^{0}$ and $\hat{\mathbb{X}}^{q}$ is an unbiased estimator of $\mathbb{X}^{q}$ we get from \eqref{eq:substituted V} by using the tower rule 
\begin{equation}
    \mathbb{E}\left[V_{t}-V_{s}|\mathcal{Y}^{0}_{s}\right] = \sigma\mathbb{E}\left[\mathbb{E}\left[W^{Z}_{t}-W^{Z}_{s}|\mathcal{F}_{s}\right]|\mathcal{Y}^{0}_{s}\right] + \mathbb{E}\left[\int_{s}^{t}\e_{1}^{\top}\wt{\mathbb{X}}^{q}_{r}dr\Big\vert\mathcal{Y}^{0}_{s}\right] = 0,
\end{equation}
where we have used \eqref{filt-x} and Fubini's theorem in the second equality. 
This shows that $\tilde{V}$ is a $\{\mathcal{Y}^{0}_{t}\}_{t\in[0,T]}$-martingale. 

Next we use It\^{o}'s lemma, \eqref{tilde-v} and \eqref{eq:substituted V} to get, 
 \begin{equation}
\begin{split}
    \wt{V}_{t}^{2} - \wt{V}_{s}^{2} 
    &= 2\int_{s}^{t} \sigma^{-1}\e_{1}^{\top}\wt{\mathbb{X}}^{q}_{r}\wt{V}_{r}dr + 2\int_{s}^{t}\wt{V}_{r}dW^{Z}_{r} + (t-s), \quad \textrm{for all } 0\leq s\leq t \leq T. 
\end{split}
\end{equation}
Then with applications of Fubini's Theorem and the tower rule we get for $0\leq s\leq t \leq T$,
\begin{equation} \label{f1}
\begin{aligned}
    &\mathbb{E}\left[\wt{V}_{t}^{2} - \wt{V}_{s}^{2} \big|\mathcal{Y}^{0}_{s}\right] \\ 
    &= 2\int_{s}^{t} \sigma^{-1}\e_{1}^{\top}\mathbb{E}\left[\wt{\mathbb{X}}^{q}_{r}\wt{V}_{r}\big|\mathcal{Y}^{0}_{s}\right]dr + 2\mathbb{E}\left[\int_{s}^{t}\wt{V}_{r}dW^{Z}_{r}\Big\vert\mathcal{Y}^{0}_{s}\right] + (t-s) \\
    &= 2\int_{s}^{t} \sigma^{-1}\e_{1}^{\top}\mathbb{E}\left[\mathbb{E}\left[\wt{\mathbb{X}}^{q}_{r}\big|\mathcal{Y}^{0}_{r}\right]\tilde{V}_{r}\big|\mathcal{Y}^{0}_{s}\right]dr + 2\mathbb{E}\left[\mathbb{E}\left[\int_{s}^{t}\wt{V}_{r}dW^{Z}_{r}\Big\vert\mathcal{F}_{s}\right]\Big\vert\mathcal{Y}^{0}_{s}\right] + (t-s) \\
    &= (t-s),
\end{aligned}
\end{equation}
where in the last equality we have used the fact that $W^{Z}$ is an $(\mathcal F_t)_{t\in[0,T]}$-Brownian motion and the fact that $\wt{\mathbb{X}}$ has zero mean by \eqref{filt-x}. 

From \eqref{f1} it follows that $\tilde{V}$ is a $\{\mathcal{Y}^{0}_{t}\}_{t\in[0,T]}$-Brownian motion by Lévy’s Characterization theorem. 
\end{proof}

Recall the dynamics of $\mathbb{X}^q$ from \eqref{eq:X SDE} and that $\hat{\mathbb{X}}^q$ was defined in \eqref{filt-x}. We introduce the $\{\mathcal{Y}^{0}_{t}\}_{t\in[0,T]}$-progressively measurable process, 
\begin{equation} \label{def:Lambda}
    \Lambda_{t} := \hat{\mathbb{X}}^{q}_{t} - X_{0} - \int_{0}^{t}(\mathbf{A}_{s}\hat{\mathbb{X}}^{q}_{s}+\mathbf{B}_{s}q_{s})ds-\int_{0}^{t}\mathbf{C}_{s}dA_{s}, \quad 0\leq t \leq T. 
\end{equation}
\begin{lemma}\label{lemma:Lambda-martingale}
    The process $\Lambda$ is a $\{\mathcal{Y}^{0}_{t}\}_{t\in[0,T]}$-martingale.
\end{lemma}

\begin{proof}
From \eqref{eq:X SDE}, \eqref{filt-x}, \eqref{def:Lambda},  and Fubini's theorem  it follows that for any $0\leq s\leq t \leq T$ we have, 
    \begin{equation}
    \begin{aligned}
        \mathbb{E}\left[\Lambda_{t}-\Lambda_{s}|\mathcal{Y}^{0}_{s}\right] 
        &= 
        \mathbb{E}\left[\hat{\mathbb{X}^{q}_{t}}-\hat{\mathbb{X}}^{q}_{s}|\mathcal{Y}^{0}_{s}\right] - \mathbb{E}\left[\int_{s}^{t}(\mathbf{A}_{r}\hat{\mathbb{X}}^{q}_{r}+\mathbf{B}_{r}q_{r})dr+\int_{s}^{t}\mathbf{C}_{r}dA_{r}\Big\vert\mathcal{Y}^{0}_{s}\right] \\
        &= \mathbb{E}\left[\mathbb{X}^{q}_{t}-\mathbb{X}^{q}_{s}|\mathcal{Y}^{0}_{s}\right] - \int_{s}^{t}\mathbb{E}\left[\mathbf{A}_{r}\hat{\mathbb{X}}^{q}_{r} + \mathbf{B}_{r}q_{r}| \mathcal{Y}^{0}_{s}\right] dr -\mathbb{E}\left[\int_{s}^{t}\mathbf{C}_{r}dA_{r}\bigg|\mathcal{Y}^{0}_{s}\right] \\
        &= \mathbb{E}\left[\mathbb{X}^{q}_{t}-\mathbb{X}^{q}_{s}|\mathcal{Y}^{0}_{s}\right] - \int_{s}^{t}\mathbb{E}\left[\mathbf{A}_{r}\mathbb{X}^{q}_{r} + \mathbf{B}_{r}q_{r}| \mathcal{Y}^{0}_{s}\right] dr -\mathbb{E}\left[\int_{s}^{t}\mathbf{C}_{r}dA_{r}\bigg|\mathcal{Y}^{0}_{s}\right] \\
        &= \mathbb{E}\bigg[\int_{s}^{t}(\mathbf{A}_{r}\mathbb{X}^{q}_{r} + \mathbf{B}_{r}q_{r})dr +\int_{s}^{t}\mathbf{C}_{r}dA_{r} \\
        &\qquad + \int_{s}^{t}\mathbf{D}_{r}dW^{\theta}_{r} + \int_{s}^{t} \mathbf{E}_{r}dW^{Z}_{r} \Big| \mathcal{Y}^{0}_{s}\bigg] \\
        &\qquad - \int_{s}^{t}\mathbb{E}\left[\mathbf{A}_{r}\mathbb{X}^{q}_{r} + \mathbf{B}_{r}q_{r}| \mathcal{Y}^{0}_{s}\right] dr -\mathbb{E}\left[\int_{s}^{t}\mathbf{C}_{r}dA_{r}\bigg|\mathcal{Y}^{0}_{s}\right]\\
        &= 0,
        \end{aligned}
    \end{equation}
where we have used the tower rule in the last equality. This concludes the result. 
\end{proof}

\begin{lemma}
\label{lemma:reduced-observation-filtration}
Let $\mu$ be the unique solution of
\begin{equation}
\label{eq:known-toxicity-component}
\begin{split}
    d\mu_t&=a_t\mu_tdt+c_tdA_t,\\
    \mu_0&=\theta_0,
\end{split}
\end{equation}
and define
\begin{equation}
\label{eq:reduced-signal-observation}
\begin{split}
    \mathfrak t_t&:=\theta^0_t-\mu_t,\\
    \mathfrak Z_t&:=Z^0_t-\int_0^t\mu_sds.
\end{split}
\end{equation}
Let $\mathcal H_t$ be the usual augmentation of $\sigma(\mathfrak Z_s:0\leq s\leq t)$ and $\widehat{\mathfrak t}_t:=\mathbb E[\mathfrak t_t\mid\mathcal H_t]$. Then $\Lambda$ is an $(\mathcal H_t)_{t\in[0,T]}$-martingale, and
\begin{equation}
\label{eq:conditional-error-covariance}
\begin{split}
    \mathbb E\left[\wt{\mathbb X}^q_t(\wt{\mathbb X}^q_t)^\top\mid\mathcal Y^0_t\right]
    &=\Sigma_t,\quad 0\leq t\leq T.
\end{split}
\end{equation}
\end{lemma}

\begin{proof}
Subtracting \eqref{eq:known-toxicity-component} from the dynamics of $\theta^0$ and using \eqref{def:Z} gives
\begin{equation}
\label{eq:reduced-signal-observation-dynamics}
\begin{split}
    d\mathfrak t_t&=a_t\mathfrak t_tdt+d_tdW^\theta_t,\qquad \mathfrak t_0=0,\\
    d\mathfrak Z_t&=\mathfrak t_tdt+\sigma dW^Z_t,\qquad \mathfrak Z_0=0.
\end{split}
\end{equation}
By the mutual independence assumption on $(\ol M,A,W^Z,W^\theta)$, the pair $(\mathfrak t,\mathfrak Z)$ is independent of $(\ol M,A)$. Moreover, \eqref{def:reference-filtration} and \eqref{eq:reduced-signal-observation} imply that $\mathcal Y^0_t$ is the usual augmentation of $\sigma(\ol M_s,A_s,\mathfrak Z_s:0\leq s\leq t)$. Hence
\begin{equation}
\label{eq:reduced-conditional-expectation}
\begin{split}
    \mathbb E[\mathfrak t_t\mid\mathcal Y^0_t]
    &=\mathbb E[\mathfrak t_t\mid\mathcal H_t]=\widehat{\mathfrak t}_t,
    \qquad \text{and} \qquad
    \hat\theta^0_t=\mu_t+\widehat{\mathfrak t}_t.
\end{split}
\end{equation}
For the first component, we use \eqref{def:Lambda}, \eqref{eq:Y SDE}, \eqref{eq:control-shifts}, \eqref{eq:filter-shift}, \eqref{eq:known-toxicity-component}, and \eqref{eq:reduced-conditional-expectation}. For the second component, we additionally use \eqref{def:X}, \eqref{inv}, \eqref{inov-proc}, Proposition \ref{prop:fixed-filtration}, and \eqref{eq:reduced-signal-observation}. This gives
\begin{equation}
\label{eq:Lambda-reduced-filtration}
\begin{split}
    \Lambda^{(1)}_t
    &=\hat\theta^q_t-\theta_0
    -\int_0^t\left(a_s\hat\theta^q_s+b_sq_s\right)ds
    -\int_0^tc_sdA_s\\
    &=\hat\theta^0_t-\theta_0
    -\int_0^ta_s\hat\theta^0_sds
    -\int_0^tc_sdA_s
    +\Delta\theta^q_t
    -\int_0^t\left(a_s\Delta\theta^q_s+b_sq_s\right)ds\\
    &=\hat\theta^0_t-\theta_0
    -\int_0^ta_s\hat\theta^0_sds
    -\int_0^tc_sdA_s\\
    &=\widehat{\mathfrak t}_t-\int_0^ta_s\widehat{\mathfrak t}_sds
    +\mu_t-\theta_0-\int_0^ta_s\mu_sds-\int_0^tc_sdA_s\\
    &=\widehat{\mathfrak t}_t-\int_0^ta_s\widehat{\mathfrak t}_sds,\\
    \Lambda^{(2)}_t
    &=X^q_t-x-\int_0^t\left(\hat\theta^q_s+q_s\right)ds\\
    &=Z^q_t-\int_0^t\hat\theta^q_sds=V_t,\\
    V_t
    &=Z^0_t+\Delta Z^q_t
    -\int_0^t\left(\hat\theta^0_s+\Delta\theta^q_s\right)ds\\
    &=Z^0_t-\int_0^t\hat\theta^0_sds\\
    &=\mathfrak Z_t+\int_0^t\mu_sds
    -\int_0^t\left(\mu_s+\widehat{\mathfrak t}_s\right)ds\\
    &=\mathfrak Z_t-\int_0^t\widehat{\mathfrak t}_sds.
\end{split}
\end{equation}
Thus \eqref{eq:Lambda-reduced-filtration} shows that $\Lambda$ is $(\mathcal H_t)_{t\in[0,T]}$-adapted. Since $\mathcal H_t\subseteq\mathcal Y^0_t$ and $\Lambda$ is a $\mathcal Y^0$-martingale by Lemma \ref{lemma:Lambda-martingale}, the tower rule shows that $\Lambda$ is also an $\mathcal H$-martingale.

It remains to show \eqref{eq:conditional-error-covariance}. By \eqref{eq:filter-shift} and \eqref{eq:reduced-conditional-expectation},
\begin{equation}
\label{eq:reduced-filtering-error}
\begin{split}
\wt{\mathbb X}^q_t
&=\e_1\left(\mathfrak t_t-\widehat{\mathfrak t}_t\right).
\end{split}
\end{equation}
By \eqref{eq:reduced-signal-observation-dynamics}, $\mathfrak t_t$ and $(\mathfrak Z_s)_{0\leq s\leq t}$ are jointly Gaussian. Indeed, the Gaussian residual $\mathsf{t}_t-\widehat{\mathsf{t}}_t$ is orthogonal to, and therefore independent of, the Gaussian space generated by $(\mathcal{Z}s){0\leq s\leq t}$, so its conditional second moment given $\mathcal{H}_t$ is deterministic. Hence
$
\mathbb E\left[
\left(\mathfrak t_t-\widehat{\mathfrak t}_t\right)^2
\mid\mathcal H_t\right]
$
is deterministic. Moreover, $\mathcal Y^0_t=\mathcal H_t\vee\sigma(\ol M_s,A_s:0\leq s\leq t)$, and the latter sigma algebra is independent of $(\mathfrak t,\mathfrak Z)$. Therefore
\begin{equation}
\label{eq:reduced-error-conditional-variance}
\begin{split}
\mathbb E\left[
\left(\mathfrak t_t-\widehat{\mathfrak t}_t\right)^2
\mid\mathcal Y^0_t\right]
&=\mathbb E\left[
\left(\mathfrak t_t-\widehat{\mathfrak t}_t\right)^2
\mid\mathcal H_t\right]=\mathbb E\left[
\left(\mathfrak t_t-\widehat{\mathfrak t}_t\right)^2\right].
\end{split}
\end{equation}
It now follows from \eqref{eq:reduced-filtering-error} and the definition of $\Sigma$ in \eqref{cov} that
\begin{equation}
\label{eq:conditional-error-covariance-calculation}
\begin{split}
\mathbb E\left[
\wt{\mathbb X}^q_t(\wt{\mathbb X}^q_t)^\top
\mid\mathcal Y^0_t\right]
&=\e_1\e_1^\top
\mathbb E\left[
\left(\mathfrak t_t-\widehat{\mathfrak t}_t\right)^2
\mid\mathcal Y^0_t\right]
=\mathbb E\left[
\wt{\mathbb X}^q_t(\wt{\mathbb X}^q_t)^\top\right]
=\Sigma_t.
\end{split}
\end{equation}

\end{proof}

We recall that $\e_1$ and $\Sigma_t$ were defined in \eqref{e1} and \eqref{cov}, respectively.  
\begin{proposition}
There exists a $\{\mathcal{Y}^{0}_{t}\}_{t\in[0,T]}$-progressively measurable process $\gamma =(\gamma_t)_{t\in [0,T]}$ such that
\begin{equation} \label{eq:FKK}
    \Lambda_{t} = \int_{0}^{t} \gamma_{s} \sigma^{-2} dV_{s}, \quad \textrm{for all } \ 0\leq t \leq T. 
\end{equation}
Moreover, $\gamma$ is given by, 
    \begin{equation}\label{gam-rep} 
        \gamma_{t} = \Sigma_{t}\e_{1} + \sigma E_{t}, \quad 0\leq t\leq T. 
    \end{equation}
\end{proposition}
\begin{proof}
Since \(\mathcal Y^0\) also contains the independent observations \((\overline M,A)\), it is not the natural filtration of \(Z^0\), so the Fujisaki–Kallianpur–Kunita theorem cannot be applied directly on \(\mathcal Y^0\). By Lemma \ref{lemma:reduced-observation-filtration}, $\Lambda$ is an $\mathcal H$-martingale. Applying the Fujisaki-Kallianpur-Kunita theorem to the reduced observation $\mathfrak Z$ in \eqref{eq:reduced-signal-observation}, whose innovation is $V$ by \eqref{eq:Lambda-reduced-filtration}, shows that there exists an $\mathcal H$-progressively measurable process $\gamma$ such that \eqref{eq:FKK} holds. Since $\mathcal H_t\subseteq\mathcal Y^0_t$, the process $\gamma$ is also $\mathcal Y^0$-progressively measurable.
Let $(\zeta_t)_{t\in[0,T]}$ be a fixed but arbitrary square-integrable, $\{\mathcal Y^0_t\}_{t\in[0,T]}$-progressively measurable process. Define
    \begin{equation} \label{f33} 
        \eta_{t} = \int_{0}^{t} \zeta_{s} \sigma^{-2} dV_{s}, \quad 0\leq t\leq T.
    \end{equation}
   From Lemma \ref{lemma:V tilde} it follows that $\eta$ is a $\{\mathcal{Y}^{0}_{t}\}_{t\in[0,T]}$-martingale. 
    From \eqref{eq:FKK} and \eqref{f33} we have 
    \begin{equation} \label{eq:Lambdalambdaexpression}
        \mathbb{E}\left[\Lambda_{t}\eta_{t}\right] = \mathbb{E}\left[\int_{0}^{t}\gamma_{s}\sigma^{-2}\zeta_{s}ds\right]
    \end{equation} 
    Plugging in \eqref{def:Lambda} we get, 

    \begin{equation} \label{eq:Lambdaetaexpression}
        \mathbb{E}\left[\Lambda_{t}\eta_{t}\right] = \mathbb{E}\left[\hat{\mathbb{X}}^{q}_{t}\eta_{t}\right] - \mathbb{E}\left[\int_{0}^{t}\left(A_{s}\hat{\mathbb{X}}^{q}_{s} + B_{s}q_{s}\right)\eta_{t}ds\right] -\mathbb{E}\left[\int_{0}^{t}C_{s}\eta_{t}dA_{s}\right]. 
    \end{equation}
    Using the martingale property of $\eta$ and \eqref{filt-x}, we have for $0\leq s\leq t \leq T$, 
    \begin{equation}\label{rr1}
         \mathbb{E}\left[\hat{\mathbb{X}}^{q}_{s}\eta_{t}\right] = \mathbb{E}\left[\mathbb{E}\left[\hat{\mathbb{X}}^{q}_{s}\eta_{t}|\mathcal{Y}^{0}_{s}\right]\right]  
=\mathbb{E}\left[\mathbb{E}\left[\mathbb{X}^{q}_{s}|\mathcal{Y}^{0}_{s}\right]\mathbb{E}\left[\eta_{t}|\mathcal{Y}^{0}_{s}\right]\right] 
        = \mathbb{E}\left[\mathbb{X}^{q}_{s}\eta_{s}\right],
    \end{equation}
    and 
    \begin{equation} \label{rr2}
        \mathbb{E}\left[\mathbf{B}_{s}q_{s}\eta_{t}\right] = \mathbb{E}\left[(\mathbf{B}_{s}q_{s})\eta_{s}\right].
    \end{equation}
    Using integration by parts we get
    \begin{equation}
        \mathbb{E}\left[\eta_{t}\int_{0}^{t}\mathbf{C}_{s}dA_{s}\right] 
        =\mathbb{E}\left[\int_{0}^{t}\eta_{s}\mathbf{C}_{s}dA_{s} + \int_{0}^{t}\left(\int_{0}^{s}\mathbf{C}_{r}dA_{r}\right)d\eta_{s}\right] 
    \end{equation}
    because $A$ is of finite variation. Since $\eta$ is a martingale this implies that
    \begin{equation} \label{rr3}
        \mathbb{E}\left[\int_{0}^{t}\eta_{t}\mathbf{C}_{s}dA_{s}\right] 
        =\mathbb{E}\left[\int_{0}^{t}\eta_{s}\mathbf{C}_{s}dA_{s}\right].
    \end{equation}
    Plugging in  \eqref{rr1}, \eqref{rr2} and \eqref{rr3} into \eqref{eq:Lambdaetaexpression} we get, 
    \begin{equation}
        \mathbb{E}\left[\Lambda_{t}\eta_{t}\right] = \mathbb{E}\left[\hat{\mathbb{X}}^{q}_{t}\eta_{t}\right] - \mathbb{E}\left[\int_{0}^{t}\left(\mathbf{A}_{s}\hat{\mathbb{X}}^{q}_{s} + \mathbf{B}_{s}q_{s}\right)\eta_{s}ds\right] -\mathbb{E}\left[\int_{0}^{t}\mathbf{C}_{s}\eta_{s}dA_{s}\right].
    \end{equation}
 From \eqref{f33} and \eqref{eq:substituted V} we get
    \begin{equation} \label{rr4} 
        d\eta_{t} = \zeta_{t} \sigma^{-2} \left(\e_{1}^{\top}\wt {\mathbb{X}}^{q}_{t}dt + \sigma dW^{Z}_{t}\right).
    \end{equation}
From  \eqref{eq:X SDE}, \eqref{rr4} and by using integration by parts it follows that, 
    \begin{equation} \label{rr5}
    \begin{split}
        &\mathbb{E}\left[\mathbb{X}^{q}_{t}\eta_{t}\right] \\
        &= \mathbb{E}\Bigg[\int_{0}^{t}\left(\mathbf{A}_{s}\mathbb{X}_{s} + \mathbf{B}_{s} q_{s}\right)\eta_{s}ds +\int_{0}^{t}\mathbf{C}_{s}\eta_{s}dA_{s} \\
        &\hspace{1.3cm}+ \int_{0}^{t}\mathbb{X}^{q}_{s}(\wt{\mathbb{X}}_{s}^{q})^{\top}\e_{1}\sigma^{-2} \zeta_{s}ds + \int_{0}^{t} \mathbf{E}_{s} \sigma^{-1} \zeta_{s} ds \Bigg].
    \end{split}
    \end{equation} 
  Using \eqref{filt-x}, the conditional covariance identity \eqref{eq:conditional-error-covariance}, and the tower rule, and the fact that $\wt{\mathbb{X}}$ has zero mean we get
    \begin{equation}\label{rr6} 
    \begin{aligned}
        &\mathbb{E}\left[\int_{0}^{t}\mathbb{X}_{s}(\wt{\mathbb{X}}^{q}_{s})^{\top}\e_{1}\sigma^{-2} \zeta_{s}ds\right] \\
        &=\mathbb{E}\left[\int_{0}^{t}(\wt{\mathbb{X}}^{q}_{s})(\wt{\mathbb{X}}^{q}_{s})^{\top}\e_{1}\sigma^{-2} \zeta_{s}ds + \int_{0}^{t}(\hat{\mathbb{X}}^{q}_{s})(\wt{\mathbb{X}}^{q}_{s})^{T}\e_{1}\sigma^{-2} \zeta_{s}ds\right] \\
        &= \mathbb{E}\left[\int_{0}^{t}\Sigma_{s}\e_{1}\sigma^{-2} \zeta_{s}ds\right].
    \end{aligned}
    \end{equation}
Plugging in \eqref{rr6} into \eqref{rr5} and using again the fact that $\wt{\mathbb{X}} \perp \mathcal{Y}^{0}$ and that $\wt{\mathbb{X}}$ has zero mean we get, 
    \begin{equation}
        \mathbb{E}\left[\hat{\mathbb{X}}^q_{t}\eta_{t}\right] 
        = \mathbb{E}\left[\int_{0}^{t}\left(\mathbf{A}_{s}\mathbb{X}^{q}_{s} + \mathbf{B}_{s} q_{s} \right)\eta_{s}ds +\int_{0}^{t}\mathbf{C}_{s}\eta_{s}dA_{s} + \int_{0}^{t}\Sigma_{s}\e_{1}\sigma^{-2} \zeta_{s}ds + \int_{0}^{t} \mathbf{E}_{s} \sigma^{-1} \zeta_{s}ds\right].
    \end{equation}
    Using this as well as \eqref{eq:Lambdalambdaexpression} and \eqref{eq:Lambdaetaexpression} yields
    \begin{equation}
        \gamma_{t} = \Sigma_{t}\e_{1} + \sigma \mathbf{E}_{t},
    \end{equation}
    since $\zeta$ was chosen arbitrarily. This completes the proof. 
\end{proof}
\begin{proof}[Proof of Theorem \ref{thm:filtering processes}] 
Using \eqref{def:Lambda}, \eqref{eq:FKK} with \eqref{gam-rep} the SDE for $\hat{\mathbb{X}}^{q}$ we get   
\begin{equation}  \label{gg1} 
\begin{split}
    d\hat{\mathbb{X}}^{q}_{t} =
    &\left(\Sigma_{t}\e_{1}+\sigma \mathbf{E}_{t}\right)\sigma^{-2}\e_{1}^{\top}\wt{\mathbb{X}}^{q}_{t}dt + \left(\Sigma_{t}\e_{1}+\sigma \mathbf{E}_{t}\right)\sigma^{-1}dW^{Z}_{t} \\
    &+ \left(\mathbf{A}_{t}\hat{\mathbb{X}}^{q}_{t}+\mathbf{B}_{t}q_{t}\right)dt +\mathbf{C}_{t}dA_{t}.
\end{split}
\end{equation}
Using \eqref{gg1} and \eqref{eq:substituted V} we recover \eqref{eq:Xarrowhat SDE}. 
By subtracting \eqref{gg1} from \eqref{eq:X SDE}, recalling \eqref{filt-x} we get, 
\begin{equation} \label{eq:Xtilde SDE}
    d\wt{\mathbb{X}}_{t} = \left(\mathbf{A}_{t} - \left(\Sigma_{t}\e_{1}+\sigma \mathbf{E}_{t}\right)\sigma^{-2}\e_{1}^{\top}\right)\wt{\mathbb{X}}^{q}_{t}dt - \Sigma_{t}\e_{1}\sigma^{-1}dW^{Z}_{t} + \mathbf{D}_{t}dW^{\theta}_{t}.
\end{equation}
Next we can use  \eqref{eq:Xtilde SDE}, integration by parts and Fubini's theorem in order to derive the dynamics of $\Sigma$ in \eqref{cov}, recalling that $\wt{\mathbb{X}}^{q}_0=0$, 
\begin{equation}
\begin{aligned}
    \Sigma_{t} &=  \mathbb{E}\bigg[\int_{0}^{t} \left(\mathbf{A}_{s} - \left(\Sigma_{s}\e_{1}+\sigma \mathbf{E}_{s}\right)\sigma^{-2}\e_{1}^{\top}\right)\wt{\mathbb{X}}^{q}_{s}\left(\wt{\mathbb{X}}^{q}_s\right)^{\top}ds \\
    &\qquad + \int_{0}^{t} \wt{\mathbb{X}}_{s}\left(\wt{\mathbb{X}}^{q}_{s}\right)^{\top}\left(\mathbf{A}_{s} - \left(\Sigma_{s}\e_{1}+\sigma \mathbf{E}_{s}\right)\sigma^{-2}\e_{1}^{\top}\right)^{\top}ds \\
    &\qquad + \int_{0}^{t} \sigma^{-2}\Sigma_{s}\e_{1}\e_{1}^{\top}\Sigma_{s} ds + \int_{0}^{t}\mathbf{D}_{s}\mathbf{D}_{s}^{\top}ds\bigg] \\
    &=  \int_{0}^{t} \left(\mathbf{A}_{s} - \left(\Sigma_{s}\e_{1}+\sigma \mathbf{E}_{s}\right)\sigma^{-2}\e_{1}^{\top}\right)\Sigma_{s}ds 
    + \int_{0}^{t} \Sigma_{s}\left(\mathbf{A}_{s} - \left(\Sigma_{s}\e_{1}+\sigma \mathbf{E}_{s}\right)\sigma^{-2}\e_{1}^{\top}\right)^{\top}ds \\
    &\quad + \int_{0}^{t} \sigma^{-2}\Sigma_{s}\e_{1}\e_{1}^{\top}\Sigma_{s} ds + \int_{0}^{t}\mathbf{D}_{s}\mathbf{D}_{s}^{\top}ds \\
    &=   \int_{0}^{t} \left(\mathbf{A}_{s} - \sigma^{-1} \mathbf{E}_{s} \e_{1}^{\top}\right)\Sigma_{s}ds 
    + \int_{0}^{t} \Sigma_{s}\left(\mathbf{A}_{s} - \sigma^{-1} \mathbf{E}_{s} \e_{1}^{\top}\right)^{\top}ds \\
    &\quad- \int_{0}^{t} \sigma^{-2}\Sigma_{s}\e_{1}\e_{1}^{\top}\Sigma_{s} ds + \int_{0}^{t}\mathbf{D}_{s}\mathbf{D}_{s}^{\top}ds. 
\end{aligned}
\end{equation}
Note that the above equation coincides with the Riccati equation   \eqref{eq:Sigma ODE}. After applying a time change, Proposition 4.2 in \cite{lim2001lqcbsde} shows that this Riccati equation has a unique, positive semi-definite solution.

\end{proof} 

\begin{proof}[Proof of Corollary \ref{thm-filt}]
The equation for $\hat X^q$ in \eqref{eq:Xhat SDE} follows directly from \eqref{x-vec}, \eqref{x-th-hat} and \eqref{eq:Xarrowhat SDE}. The following equation for $\hat{\theta}$ is derived by the same reasoning: 
    \begin{equation} \label{eq:thetahat SDE}
        \hat{\theta}_{t} = \theta_{0} + \int_{0}^{t}\kappa^{(1)}_{s}dV_{s} + \int_{0}^{t} \left(a_{s}\hat{\theta}_{s}+b_{s}q_{s}\right)ds +\int_{0}^{t}c_{s}dA_{s}.
    \end{equation}
Note that  \eqref{eq:thetahat} solves \eqref{eq:thetahat SDE}.
\end{proof}

\section{Proof of Theorem \ref{thm:optimal control}} \label{sec:control proof}
In this appendix we prove Theorem \ref{thm:optimal control}. We begin with the proof of Proposition \ref{prop:Phi}.
We then proceed to use a variational approach in the spirit of \citep{bank2017hedging,neuman2022optimalsignaltrading}. 

\begin{proof}[Proof of Proposition~\ref{prop:Phi}]
For matrices, write $\|B\|:=\max_j\sum_i|B_{ij}|$ for the
maximum column sum norm, which satisfies $\|BC\|\leq\|B\|\|C\|$.
For scalar functions, write
$\|f\|_{\infty}:=\sup_{0\leq t\leq T}|f_t|$.
Using that $a$, $b$, and $b'$ are bounded by definition,
$\|r\|_{\infty}\leq
\|b'\|_{\infty}T e^{T\|a\|_{\infty}}+\|b\|_{\infty}$.
Since the elements of \eqref{def:L} are bounded, it follows that
$\sup_{0\leq t\leq T}\|L_t\|<\infty$.
Set $\Phi^{(0)}_t=I$, and for $n\geq0$ define
\begin{equation}\label{eq:Phi_iter}
\begin{split}
    \Phi^{(n+1)}_t
    =I+\int_0^t L_s\Phi^{(n)}_s\,ds,
    \qquad 0\leq t\leq T.
\end{split}
\end{equation}
We would like to show by induction
\begin{equation}
\label{eq:inductive_bound}
    \|\Phi^{(n+1)}_t-\Phi^{(n)}_t\| 
    \leq 
    \frac{\left(t\sup_{0\leq u\leq T}\|L_u\|\right)^{n+1}}
    {(n+1)!}.
\end{equation}
For $n=0$, \eqref{eq:Phi_iter} gives
$\|\Phi^{(1)}_t-\Phi^{(0)}_t\|
\leq t\sup_{0\leq u\leq T}\|L_u\|$.
If \eqref{eq:inductive_bound} holds for $n-1$, then \eqref{eq:Phi_iter} gives
\begin{equation}\label{eq:Phi_induction}
\begin{split}
    \|\Phi^{(n+1)}_t-\Phi^{(n)}_t\|
    &\leq \int_0^t
    \|L_s\|\,\|\Phi^{(n)}_s-\Phi^{(n-1)}_s\|\,ds\\
    &\leq \int_0^t
    \frac{s^n\left(\sup_{0\leq u\leq T}\|L_u\|\right)^{n+1}}
    {n!}\,ds\\
    &=\frac{\left(t\sup_{0\leq u\leq T}\|L_u\|\right)^{n+1}}
    {(n+1)!}
\end{split}
\end{equation}
where the second inequality follows from the aforementioned induction hypothesis. We have shown by induction \eqref{eq:inductive_bound}, hence the iterates converge uniformly on $[0,T]$.
Passing to the limit in \eqref{eq:Phi_iter} and using the fundamental theorem of calculus shows that $\Phi=\lim_{n\rightarrow\infty}\Phi^{(n)}$ satisfies
\eqref{def:Phi} almost everywhere
and is absolutely continuous.
For any two absolutely continuous solutions $\Phi$ and
$\widetilde{\Phi}$ of \eqref{def:Phi}, integration gives
$\|\Phi_t-\widetilde{\Phi}_t\|
\leq \sup_{0\leq t\leq T}\|L_t\|
\int_0^t\|\Phi_s-\widetilde{\Phi}_s\|\,ds$.
Uniqueness therefore follows from Gronwall's inequality.
By \eqref{def:L}, $\operatorname{tr}(L_t)=0$.
Liouville's formula applied to \eqref{def:Phi} yields
\begin{equation}\label{eq:Phi_det}
\begin{split}
    \det\Phi_t
    =\exp\left(\int_0^t\operatorname{tr}(L_s)\,ds\right)
    =1,
    \qquad 0\leq t\leq T.
\end{split}
\end{equation}
Thus \eqref{eq:Phi_det} implies that $\Phi_t$ is invertible
for every $t\in[0,T]$. Continuity of $\Phi$ and of matrix
inversion, together with compactness of $[0,T]$, gives
\eqref{eq:Phi_bound} by the extreme value theorem.
\end{proof}

We now prove the essential convexity property for the cost functional \eqref{eq:cost Xhat}. First, recall that the innovation process $V$ was defined in \eqref{inov-proc}. Thanks to Lemma \ref{lemma:V tilde} we can introduce the following square-integrable $\mathcal{Y}^{0}$-martingales.
\begin{equation} \label{v12} 
    V^{(1)}_{t} = \int_{0}^{t}\kappa^{(1)}_{s} dV_{s} 
    \qquad \text{and} \qquad 
    V^{(2)}_{t} = \int_{0}^{t} (\kappa^{(2)}_{s} + 1) dV_{s},\quad 0\leq t \leq T. 
\end{equation} 
\begin{lemma} \label{lemma:strictly convex}
    The cost functional  $q \mapsto \mathcal{C}(q;p,x)$ in \eqref{eq:cost Xhat} is strictly convex on $\mathcal{A}$.
\end{lemma}
\begin{proof}
    The proof uses ideas from the proof of Lemma 10.1 in \cite{neuman2023crowd}. Integration by parts applied to the term $\hat{X}^{q}_{T}\left(\alpha\hat{X}^{q}_{T}-P_{T}\right)$ together with \eqref{eq:Xhat SDE} yields
    \begin{equation} \label{eq:terminal_int_parts}
    \begin{split}
        -P_{T}\hat{X}^{q}_{T} + \alpha\left(\hat{X}^{q}_{T}\right)^{2} 
        &= -px + \alpha x^{2} - \int_{0}^{T}\hat{X}^{q}_{t}dP_{t} - \int_{0}^{T}P_{t}dV^{(2)}_{t} \\
        &\quad- \int_{0}^{T}P_{t}\left(\hat{\theta}^{q}_{t}+q_{t}\right)dt 
        + 2\alpha\int_{0}^{T}\hat{X}^{q}_{t}dV^{(2)}_{t} \\
        &\quad+ 2\alpha\int_{0}^{T}\hat{X}^{q}_{t}\left(\hat{\theta}^{q}_{t}+q_{t}\right)dt + \alpha[\hat{X}^{q}]_{T}
    \end{split}
    \end{equation}
    since $W^{Z}$ and $\overline{M}$ are independent.
    Substituting \eqref{eq:terminal_int_parts} into $\eqref{eq:cost Xhat}$ then gives us  
    \begin{equation} \label{cost3} 
    \begin{aligned}
        \mathcal{C}(q;p,x) &= - xp + \alpha x^{2}  \\
        &\quad+ \mathbb{E}_{p,x}\bigg[\int_{0}^{T}\left(-P_{t}\hat{\theta}^{q}_{t} + 2\alpha\hat{X}^{q}_{t}\left(\hat{\theta}^{q}_{t} + q_{t}\right) + Y^{q}_{t}q_{t} + \frac{1}{2}\epsilon q_{t}^{2}\right)dt \\
        &\qquad\qquad+ \int_{0}^{T}\left(2\alpha\hat{X}^{q}_{t}-P_{t}\right)dV^{(2)}_{t} - \int_{0}^{T}\hat{X}^{q}_{t}dP_{t} + \alpha[\hat{X}^{q}]_{T}\bigg]. 
    \end{aligned}
    \end{equation}
   For convenience we omit the constants and the initial values notation in \eqref{cost3} and define,
    \begin{equation}\label{cost-tl} 
    \begin{split}
        \wt{\mathcal{C}}(q) &= \mathbb{E}\bigg[\int_{0}^{T}\left(-P_{t}\hat{\theta}^{q}_{t} + 2\alpha\hat{X}^{q}_{t}\left(\hat{\theta}^{q}_{t} + q_{t}\right) + Y^{q}_{t}q_{t} + \frac{1}{2}\epsilon q_{t}^{2}\right)dt \\
        &\qquad\quad+ \int_{0}^{T}\left(2\alpha\hat{X}^{q}_{t}-P_{t}\right)dV^{(2)}_{t} - \int_{0}^{T}\hat{X}^{q}_{t}dP_{t} + \alpha[\hat{X}^{q}]_{T}\bigg].
    \end{split}
    \end{equation}
Let $q, u \in \mathcal{A}$ such that $q \neq u$ $d\mathbb{P}\otimes ds$-a.e. on $\Omega\times[0,T]$. By Proposition \ref{prop:fixed-filtration}, $\mathcal{A}$ is a vector space and thus $\delta q+(1-\delta)u\in\mathcal A$. Moreover, \eqref{inov-proc} shows that the innovation process $V$, and hence $V^{(1)}$ and $V^{(2)}$, is common to all three controls.
    \begin{equation} \label{eq:Cstar condition}
        \wt{\mathcal{C}}\left(\delta q + (1-\delta)u\right) - \delta\wt{\mathcal{C}}(q) - (1-\delta)\wt{\mathcal{C}}(u) < 0, \quad \textrm{for all }\delta\in(0,1),
    \end{equation}
 which will prove the convexity of \eqref{eq:cost Xhat}. 
From \eqref{thm-filt} and \eqref{eq:thetahat} it follows that $\hat{\theta}^{\delta q + (1-\delta)u}_{t} = \delta\hat{\theta}^{q}_{t} + (1-\delta)\hat{\theta}^{u}_{t}$ and $\hat{X}^{\delta q + (1-\delta)u}_{t} = \delta \hat{X}^{q}_{t} + (1-\delta) \hat{X}^{u}_{t}$. We therefore get from \eqref{cost-tl} and bilinearity of quadratic variation, 
    \begin{equation}\label{eq:Cstar diff}
    \begin{split}
        &\wt{\mathcal{C}}\left(\delta q+(1-\delta)u\right)
        -\delta\wt{\mathcal{C}}(q)
        -(1-\delta)\wt{\mathcal{C}}(u)
        \\
        &=-\delta(1-\delta)\mathbb{E}\Bigg[
        \int_0^T\Bigg(
        2\alpha
        \left(\hat{X}_t^q-\hat{X}_t^u\right)
        \left(
        (\hat{\theta}_t^q+q_t)
        -(\hat{\theta}_t^u+u_t)
        \right)
        \\
        &\qquad\qquad\qquad\quad
        +\left(Y_t^q-Y_t^u\right)(q_t-u_t)
        +\frac{1}{2}\epsilon(q_t-u_t)^2
        \Bigg)\,dt
        \\
        &\qquad\qquad\qquad\quad
        +\alpha\left(
        [\hat{X}^q]_T
        -2[\hat{X}^q,\hat{X}^u]_T
        +[\hat{X}^u]_T
        \right)
        \Bigg].
    \end{split}
    \end{equation}
    Clearly the third term in the integral on the right-hand side of \eqref{eq:Cstar diff} is always strictly positive. Using \eqref{eq:Xhat SDE} and integration by parts on the term $\big(\hat{X}^{q}_{T}-\hat{X}^{u}_{T}\big)^{2}$ yields
    \begin{equation} \label{dd1} 
    \begin{split}
        &\int_{0}^{T} 2\alpha\left(\hat{X}^{q}_{t}-\hat{X}^{u}_{t}\right)\left(\left(\hat{\theta}^{q}_{t} + q_{t}\right) - \left(\hat{\theta}^{u}_{t} + u_{t}\right)\right)dt \\
        &=\alpha\left(\hat{X}^{q}_{T}-\hat{X}^{u}_{T}\right)^{2} - \alpha\left[\hat{X}^{q}-\hat{X}^{u}\right]_{T} \\
        &=\alpha\left(\hat{X}^{q}_{T}-\hat{X}^{u}_{T}\right)^{2} - \alpha\left([\hat{X}^q]_T-2[\hat{X}^q,\hat{X}^u]_T +[\hat{X}^u]_T \right).
    \end{split}
    \end{equation}
    Similarly, using \eqref{def:Y} and integration by parts applied to the term $\left(Y^{q}_{T}-Y^{u}_{T}\right)^{2}$ further yields
    \begin{equation}\label{dd2}
        \int_{0}^{T}\left(Y^{q}_{t} - Y^{u}_{t}\right)\left(q_{t} - u_{t}\right)dt = \frac{1}{2\lambda}\left(Y^{q}_{T} - Y^{u}_{T}\right)^{2} + \frac{\beta}{\lambda}\int_{0}^{T}\left(Y^{q}_{t} - Y^{u}_{t}\right)^{2}dt > 0. 
    \end{equation}
From \eqref{eq:Cstar diff}, \eqref{dd1}, and \eqref{dd2}, and noting that $\alpha(\hat{X}^{q}_{T}-\hat{X}^{u}_{T})^{2}\geq0$, we get \eqref{eq:Cstar condition}. This concludes the result. 
\end{proof}

For the sake of readability, we omit the use of the initial values notation in the cost functional \eqref{eq:cost Xhat} for the remainder of this section and refer to it as $\mathcal{C}(q)$. By Lemma \ref{lemma:strictly convex} and Proposition 1.2 from Chapter 2 of \cite{ekeland1999convex}, $\mathcal{C}(\cdot)$ admits a unique minimiser $q^* \in \mathcal A$ if the Gateaux derivative satisfies (see e.g. Proposition 2.1 in Chapter II.1 of \cite{ekeland1999convex}) 
\begin{equation} \label{def:Gateaux}
    \langle\mathcal{C}'(q^*), \eta\rangle := \underset{\delta\rightarrow0}{\lim}\left(\frac{\mathcal{C}(q^*+\delta\eta)-\mathcal{C}(q^*)}{\delta}\right) =0, \quad \textrm{for all }\eta\in\mathcal{A}. 
\end{equation}

In the following lemma we derive an explicit expression for the Gateaux derivative. 
\begin{lemma} \label{lemma:Gateaux}
    Let $\mathcal{C}$ be the cost functional in \eqref{eq:cost Xhat}. Then, for any $\eta\in\mathcal{A}$ we have
    \begin{equation} \label{eq:gateaux derivative of C}
    \begin{split}
     \langle\mathcal{C}'(q), \eta\rangle 
    &= \mathbb{E}\bigg[\int_{0}^{T}\eta_{s}\bigg(P_{s}+\epsilon q_{s}+Y^{q}_{s}+\int_{s}^{T}e^{-\beta(t-s)}\lambda q_{t}dt \\
    &\qquad-b_{s} P_{T}\int_{s}^{T} \exp\left(\int_{s}^{t}a_{r}dr\right)dt-P_{T} \\
    &\qquad+2\alpha\left(b_{s}\hat{X}^{q}_{T}\int_{s}^{T} \exp\left(\int_{s}^{t}a_{r}dr\right)dt+\hat{X}^{q}_{T}\right)\bigg)ds\bigg].
\end{split}
\end{equation}
\end{lemma}
\begin{proof}
Let $\eta\in\mathcal{A}$. Proposition \ref{prop:fixed-filtration} gives that $\mathcal{A}$ is a vector space and thus $q+\delta\eta\in\mathcal A$ for every $\delta\in\mathbb R$. From  \eqref{eq:thetahat}, we have 
\begin{equation} \label{eq:thetahat iota eta}
\begin{split}
    \hat{\theta}_{t}^{q+\delta\eta} &
    = \theta_{0}\exp\left(\int_{0}^{t}a_{r}dr\right) + \int_{0}^{t}\exp\left(\int_{s}^{t}a_{r}dr\right)\kappa^{(1)}_{s}dV_{s} \\
    &\quad+\int_{0}^{t}b_{s}\exp\left(\int_{s}^{t}a_{r}dr\right)\left(q_{s}+\delta\eta_{s}\right)ds + \int_{0}^{t}c_{s}\exp\left(\int_{s}^{t}a_{r}dr\right)dA_{s} \\
    &= \hat{\theta}^{q}_{t} + \delta \int_{0}^{t} b_{s}\exp\left(\int_{s}^{t}a_{r}dr\right) \eta_{s} ds.
\end{split}
\end{equation}
From \eqref{eq:thetahat iota eta} and \eqref{eq:Xhat SDE}, we get that
\begin{equation} \label{eq:Xhat iota eta}
\begin{split}
    \hat{X}^{q+\delta\eta}_{t} 
    &= x + \int_{0}^{t} \left(\kappa^{(2)}_{s} + 1 \right)dV_{s} + \int_{0}^{t} \left(\hat{\theta}^{q+\delta\eta}_{s}+q_{s} + \delta\eta_{s}\right)ds \\
    &= x + \int_{0}^{t} \left(\kappa^{(2)}_{s} +1\right) dV_{s} \\
    &\quad+ \int_{0}^{t} \left(\hat{\theta}_{s}^{q} + \delta  \int_{0}^{s} b_{r}\exp\left(\int_{r}^{s}a_{u}du\right) \eta_{r} dr  + q_{s} + \delta \eta_{s}\right)ds \\
    &= \hat{X}^{q}_{t} + \delta \int_{0}^{t} \left(   \int_{0}^{s} b_{r}\exp\left(\int_{r}^{s}a_{u}du\right) \eta_{r} dr + \eta_{s} \right) ds.
\end{split}
\end{equation}
From \eqref{def:Y} it follows that
\begin{equation} \label{eq:Y iota eta}
    Y^{q+\delta\eta}_{t} = Y^{q}_{t} + \delta\int_{0}^{t}e^{-\beta(t-s)}\lambda\eta_{s}ds.
\end{equation}
Applying \eqref{eq:thetahat iota eta}, \eqref{eq:Xhat iota eta} and \eqref{eq:Y iota eta} to \eqref{eq:cost Xhat}, gives
\begin{equation}
\begin{split}
    &\mathcal{C}(q+\delta\eta) \\
    &= \mathbb{E}\bigg[\int_{0}^{T}\left(P_{t}(q_{t}+\delta\eta_{t}) + Y^{q+\delta\eta}_{t}(q_{t}+\delta\eta_{t}) + \frac{1}{2}\epsilon(q_{t}+\delta\eta_{t})^{2}\right)dt \\
    &\qquad- P_{T}\hat{X}^{q+\delta\eta}_{T} + \alpha\left(\hat{X}^{q+\delta\eta}_{T}\right)^{2}\bigg] \\
    &= \mathbb{E}\bigg[\int_{0}^{T}\left(P_{t}(q_{t}+\delta\eta_{t}) + \left(Y^{q}_{t} + \delta\int_{0}^{t}e^{-\beta(t-s)}\lambda\eta_{s}ds\right)(q_{t}+\delta\eta_{t}) + \frac{1}{2}\epsilon(q_{t}+\delta\eta_{t})^{2}\right)dt \\
    &\qquad- P_{T}\left(\hat{X}^{q}_{T} + \delta \int_{0}^{T} \left( \int_{0}^{t} b_{s}\exp\left(\int_{s}^{t}a_{r}dr\right) \eta_{s} ds + \eta_{t} \right) dt\right) \\
    &\qquad+ \alpha\left(\hat{X}^{q}_{T} + \delta \int_{0}^{T} \left( \int_{0}^{t} b_{s}\exp\left(\int_{s}^{t}a_{r}dr\right) \eta_{s} ds + \eta_{t} \right) dt\right)^{2}\bigg].
     \end{split}
\end{equation}
We therefore obtain, 
 \begin{equation}
\begin{split}
    &\frac{\mathcal{C}(q+\delta\eta)-\mathcal{C}(q)}{\delta} \\
    &= \mathbb{E}\bigg[\int_{0}^{T}\bigg(P_{t}\eta_{t} + \frac{1}{2}\epsilon\left(2q_{t}\eta_{t}\right) 
    + Y^{q}_{t}\eta_{t}  + q_{t}\int_{0}^{t}e^{-\beta(t-s)}\lambda\eta_{s}ds\bigg)dt \\
    &\qquad- P_{T}\left(\int_{0}^{T} \left(\int_{0}^{t} b_{s} \exp\left(\int_{s}^{t}a_{r}dr\right) \eta_{s} ds + \eta_{t} \right) dt\right) \\
    &\qquad+ \delta\bigg(\int_{0}^{T}\frac{1}{2}\epsilon\eta_{t}^{2} + \eta_{t}\int_{0}^{t}e^{-\beta(t-s)}\lambda\eta_{s}dsdt\bigg) \\
    &\qquad+ \alpha\bigg(  
    2\hat{X}^{q}_{T}\int_{0}^{T} \int_{0}^{t} b_{s}\exp\left(\int_{s}^{t}a_{r}dr\right)\eta_{s}dsdt+2\hat{X}^{q}_{T}\int_{0}^{T}\eta_{t}dt \\
    &\quad\qquad+\delta\bigg(\left(\int_{0}^{T} \int_{0}^{t} b_{s}\exp\left(\int_{s}^{t}a_{r}dr\right)\eta_{s}dsdt\right)^{2} + \left(\int_{0}^{T}\eta_{t}dt\right)^{2} \\ 
    &\quad\qquad+ 2\left(\int_{0}^{T} \int_{0}^{t}b_{s}\exp\left(\int_{s}^{t}a_{r}dr\right)\eta_{s}dsdt\right)\left(\int_{0}^{T}\eta_{t}dt\right)\bigg)\bigg)\bigg] 
\end{split}
\end{equation}
By taking $\dl \rr 0$ we derive the Gateaux derivative, 
\begin{equation}
\begin{split}
    \langle\mathcal{C}'(q), \eta\rangle
    &= \mathbb{E}\bigg[\int_{0}^{T}\bigg(P_{t}\eta_{t} + \epsilon q_{t}\eta_{t} 
    + Y^{q}_{t}\eta_{t} + q_{t}\int_{0}^{t}e^{-\beta(t-s)}\lambda\eta_{s}ds\bigg)dt \\
    &\qquad- P_{T}\left(\int_{0}^{T} \left(  \int_{0}^{t} b_{s} \exp\left(\int_{s}^{t}a_{r}dr\right) \eta_{s} ds + \eta_{t} \right) dt\right) \\
    &\qquad+ 2\alpha\bigg(\hat{X}^{q}_{T}\int_{0}^{T} \int_{0}^{t} b_{s} \exp\left(\int_{s}^{t}a_{r}dr\right)\eta_{s}dsdt+\hat{X}^{q}_{T}\int_{0}^{T}\eta_{t}dt\bigg)\bigg].
\end{split}
\end{equation}
We get \eqref{eq:gateaux derivative of C} by applications of Fubini's theorem. 
\end{proof}

Before we state Lemma \ref{lemma:FBSDE}, which characterises the optimal trading rate as a solution to a system of FBSDEs, we introduce the following auxiliary processes:
\begin{equation} \label{def:Mtilde}
    \widetilde{M}_{t} := \mathbb{E}\left[\left(\int_{0}^{T}\exp\left(\int_{0}^{s}a_{r}dr\right)ds\right)\left(2\alpha\hat{X}^{q*}_{T}-P_{T}\right)\Big\vert\mathcal{Y}^{0}_{t}\right], \quad 0\leq t \leq T, 
\end{equation}
\begin{equation} \label{def:Ntilde} 
\begin{split}
    \widetilde{N}_{t} := \mathbb{E}\left[\int_{0}^{T}e^{-\beta s}\lambda q^{*}_{s}ds\Big\vert\mathcal{Y}^{0}_{t}\right], \quad 0\leq t \leq T, 
\end{split}
\end{equation}
and
\begin{equation} \label{def:Rtilde} 
    \widetilde{R}_{t} := 
    \mathbb{E}\left[2\alpha\hat{X}^{q*}_{T}-P_{T}|\mathcal{Y}^{0}_{t}\right],\quad 0\leq t \leq T. 
\end{equation}
Note that from \eqref{def:admissset}, \eqref{eq:Xhat SDE}, \eqref{eq:thetahat} and since the functions $a,b,c,d$ are bounded by assumption, it follows that the processes $\widetilde{M}, \widetilde{N}$ and $\widetilde{R}$ are square-integrable $\mathcal{Y}^{0}$-martingales.

We further define
\begin{equation}
    R_{t} = \widetilde{R}_{t} + N_{t} + M_{t},\quad 0\leq t \leq T.
\end{equation}
where
\begin{equation}
\begin{split}
    M_{t} 
    &= \int_{0}^{t}b_{s}\exp\left(-\int_{0}^{s}a_{r}dr\right)\left(d\widetilde{M}_{s} - \int_{0}^{s}\exp\left(\int_{0}^{r}a_{u}du\right)drd\widetilde{R}_{s}\right),\quad 0\leq t \leq T, 
\end{split}
\end{equation}
and
\begin{equation}
    N_{t} = \int_{0}^{t}e^{\beta s}d\widetilde{N}_{s},\quad 0\leq t \leq T.
\end{equation}
In addition, we introduce the two auxiliary square-integrable processes: 
\begin{equation} \label{def:Gamma}
    \Gamma^{q*}_{t} := e^{\beta t}\left(\widetilde{N}_{t} - \int_{0}^{t}e^{-\beta s}\lambda q^{*}_{s}ds\right),\quad 0\leq t \leq T,
\end{equation}
and 
\begin{equation} \label{def:Psi} 
    \Psi^{q*}_{t} = b_{t}\exp\left(-\int_{0}^{t}a_{r}dr\right)\left(\widetilde{M}_{t}-\int_{0}^{t}\exp\left(\int_{0}^{s}a_{r}dr\right)ds\widetilde{R}_{t}\right),\quad 0\leq t \leq T. 
\end{equation}

\begin{lemma} \label{lemma:FBSDE}
    A control $q^{*}\in\mathcal{A}$ is the unique minimiser of the cost functional \eqref{eq:cost Xhat} if and only if $(\hat{X}^{q*}, \hat{\theta}^{q*}, Y^{q*}, q^{*}, \Gamma^{q*}, \Psi^{q*})$ satisfy the following linear forward-backward SDE system
\begin{equation} \label{eq:FBSDE}
    \begin{dcases}
        &d\hat{X}^{q*}_{t} = dV^{(2)}_{t} + \left(\hat{\theta}^{q*}_{t} + q^{*}_{t}\right)dt,  
          \\
        &d\hat{\theta}^{q*}_{t} = dV^{(1)}_{t} + \left(a_{t}\hat{\theta}^{q*}_{t} + b_{t}q^{*}_{t}\right)dt +c_{t}dA_{t}, 
        \\
        &dY^{q*}_{t} = -\beta Y^{q*}_{t}dt + \lambda q^{*}_{t} dt, 
         \\
        &dq^{*}_{t} = -\frac{1}{\epsilon}\left(dA_{t}-\beta Y^{q*}_{t}dt + \beta \Gamma^{q*}_{t}dt  - a_{t}\Psi^{q*}_{t}dt + r_{t}\widetilde{R}_{t}dt + dR_{t} +d\overline{M}_{t }\right), \\
        &d\Gamma^{q*}_{t} = \beta \Gamma^{q*}_{t}dt - \lambda q^{*}_{t} dt + dN_{t},
        \\
        &d\Psi^{q*}_{t} = -a_{t}\Psi^{q*}_{t}dt + dM_{t} + r_{t}\widetilde{R}_{t}dt,  
    \end{dcases}
\end{equation}
with the initial and terminal conditions: 
$$(\hat{X}_0^{q*}, \hat{\theta}^{q*}_0, Y^{q*}_0)=(x,\theta_0,0), \qquad  (q^{*}_T,\Gamma_T^{q*},\Psi_T^{q*}) = \left( -\frac{1}{\epsilon}(Y^{q*}_{T}+2\alpha\hat{X}^{q*}_{T}),0,0 \right).$$
\end{lemma}
\begin{proof}
    By Proposition 2.1 in Chapter II.1 of \cite{ekeland1999convex}, the unique minimising $q^{*}$ is that which satisfies
    \begin{equation}
        \langle\mathcal{C}'(q*), \eta\rangle = \underset{\delta\rightarrow0}{\lim}\left(\frac{\mathcal{C}(q^{*}+\delta\eta)-\mathcal{C}(q^{*})}{\delta}\right) = 0
    \end{equation}for any $\eta\in\mathcal{A}$. Using Lemma \ref{lemma:Gateaux}, we see that we have a first order condition on the optimal unwind policy $q^{*}$, which is that
\begin{equation} \label{eq:first order condition}
\begin{split}
    & \mathbb{E}\bigg[\int_{0}^{T}\eta_{s}\bigg(P_{s}+\epsilon q^{*}_{s}+Y^{q*}_{s}+\int_{s}^{T}e^{-\beta(t-s)}\lambda q^{*}_{t}dt \\
    &\qquad-b_{s}P_{T}\int_{s}^{T}\exp\left(\int_{s}^{t}a_{r}dr\right)dt-P_{T} \\
    &\qquad+2\alpha\left(b_{s}\hat{X}^{q*}_{T}\int_{s}^{T}\exp\left(\int_{s}^{t}a_{r}dr\right)dt+\hat{X}^{q*}_{T}\right)\bigg)ds\bigg] = 0
\end{split}
\end{equation}
for all $\eta\in\mathcal{A}$. 

We now use this first order condition to derive a linear FBSDE system which the optimal control and the corresponding state variables satisfy. As the optimal control is unique, the proof follows from our sufficiency argument, but we additionally include a necessity argument which includes the derivation of the FBSDE \eqref{eq:FBSDE}.

\textit{Necessity:} 
By the optional projection theorem, as well as Fubini's theorem, we have
\begin{equation} \label{eq:1st opt proj}
\begin{split}
    &\mathbb{E}\bigg[\int_{0}^{T}\eta_{s}\mathbb{E}\bigg[P_{s}+\epsilon q^{*}_{s}+Y^{q*}_{s}+\int_{s}^{T}e^{-\beta(t-s)}\lambda q^{*}_{t}dt \\
    &\quad-b_{s}\left(\int_{s}^{T}\exp\left(\int_{s}^{t}a_{r}dr\right)dt\right)P_{T}-P_{T} \\
    &\quad+2\alpha\left(b_{s}\hat{X}^{q*}_{T}\int_{s}^{T}\exp\left(\int_{s}^{t}a_{r}dr\right)dt+\hat{X}^{q*}_{T}\right)\bigg\vert\mathcal{F}_{s}\bigg]ds\bigg] = 0
\end{split}
\end{equation}
Then, using the tower property to condition on $\mathcal{Y}^{0}_{s}\subset\mathcal{F}_{s}$ for $s\in[0,T]$, we have
\begin{equation} \label{eq:2nd opt proj}
\begin{split}
    &\mathbb{E}\bigg[\int_{0}^{T}\eta_{s}\mathbb{E}\bigg[P_{s}+\epsilon q^{*}_{s}+Y^{q*}_{s}+\mathbb{E}\left[\int_{s}^{T}e^{-\beta(t-s)}\lambda q^{*}_{t}dt\Big\vert\mathcal{Y}^{0}_{s}\right] \\
    &\quad+\mathbb{E}\bigg[-b_{s}\left(\int_{s}^{T}\exp\left(\int_{s}^{t}a_{r}dr\right)dt\right)P_{T}-P_{T} \\
    &\quad+2\alpha\left(b_{s}\hat{X}^{q*}_{T}\int_{s}^{T}\exp\left(\int_{s}^{t}a_{r}dr\right)dt+\hat{X}^{q*}_{T}\right)\Big\vert\mathcal{Y}^{0}_{s}\bigg]\bigg\vert\mathcal{F}_{s}\bigg]ds\bigg] = 0,
\end{split}
\end{equation}
for all $\eta\in\mathcal{A}$. By using both Fubini's theorem as well as the tower property again, and then varying over $\eta$ implies that, 
\begin{equation} \label{eq:vary eta}
\begin{split}
    &P_{s}+\epsilon q^{*}_{s}+Y^{q*}_{s}+\mathbb{E}\left[\int_{s}^{T}e^{-\beta(t-s)}\lambda q^{*}_{t}dt\Big\vert\mathcal{Y}^{0}_{s}\right] \\
    &+\mathbb{E}\bigg[2\alpha \hat{X}^{q*}_{T}-P_{T} -b_{s}\left(\int_{s}^{T}\exp\left(\int_{s}^{t}a_{r}dr\right)dt\right)P_{T}\\
    &\qquad +2\alpha\left(b_{s}\hat{X}^{q*}_{T}\int_{s}^{T}\exp\left(\int_{s}^{t}a_{r}dr\right)dt\right)\Big\vert\mathcal{Y}^{0}_{s}\bigg] = 0, \quad  d\mathbb{P}\otimes ds\text{-a.e}.
\end{split}
\end{equation}
Recall the definitions of $\wt R$, $\Gamma^{q*}$ and $\Psi^{q*}$  in \eqref{def:Rtilde}, \eqref{def:Gamma} and \eqref{def:Psi}, respectively.  
Then we can rewrite \eqref{eq:vary eta} as follows
\begin{equation} \label{eq:introduced auxiliaries}
\begin{split}
    &P_{s} + \epsilon q^{*}_{s} +Y^{q*}_{s} + \Gamma^{q*}_{s} + \Psi^{q*}_{s} + \widetilde{R}_{s} = 0, 
    \qquad d\mathbb{P}\otimes ds\text{-a.e} \,\, \text{ on } \,\, \Omega\times[0,T].
\end{split}
\end{equation}
Using now the dynamics of $Y^{q*}_{s}$ given in \eqref{def:Y} and noting from \eqref{def:Gamma} it follows that $\Gamma^{q*}_{s}$ satisfies 
\begin{equation} \label{eq:gamma BSDE}
    d\Gamma^{q*}_{t} = \beta \Gamma^{q*}_{t}dt + e^{\beta t}d\widetilde{N}_{t} - \lambda q^{*}_{t} dt, \quad \textrm{for all }   0 \leq  t < T,   \quad \Gamma^{q*}_{T}=0. 
\end{equation}
From \eqref{def:Psi} it follows that $\Psi^{q*}_{s}$ satisfies, 
\begin{equation} \label{eq:psi BSDE}
\begin{split}
    d\Psi_t^{q^*}
    ={}&
    -a_t\Psi_t^{q^*}\,dt \\
    &+
    b_t
    \exp\left(-\int_0^t a_r\,dr\right)
    \left[
    d\widetilde{M}_t
    -
    \left(
    \int_0^t
    \exp\left(\int_0^s a_r\,dr\right)ds
    \right)
    d\widetilde{R}_t
    \right] \\
    &+
    \left[
    b'_t
    \int_t^T
    \exp\left(\int_t^s a_r\,dr\right)ds
    -b_t
    \right]
    \widetilde{R}_t\,dt \\
    ={}&
    -a_t\Psi_t^{q^*}\,dt
    +dM_t
    +r_t\widetilde{R}_t\,dt,
    \qquad 0\leq t<T,
    \qquad \Psi_T^{q^*}=0.
\end{split}
\end{equation}
By plugging in \eqref{eq:gamma BSDE} and \eqref{eq:psi BSDE} into \eqref{eq:introduced auxiliaries}, we find that $q^{*}_{t}$ satisfies, \begin{equation} \label{eq:q* BSDE}
\begin{split}
    dq_t^*
    ={}&
    -\frac{1}{\epsilon}
    \left(
    dP_t
    -\beta Y_t^{q^*}\,dt
    +\beta\Gamma_t^{q^*}\,dt
    -a_t\Psi_t^{q^*}\,dt
    +r_t\widetilde{R}_t\,dt
    +dR_t
    \right), \\
    q_T^*
    ={}&
    -\frac{1}{\epsilon}
    \left(
    Y_T^{q^*}
    +2\alpha\widehat{X}_T^{q^*}
    \right).
\end{split}
\end{equation}
From \eqref{def:Y}, \eqref{eq:Xhat SDE}, \eqref{eq:gamma BSDE}, \eqref{eq:psi BSDE} and \eqref{eq:q* BSDE} it follows that if \eqref{def:Gateaux} is satisfied, then $(\hat{X}^{q*}, \hat{\theta}^{q*}, Y^{q*}, q*, \Gamma^{q*}, \Psi^{q*})$ satisfies \eqref{eq:FBSDE}.

\textit{Sufficiency:} Let $(\hat{X}^{q*}, \hat{\theta}^{q*}, Y^{q*}, q*, \Gamma^{q*}, \Psi^{q*})$ be the solution to \eqref{eq:FBSDE}. Then by reverting the steps in \eqref{eq:introduced auxiliaries}--\eqref{eq:q* BSDE} it follows that $q^{*}$ satisfies 
\begin{equation} \label{eq:q* solution}
    \epsilon q^{*}_{t} = -P_{t}  - Y^{q*}_{t} - \Gamma^{q*}_{t} - \Psi^{q*}_{t} - \widetilde{R}_{t}, \quad 0\leq t \leq T.   
\end{equation}
Substituting this into the left-hand side of first order condition \eqref{eq:first order condition} and using the definitions of $\Gamma^{q*}, \Psi^{q*}, \widetilde{M}, \widetilde{N}$ and $\widetilde{R}$ in \eqref{def:Mtilde}--\eqref{def:Psi} we get,
\begin{equation} \label{gfr1} 
\begin{split}
    &\quad\mathbb{E}\bigg[\int_{0}^{T}\eta_{s}\bigg(-e^{\beta s}\left(\widetilde{N}_{s}-\int_{0}^{s}e^{-\beta t }\lambda q^{*}_{t}dt\right)+\int_{s}^{T}e^{-\beta (t-s)}\lambda q^{*}_{t}dt \\
    &\qquad\qquad\quad-b_{s}\exp\left(-\int_{0}^{s}a_{r}dr\right)\left(\widetilde{M}_{s}-\widetilde{R}_{s}\int_{0}^{s}\exp\left(\int_{0}^{t}a_{r}dr\right)dt\right) - \widetilde{R}_{s} \\
    &\qquad\qquad\quad+\left(b_{s}\int_{s}^{T}\exp\left(\int_{s}^{t}a_{r}dr\right)dt+1\right)\left(2\alpha\hat{X}^{q*}_{T}-P_{T}\right)\bigg)ds\bigg] \\ 
    &=\mathbb{E}\bigg[\int_{0}^{T}\eta_{s}\bigg(-e^{\beta s}\left(\widetilde{N}_{s}-\int_{0}^{T}e^{-\beta t }\lambda q^{*}_{t}dt\right) -\widetilde{R}_{s}b_{s}\int_{s}^{T}\exp\left(\int_{s}^{t}a_{r}dr\right)dt  \\
    &\qquad\qquad\qquad-\widetilde{R}_{s}+ \left(b_{s}\int_{s}^{T}\exp\left(\int_{s}^{t}a_{r}dr\right)dt+1\right)\left(2\alpha\hat{X}^{q*}_{T}-P_{T}\right)\bigg)ds\bigg] \\
    &=\mathbb{E}\bigg[\int_{0}^{T}\eta_{s}\bigg(-e^{\beta s}\left(\widetilde{N}_{s}-\widetilde{N}_{T}\right) 
    \\
    &\qquad\qquad\qquad-\left(\widetilde{R}_{s}-\widetilde{R}_{T}\right)\left(b_{s}\int_{s}^{T}\exp\left(\int_{s}^{t}a_{r}dr\right)dt+1\right)\bigg)ds\bigg].
\end{split}
\end{equation}
By using Fubini's theorem and the tower property (conditioning on $\mathcal{Y}^{0}_{s}$) together with the martingale property of $\widetilde{M}, \widetilde{N}$ and $\widetilde{R}$ on \eqref{gfr1}, it follows that 
\begin{equation}
\begin{split}
    &=\mathbb{E}\bigg[\int_{0}^{T}\eta_{s}\bigg(-e^{\beta s}\left(\widetilde{N}_{s}-\mathbb{E}\left[\widetilde{N}_{T}|\mathcal{Y}^{0}_{s}\right]\right) \\
    &\qquad-\left(\widetilde{R}_{s}-\mathbb{E}\left[\widetilde{R}_{T}|\mathcal{Y}^{0}_{s}\right]\right)\left(b_{s}\int_{s}^{T}\exp\left(\int_{s}^{t}a_{r}dr\right)dt+1\right)\bigg)ds\bigg]
    = 0,
\end{split}
\end{equation}
This verifies \eqref{eq:first order condition}. 
\end{proof}

\begin{proof}[Proof of Theorem \ref{thm:optimal control}]
From Lemmas \ref{lemma:strictly convex} and \ref{lemma:FBSDE} it follows that in order to prove Theorem \ref{thm:optimal control} we only need to solve the system \eqref{eq:FBSDE} and to verify that the solution is admissible. Since the derivation of the solution to \eqref{eq:FBSDE} is long and involved, we divide the proof into several steps.

\textbf{Step 1: Matrix form of \eqref{eq:FBSDE}.}
Recall that $P=p+A+\overline{M}$. We define:  
\begin{equation} \label{sol-mat} 
    \mathbf{X}^{q*}_{t} = 
    \begin{pmatrix}
        \hat{X}^{q*}_{t} \\
        \hat{\theta}^{q*}_{t} \\
        Y^{q*}_{t} \\ 
        q^{*}_{t} \\
        \Gamma^{q*}_{t} \\
        \Psi^{q*}_{t} \\
        \widetilde{R}_{t} \\
        P_{t}
    \end{pmatrix}
    \qquad \text{and} \qquad
    \mathbf{M}_{t} = 
    \begin{pmatrix}
        V^{(2)}_{t} \\
        V^{(1)}_{t} +\int_{0}^{t}c_{s}dA_{s} \\
        0 \\ 
        -\frac{1}{\epsilon}\left(A_{t} + \overline{M}_{t}+R_{t}\right) \\
        N_{t} \\
        M_{t} \\
        \widetilde{R}_{t} \\
        A_{t} + \overline{M}_{t}
    \end{pmatrix},
    \qquad 0\leq t \leq T, 
\end{equation}
and recall the definition of $(\mathbf{L}_{t})_
{t\in[0,T]}$ as given in \eqref{def:L}. Using this notation, we can rewrite the  system \eqref{eq:FBSDE} as follows,
\begin{equation} \label{eq:matrix FBSDE}
    d\mathbf{X}^{q*}_{t} = \mathbf{L}_{t}\mathbf{X}^{q*}_{t}dt + d\mathbf{M}_{t},
\end{equation}
with the initial conditions $\mathbf{X}^{q*, 1}_{0}=x$, $\mathbf{X}^{q*, 2}_{0}=\theta_{0}$, $\mathbf{X}^{q*, 3}_{0}=0$ and $\mathbf{X}^{q*, 8}_{0}=p$ as well as the terminal conditions
\begin{equation} \label{terminal} 
\begin{aligned} 
    \begin{pmatrix}
        \frac{2\alpha}{\epsilon} & 0 & \frac{1}{\epsilon} & 1 & 0 & 0 & 0 & 0
    \end{pmatrix}
    \mathbf{X}^{q*}_{T} &= 0, \\ 
     \begin{pmatrix}
        0 & 0 & 0 & 0 & 1 & 0 & 0 & 0
    \end{pmatrix}
    \mathbf{X}^{q*}_{T} &= 0,  \\ 
     \begin{pmatrix}
        0 & 0 & 0 & 0 & 0 & 1 & 0 & 0
    \end{pmatrix}
    \mathbf{X}^{q*}_{T} & = 0,  \\ 
    \begin{pmatrix}
        -2\alpha & 0 & 0 & 0 & 0 & 0 & 1 & 1
    \end{pmatrix}\
    \mathbf{X}^{q*}_{T} &= 0.
    \end{aligned} 
\end{equation}
The unique solution to \eqref{eq:matrix FBSDE} can be expressed as
\begin{equation} \label{eq:matrix FBSDE solution}
    \mathbf{X}^{q*}_{T} = \Phi_{T}\Phi^{-1}_{t}\mathbf{X}^{q*}_{t} + \int_{t}^{T}\Phi_{T}\Phi^{-1}_{s}d\mathbf{M}_{s}, \quad 0\leq t \leq T. 
\end{equation}
where $\Phi$ 
is the unique solution of \eqref{def:Phi}. This follows from integration by parts, It\^o's lemma, \eqref{eq:matrix FBSDE} and \eqref{def:Phi} which together give that
\begin{equation}
\begin{split}
    d\left(\Phi^{-1}_{t}\mathbf{X}^{q*}_{t}\right) = \Phi_{t}^{-1}\mathbf{L}_{t}\mathbf{X}^{q*}_{t}dt + \Phi^{-1}_{t}d\mathbf{M}_{t} - \Phi_{t}^{-1}\mathbf{L}_{t}\mathbf{X}^{q*}_{t}dt.
\end{split}
\end{equation}
Then rearrange to solve for $\mathbf{X}^{q*}_{T}$.
Using \eqref{s-mat} we note that \eqref{eq:matrix FBSDE solution} can be written as
\begin{equation} \label{eq:FBSDE representation}
    \mathbf{X}^{q*}_{T} = S(t)\mathbf{X}^{q*}_{t} + \int_{t}^{T} S(s)d\mathbf{M}_{s}, \quad 0\leq t \leq T. 
\end{equation}
\textbf{Step 2: Application of the terminal conditions.}

From  \eqref{eq:first terminal}, \eqref{sol-mat}, \eqref{terminal} and \eqref{eq:FBSDE representation} we get that, 
\begin{equation} \label{ggt1} 
\begin{split}
    0 
    &= G(t)\mathbf{X}^{q}_{t} + \int_{t}^{T}G(s)d\mathbf{M}_{s} \\
    &= G_{1}(t)\hat{X}^{q}_{t} + G_{2}(t)\hat{\theta}^{q}_{t} + G_{3}(t)Y^{q}_{t} + G_{4}(t)q_{t} + G_{5}(t)\Gamma^{q}_{t} + G_{6}(t)\Psi^{q}_{t} \\
    &\quad + G_{7}(t)\widetilde{R}_{t} + G_{8}(t)P_{t} + \int_{t}^{T}G(s)d\mathbf{M}_{s}.
\end{split}
\end{equation}
Using Assumption \ref{assum:nonzeros}(i), \eqref{sol-mat} and by taking conditional expectation on both sides of \eqref{ggt1} we get for all $0\leq t \leq T$, 
\begin{equation} \label{eq:q all vars}
\begin{split}
    q_{t} 
    &= -\frac{G_{1}(t)}{G_{4}(t)}\hat{X}^{q}_{t} - \frac{G_{2}(t)}{G_{4}(t)}\hat{\theta}^{q}_{t} - \frac{G_{3}(t)}{G_{4}(t)}Y^{q}_{t} - \frac{G_{5}(t)}{G_{4}(t)}\Gamma^{q}_{t} - \frac{G_{6}(t)}{G_{4}(t)}\Psi^{q}_{t} \\
    &\quad - \frac{G_{7}(t)}{G_{4}(t)}\widetilde{R}_{t}  - \frac{G_{8}(t)}{G_{4}(t)}P_{t} -  \mathbb{E}\bigg[\int_{t}^{T}\left(c_{s}\frac{G_{2}(s)}{G_{4}(t)}-\frac{1}{\epsilon}\frac{G_{4}(s)}{G_{4}(t)} + \frac{G_{8}(s)}{G_{4}(t)}\right)dA_{s}\Big\vert\mathcal{Y}^{0}_{t}\bigg], 
\end{split}
\end{equation}
where we have used the fact that $\wt N$, $\wt R$, $\wt M$, $V^{(1)}$ and $V^{(2)}$ are $\mathcal{Y}^{0}$-martingales (see \eqref{def:Rtilde}, \eqref{def:Ntilde}, \eqref{def:Mtilde} \eqref{v12}). 

Using the second terminal condition in \eqref{terminal} together with \eqref{eq:second terminal},  \eqref{sol-mat} and \eqref{eq:FBSDE representation} gives,
\begin{equation}
\begin{split}
    0 
    &= H(t)\mathbf{X}^{q}_{t} + \int_{t}^{T}H(s)d\mathbf{M}_{s} \\
    &= H_{1}(t)\hat{X}^{q}_{t} + H_{2}(t)\hat{\theta}^{q}_{t} + H_{3}(t)Y^{q}_{t} + H_{4}(t)q_{t} + H_{5}(t)\Gamma^{q}_{t} + H_{6}(t)\Psi^{q}_{t} \\
    & \quad + H_{7}(t)\widetilde{R}_{t} + H_{8}(t)P_{t} + \int_{t}^{T}H(s)d\mathbf{M}_{s}.
\end{split}
\end{equation}
From Assumption \ref{assum:nonzeros}(i) and by a similar argument as in \eqref{eq:q all vars} we get that, 
\begin{equation} \label{eq:gamma all vars}
\begin{split}
    \Gamma^{q}_{t} 
    &= -\frac{H_{1}(t)}{H_{5}(t)}\hat{X}^{q}_{t} - \frac{H_{2}(t)}{H_{5}(t)}\hat{\theta}^{q}_{t} - \frac{H_{3}(t)}{H_{5}(t)}Y^{q}_{t} - \frac{H_{4}(t)}{H_{5}(t)}q_{t} - \frac{H_{6}(t)}{H_{5}(t)}\Psi^{q}_{t} \\
    &\quad -\frac{H_{7}(t)}{H_{5}(t)}\widetilde{R}_{t} - \frac{H_{8}(t)}{H_{5}(t)}P_{t} \\
    &\quad -  \mathbb{E}\bigg[\int_{t}^{T}\left(c_{s}\frac{H_{2}(s)}{H_{5}(t)}-\frac{1}{\epsilon}\frac{H_{4}(s)}{H_{5}(t)} + \frac{H_{8}(s)}{H_{5}(t)} \right)dA_{s}\Big\vert\mathcal{Y}^{0}_{t}\bigg].
\end{split}
\end{equation}
Using the third terminal condition in \eqref{terminal} together with \eqref{eq:third terminal} and \eqref{eq:FBSDE representation} we get, 
\begin{equation}
\begin{split}
    0 
    &= I(t)\mathbf{X}^{q}_{t} + \int_{t}^{T}I(s)d\mathbf{M}_{s} \\
    &= I_{1}(t)\hat{X}^{q}_{t} + I_{2}(t)\hat{\theta}^{q}_{t} + I_{3}(t)Y^{q}_{t} + I_{4}(t)q_{t} + I_{5}(t)\Gamma^{q}_{t} + I_{6}(t)\Psi^{q}_{t} \\
    &\quad + I_{7}(t)\widetilde{R}_{t} + I_{8}(t)P_{t} + \int_{t}^{T}I(s)d\mathbf{M}_{s}.
\end{split}
\end{equation}
From Assumption \ref{assum:nonzeros}(i) and by a similar argument as in \eqref{eq:q all vars} it follows that, \begin{equation} \label{eq:psi all vars}
\begin{split}
    \Psi_{t} 
    &= -\frac{I_{1}(t)}{I_{6}(t)}\hat{X}^{q}_{t} - \frac{I_{2}(t)}{I_{6}(t)}\hat{\theta}^{q}_{t} - \frac{I_{3}(t)}{I_{6}(t)}Y^{q}_{t} - \frac{I_{4}(t)}{I_{6}(t)}q_{t} - \frac{I_{5}(t)}{I_{6}(t)}\Gamma^{q}_{t} \\
    &\quad - \frac{I_{7}(t)}{I_{6}(t)}\widetilde{R}_{t} - \frac{I_{8}(t)}{I_{6}(t)}P_{t} -  \mathbb{E}\bigg[ \int_{t}^{T}\left( c_{s}\frac{I_{2}(s)}{I_{6}(t)}-\frac{1}{\epsilon}\frac{I_{4}(s)}{I_{6}(t)} + \frac{I_{8}(s)}{I_{6}(t)}\right)dA_{s}\Big\vert\mathcal{Y}^{0}_{t}\bigg].
\end{split}
\end{equation}
Finally, using the fourth terminal condition in \eqref{terminal} together with \eqref{eq:fourth terminal} and \eqref{eq:FBSDE representation} we get, 
\begin{equation}
\begin{split}
    0 
    &= J(t)\mathbf{X}^{q}_{t} + \int_{t}^{T}J(s)d\mathbf{M}_{s} \\
    &= J_{1}(t)\hat{X}^{q}_{t} + J_{2}(t)\hat{\theta}^{q}_{t} + J_{3}(t)Y^{q}_{t} + J_{4}(t)q_{t} + J_{5}(t)\Gamma^{q}_{t} + J_{6}(t)\Psi^{q}_{t} \\
    &+ J_{7}(t)\widetilde{R}_{t} + J_{8}(t)P_{t} + \int_{t}^{T}J(s)d\mathbf{M}_{s}.
\end{split}
\end{equation}
From Assumption \ref{assum:nonzeros}(i) and by a similar argument as in \eqref{eq:q all vars} it follows that, \begin{equation} \label{eq:Rtilde all vars}
\begin{split}
    \widetilde{R}_{t} 
    &= -\frac{J_{1}(t)}{J_{7}(t)}\hat{X}^{q}_{t} - \frac{J_{2}(t)}{J_{7}(t)}\hat{\theta}^{q}_{t} - \frac{J_{3}(t)}{J_{7}(t)}Y^{q}_{t} - \frac{J_{4}(t)}{J_{7}(t)}q_{t} - \frac{J_{5}(t)}{J_{7}(t)}\Gamma^{q}_{t} \\
    &\quad - \frac{J_{6}(t)}{J_{7}(t)}\Psi_{t} - \frac{J_{8}(t)}{J_{7}(t)}P_{t} -  \mathbb{E}\bigg[\int_{t}^{T}\left(c_{s}\frac{J_{2}(s)}{J_{7}(t)}-\frac{1}{\epsilon}\frac{J_{4}(s)}{J_{7}(t)} + \frac{J_{8}(s)}{J_{7}(t)}\right)dA_{s}\Big\vert\mathcal{Y}^{0}_{t}\bigg].
\end{split}
\end{equation}
\textbf{Step 3: Derivation of the optimal control.}

Substituting \eqref{eq:gamma all vars} into \eqref{eq:psi all vars} yields an expression for $\Psi^{q}_{t}$ excluding $\Gamma^{q}_{t}$,
\begin{equation}  \label{gf} 
\begin{split}
    \Psi^{q}_{t} 
    &= \bigg(1-\frac{I_{5}(t)H_{6}(t)}{I_{6}(t)H_{5}(t)}\bigg)^{-1}  \\ 
    &\quad \times \Bigg( 
    \bigg(\frac{I_{5}(t)}{I_{6}(t)}\frac{H_{1}(t)}{H_{5}(t)}-\frac{I_{1}(t)}{I_{6}(t)}\bigg)\hat{X}^{q}_{t} 
    + \bigg(\frac{I_{5}(t)}{I_{6}(t)}\frac{H_{2}(t)}{H_{5}(t)}-\frac{I_{2}(t)}{I_{6}(t)}\bigg)\hat{\theta}^{q}_{t} \\ 
    &\qquad \ + \bigg(\frac{I_{5}(t)}{I_{6}(t)}\frac{H_{3}(t)}{H_{5}(t)}-\frac{I_{3}(t)}{I_{6}(t)}\bigg)Y^{q}_{t} 
    + \bigg(\frac{I_{5}(t)}{I_{6}(t)}\frac{H_{4}(t)}{H_{5}(t)}-\frac{I_{4}(t)}{I_{6}(t)}\bigg)q_{t} \\
    &\qquad  \ + \bigg(\frac{I_{5}(t)H_{7}(t)}{I_{6}(t)H_{5}(t)} - \frac{I_{7}(t)}{I_{6}(t)}\bigg)\widetilde{R}_{t}
    + \bigg(\frac{I_{5}(t)}{I_{6}(t)}\frac{H_{8}(t)}{H_{5}(t)}-\frac{I_{8}(t)}{I_{6}(t)}\bigg)P_{t} \\
    &\qquad \ +  \mathbb{E}\bigg[\int_{t}^{T}\bigg(\frac{I_{5}(t)}{I_{6}(t)}\bigg(c_{s}\frac{H_{2}(s)}{H_{5}(t)}-\frac{1}{\epsilon}\frac{H_{4}(s)}{H_{5}(t)}+\frac{H_{8}(s)}{H_{5}(t)}\bigg) \\
    &\qquad\qquad\qquad\quad-c_{s}\frac{I_{2}(s)}{I_{6}(t)}+\frac{1}{\epsilon}\frac{I_{4}(s)}{I_{6}(t)} - \frac{I_{8}(s)}{I_{6}(t)}\bigg)dA_{s}\Big\vert\mathcal{Y}^{0}_{t}\bigg]\Bigg).
\end{split}
\end{equation}
For convenience, we relabel the terms in \eqref{gf} as follows, 
\begin{equation} \label{eq:psi no gamma}
\begin{split}
    \Psi^{q}_{t} &= \tilde{i}^{X}(t)\hat{X}^{q}_{t} + \tilde{i}^{\theta}(t)\hat{\theta}^{q}_{t} + \tilde{i}^{Y}(t)Y^{q}_{t} + \tilde{i}^{q}(t)q_{t} \\
    &\quad+ \tilde{i}^{R}(t)\widetilde{R}_{t} + \tilde{i}^{P}(t)P_{t} + \mathbb{E}\left[\int_{t}^{T}\tilde{i}^{A}(s, t)dA_{s}\Big\vert\mathcal{Y}^{0}_{t}\right], 
\end{split}
\end{equation}
where $\tilde{i}^{X}, \tilde{i}^{\theta}, \tilde{i}^{Y}, \tilde{i}^{q}, \tilde{i}^{R}, \tilde{i}^{P}$ and $\tilde{i}^{A}$ are defined in \eqref{def:i tilde}.

Substituting \eqref{eq:psi no gamma} into \eqref{eq:gamma all vars} gives an expression for $\Gamma^{q}_{t}$ not depending on $\Psi^{q}_{t}$,
\begin{equation} \label{ghq} 
\begin{split}
    \Gamma^{q}_{t} 
    &= \bigg(\frac{-H_{6}(t)}{H_{5}(t)}\tilde{i}^{X}(t)-\frac{H_{1}(t)}{H_{5}(t)}\bigg)\hat{X}^{q}_{t} 
    + \bigg(-\frac{H_{6}(t)}{H_{5}(t)}\tilde{i}^{\theta}(t)-\frac{H_{2}(t)}{H_{5}(t)}\bigg)\hat{\theta}^{q}_{t} \\
    &\quad + \bigg(-\frac{H_{6}(t)}{H_{5}(t)}\tilde{i}^{Y}(t)-\frac{H_{3}(t)}{H_{5}(t)}\bigg)Y^{q}_{t}
    + \bigg(-\frac{H_{6}(t)}{H_{5}(t)}\tilde{i}^{q}(t)-\frac{H_{4}(t)}{H_{5}(t)}\bigg)q_{t} \\
    &\quad+ \bigg(-\frac{H_{6}(t)}{H_{5}(t)}\tilde{i}^{R}(t) - \frac{H_{7}(t)}{H_{5}(t)}\bigg)\widetilde{R}_{t}
    + \bigg(-\frac{H_{6}(t)}{H_{5}(t)}\tilde{i}^{P}(t)-\frac{H_{8}(t)}{H_{5}(t)}\bigg)P_{t} \\
    &\quad+ \mathbb{E}\left[\int_{t}^{T}\bigg(-\frac{H_{6}(t)}{H_{5}(t)}\tilde{i}^{A}(s, t) -c_{s}\frac{H_{2}(s)}{H_{5}(t)} + \frac{1}{\epsilon}\frac{H_{4}(s)}{H_{5}(t)} - \frac{H_{8}(s)}{H_{5}(t)}\bigg)dA_{s}\Big\vert\mathcal{Y}^{0}_{t}\right].
\end{split}
\end{equation}
For convenience, we relabel the terms in \eqref{ghq} as follows, 
\begin{equation} \label{eq:gamma no psi}
\begin{split}
    \Gamma^{q}_{t} &= \tilde{h}^{X}(t)\hat{X}^{q}_{t} + \tilde{h}^{\theta}(t)\hat{\theta}^{q}_{t} + 
    \tilde{h}^{Y}(t)Y^{q}_{t} + \tilde{h}^{q}(t)q_{t} \\
    &\quad+ \tilde{h}^{R}(t)\widetilde{R}_{t} + \tilde{h}^{P}(t)P_{t} + \mathbb{E}\left[\int_{t}^{T}\tilde{h}^{A}(s, t)dA_{s}\Big\vert\mathcal{Y}^{0}_{t}\right], 
\end{split}
\end{equation}
where $\tilde{h}^{X}, \tilde{h}^{\theta}, \tilde{h}^{Y}, \tilde{h}^{q}, \tilde{h}^{R}, \tilde{h}^{P}$ and $\tilde{h}^{A}$ are defined in \eqref{def:h tilde}. Substituting \eqref{eq:psi no gamma} and \eqref{eq:gamma no psi} into \eqref{eq:Rtilde all vars} gives an expression for $\widetilde{R}_{t}$ only in terms of state variables
\begin{equation} \label{ghq2} 
\begin{split}
    \widetilde{R}_{t} 
    &= \bigg(1 + \frac{J_{5}(t)}{J_{7}(t)}\tilde{h}^{R}(t) + \frac{J_{6}(t)}{J_{7}(t)}\tilde{i}^{R}(t)\bigg)^{-1} \\
    &\quad \times\Bigg(\bigg(-\frac{J_{5}(t)}{J_{7}(t)}\tilde{h}^{X}(t) - \frac{J_{6}(t)}{J_{7}(t)}\tilde{i}^{X}(t)-\frac{J_{1}(t)}{J_{7}(t)}\bigg)\hat{X}^{q}_{t} \\
    &\qquad  \ + \bigg(-\frac{J_{5}(t)}{J_{7}(t)}\tilde{h}^{\theta}(t) - \frac{J_{6}(t)}{J_{7}(t)}\tilde{i}^{\theta}(t)-\frac{J_{2}(t)}{J_{7}(t)}\bigg)\hat{\theta}^{q}_{t} \\
    &\qquad \ + \bigg(-\frac{J_{5}(t)}{J_{7}(t)}\tilde{h}^{Y}(t) - \frac{J_{6}(t)}{J_{7}(t)}\tilde{i}^{Y}(t)-\frac{J_{3}(t)}{J_{7}(t)}\bigg)Y^{q}_{t} \\
    &\qquad \ + \bigg(-\frac{J_{5}(t)}{J_{7}(t)}\tilde{h}^{q}(t) - \frac{J_{6}(t)}{J_{7}(t)}\tilde{i}^{q}(t)-\frac{J_{4}(t)}{J_{7}(t)}\bigg)q_{t} \\
    &\qquad \ + \bigg(-\frac{J_{5}(t)}{J_{7}(t)}\tilde{h}^{P}(t) - \frac{J_{6}(t)}{J_{7}(t)}\tilde{i}^{P}(t)-\frac{J_{8}(t)}{J_{7}(t)}\bigg)P_{t} \\
    &\qquad \ +\mathbb{E}\bigg[\int_{t}^{T}-c_{s}\frac{J_{2}(s)}{J_{7}(t)} + \frac{1}{\epsilon}\frac{J_{4}(s)}{J_{7}(t)} - \frac{J_{8}(s)}{J_{7}(t)} \\
    &\qquad\qquad\qquad- \frac{J_{5}(t)}{J_{7}(t)}\tilde{h}^{A}(s,t) - \frac{J_{6}(t)}{J_{7}(t)}\tilde{i}^{A}(s,t)dA_{s}\Big\vert\mathcal{Y}^{0}_{t}\bigg]\Bigg).
\end{split}
\end{equation}
We relabel the terms in \eqref{ghq2} for convenience as follows, 
\begin{equation} \label{gt1} 
    \widetilde{R}_{t} = j^{X}(t)\hat{X}^{q}_{t} + j^{\theta}(t)\hat{\theta}^{q}_{t} + j^{Y}(t)Y^{q}_{t} + j^{q}(t)q_{t} + j^{P}(t)P_{t} + \mathbb{E}\left[\int_{t}^{T}j^{A}(s,t)dA_{s}\Big\vert\mathcal{Y}^{0}_{t}\right], 
\end{equation}
where $j^{X}, j^{\theta}, j^{Y}, j^{q}, j^{P}$ and $j^{A}$ are defined in \eqref{def:j}. Next we substitute \eqref{gt1} into both \eqref{eq:psi no gamma} and \eqref{eq:gamma no psi} to obtain expressions for $\Psi^{q}_{t}$ and $\Gamma^{q}_{t}$ only in terms of state variables. We get, 
\begin{equation}
\begin{split}
    \Psi^{q}_{t} 
    &= \bigg(\tilde{i}^{X}(t)+\tilde{i}^{R}(t)j^{X}(t)\bigg)\hat{X}^{q}_{t} 
    + \bigg(\tilde{i}^{\theta}(t)+\tilde{i}^{R}(t)j^{\theta}(t)\bigg)\hat{\theta}^{q}_{t} 
    \\
    &\quad+ \bigg(\tilde{i}^{Y}(t)+\tilde{i}^{R}(t)j^{Y}(t)\bigg)Y^{q}_{t} 
    + \bigg(\tilde{i}^{q}(t)+\tilde{i}^{R}(t)j^{q}(t)\bigg)q_{t} \\
    &\quad+ \bigg(\tilde{i}^{P}(t)+\tilde{i}^{R}(t)j^{P}(t)\bigg)P_{t} \\
    &\quad+ \mathbb{E}\left[\int_{t}^{T}\tilde{i}^{A}(s,t) + \tilde{i}^{R}(t)j^{A}(s,t)dA_{s}\Big\vert\mathcal{Y}^{0}_{t}\right],
\end{split}
\end{equation}
which can be rewritten as
\begin{equation} \label{eq:psi only state}
    \Psi^{q}_{t} = i^{X}(t)\hat{X}^{q}_{t} + i^{\theta}(t)\hat{\theta}^{q}_{t} + i^{Y}(t)Y^{q}_{t} + i^{q}(t)q_{t} + i^{P}(t)P_{t} + \mathbb{E}\left[\int_{t}^{T}i^{A}(s,t)dA_{s}\Big\vert\mathcal{Y}^{0}_{t}\right],
\end{equation}
where $i^{X}, i^{\theta}, i^{Y}, i^{q}, i^{P}$ and $i^{A}$ are defined in \eqref{def:i}. We also have, 
\begin{equation}
\begin{split}
    \Gamma^{q}_{t} 
    &= \bigg(\tilde{h}^{X}(t)+\tilde{h}^{R}(t)j^{X}(t)\bigg)\hat{X}^{q}_{t} 
    + \bigg(\tilde{h}^{\theta}(t)+\tilde{h}^{R}(t)j^{\theta}(t)\bigg)\hat{\theta}^{q}_{t} 
    \\
    &\quad + \bigg(\tilde{h}^{Y}(t)+\tilde{h}^{R}(t)j^{Y}(t)\bigg)Y^{q}_{t} 
    + \bigg(\tilde{h}^{q}(t)+\tilde{h}^{R}(t)j^{q}(t)\bigg)q_{t} \\
    &\quad+ \bigg(\tilde{h}^{P}(t)+\tilde{h}^{R}(t)j^{P}(t)\bigg)P_{t} \\
    &\quad+ \mathbb{E}\left[\int_{t}^{T}\tilde{h}^{A}(s,t) + \tilde{h}^{R}(t)j^{A}(s,t)dA_{s}\Big\vert\mathcal{Y}^{0}_{t}\right],
\end{split}
\end{equation}
which can be rewritten as
\begin{equation} \label{eq:gamma only state}
    \Gamma^{q}_{t} = h^{X}(t)\hat{X}^{q}_{t} + h^{\theta}(t)\hat{\theta}^{q}_{t} + h^{Y}(t)Y^{q}_{t} + h^{q}(t)q_{t} + h^{P}(t)P_{t} + \mathbb{E}\left[\int_{t}^{T}h^{A}(s,t)dA_{s}\Big\vert\mathcal{Y}^{0}_{t}\right],
\end{equation}
where $h^{X}, h^{\theta}, h^{Y}, h^{q}, h^{P}$ and $h^{A}$ are defined in \eqref{def:h}. Now substituting \eqref{gt1}, \eqref{eq:psi only state} and \eqref{eq:gamma only state} into \eqref{eq:q all vars}, we arrive at an expression for the optimal control $q^*_{t}$ only in terms of observable state variables, 
\begin{equation} \label{tt1} 
\begin{split}
    q^*_{t} &
    = \bigg(1+\frac{G_{5}(t)}{G_{4}(t)}h^{q}(t) + \frac{G_{6}(t)}{G_{4}(t)}i^{q}(t) + \frac{G_{7}(t)}{G_{4}(t)}j^{q}(t)\bigg)^{-1} \\
    &\quad \times\Bigg(-\bigg(\frac{G_{5}(t)}{G_{4}(t)}h^{X}(t) + \frac{G_{6}(t)}{G_{4}(t)}i^{X}(t) + \frac{G_{7}(t)}{G_{4}(t)}j^{X}(t) + \frac{G_{1}(t)}{G_{4}(t)}\bigg)\hat{X}^{q^*}_{t} \\
    &\qquad \  -\bigg(\frac{G_{5}(t)}{G_{4}(t)}h^{\theta}(t) + \frac{G_{6}(t)}{G_{4}(t)}i^{\theta}(t) + \frac{G_{7}(t)}{G_{4}(t)}j^{\theta}(t) + \frac{G_{2}(t)}{G_{4}(t)}\bigg)\hat{\theta}^{q^*}_{t} \\
    &\qquad \ - \bigg(\frac{G_{5}(t)}{G_{4}(t)}h^{Y}(t) + \frac{G_{6}(t)}{G_{4}(t)}i^{Y}(t) + \frac{G_{7}(t)}{G_{4}(t)}j^{Y}(t) + \frac{G_{3}(t)}{G_{4}(t)}\bigg)Y^{q^*}_{t} \\
    &\qquad \ - \bigg(\frac{G_{5}(t)}{G_{4}(t)}h^{P}(t) + \frac{G_{6}(t)}{G_{4}(t)}i^{P}(t) + \frac{G_{7}(t)}{G_{4}(t)}j^{P}(t) + \frac{G_{8}(t)}{G_{4}(t)}\bigg)P_{t} \\
    &\qquad \ - \mathbb{E}\bigg[\int_{t}^{T}\frac{G_{5}(t)}{G_{4}(t)}h^{A}(s, t) + \frac{G_{6}(t)}{G_4(t)}i^{A}(s, t) + \frac{G_{7}(t)}{G_4(t)}j^{A}(s, t) \\
    &\qquad\qquad\qquad +c_{s}\frac{G_{2}(s)}{G_{4}(t)} - \frac{1}{\epsilon}\frac{G_{4}(s)}{G_{4}(t)} + \frac{G_{8}(s)}{G_{4}(t)}dA_{s}\Big\vert\mathcal{Y}^{0}_{t}\bigg]\Bigg).
\end{split}
\end{equation} 
We relabel  \eqref{tt1} for convenience as follows,  
\begin{equation} 
    q^*_{t} = g^{X}(t)\hat{X}^{q^*}_{t} + g^{\theta}(t)\hat{\theta}^{q^*}_{t} + g^{Y}(t)Y^{q^*}_{t} + g^{P}(t)P_{t} + \mathbb{E}\left[\int_{t}^{T}g^{A}(s,t)dA_{s}\Big\vert\mathcal{Y}^{0}_{t}\right],
\end{equation}
where $g^{X}, g^{\theta}, g^{Y}, g^{P}$ and $g^{A}$ are defined in \eqref{def:g}. This proves that $q^*$ is given by \eqref{eq:optimal q}. 

\textbf{Step 4: Admissibility:} From Proposition \ref{prop:Phi} and  Assumptions \ref{assum:nonzeros}, \ref{ass2}, and by tracking the formulas for $(g^{X}, g^{\theta}, g^{Y}, g^{P},g^{A})$ in Appendix \ref{sec:full representation of control} it follows that the deterministic coefficients in the right-hand side of \eqref{eq:optimal q} are bounded. From \eqref{eq:Xhat SDE} and \eqref{eq:thetahat}, the boundedness assumption on the coefficients of \eqref{def:theta} and standard arguments it follows that there exist constants $C_i >0$, $i=1,2,3$ such that, 
\be \label{ll1} 
\E\big[ \big(\hat \theta_t^{q^*}\big)^2\big]  \leq C_1 \left(1 + \mathbb{E}\left[\int_0^t (q^*_s)^2 ds \right] \right),\ee 
and
\be\E\big[ \big(\hat X_t^{q^*}\big)^2\big]  \leq C_2+ C_3 \E\left[\int_0^t \big( (q^*_s)^2 +\big(\hat \theta_s^{q^*}\big)^2 \big) ds \right], \ee
for all $0\leq t \leq T$. 

From \eqref{ass:P}, conditional Jensen inequality it follows that there exists a constant $C_4>0$ such that, 
\begin{equation} \label{ll2} 
    \underset{0\leq t \leq T}{\sup} \E \left[ \left(\mathbb{E}\left[\int_{t}^{T}g^{A}(s,t)dA_{s}\Big\vert\mathcal{Y}^{0}_{t}\right] \right)^2\right] \leq C\mathbb{E}\left[\left(\int_{0}^{T}|dA_{t}|\right)^{2}\right] < \infty.
\end{equation}
By applying  \eqref{ass:P}, \eqref{def:Y}, \eqref{ll1} and \eqref{ll2} together with Jensen's inequality to \eqref{eq:optimal q} we conclude that there exist constants $\wt C_1, \wt C_2>0$ such that 
\begin{equation}
    \mathbb{E}\left[(q^*_{t})^{2}\right] \leq \wt C_{1} + \wt C_{2} \int_{0}^{t} \mathbb{E}\left[(q^{*}_{s})^{2}\right] ds, \quad \textrm{for all } 0\leq t \leq T.
    \end{equation}
An application of Gronwall's lemma yields, 
\begin{equation}
    \underset{0 \leq t \leq T}{\sup} \mathbb{E}\left[(q^*_{t})^{2}\right] < \infty.
\end{equation}
Since each of the terms in \eqref{eq:optimal q} is also $\{\mathcal{Y}^{0}_{t}\}_{0\leq t \leq T}$-progressively measurable, it follows from \eqref{def:admissset} that $q^*\in\mathcal{A}$. This completes the proof.  
\end{proof}

\section{Full representation of optimal control} \label{sec:full representation of control}

In this section we present the full representations of the deterministic time-varying coefficients appearing in both Theorem \ref{thm:optimal control} and Theorem \ref{thm:optimal control full information}. In order to describe this, we first introduce various other deterministic time-varying processes, the motivation for which is apparent in the proof of Theorem \ref{thm:optimal control} in Appendix  \ref{sec:control proof}. 
Firstly, the deterministic functions $\tilde{i}^{X}, \tilde{i}^{\theta}, \tilde{i}^{Y}, \tilde{i}^{q}, \tilde{i}^{R}, \tilde{i}^{P}, \tilde{i}^{A}$ are given by
\begin{equation}\label{def:i tilde}
\begin{split}
    &\tilde{i}^{X}(t) = \bigg(\frac{I_{5}(t)H_{1}(t)}{I_{6}(t)H_{5}(t)} - \frac{I_{1}(t)}{I_{6}(t)}\bigg)\bigg(1-\frac{I_{5}(t)H_{6}(t)}{I_{6}(t)H_{5}(t)}\bigg)^{-1}, \\
    &\tilde{i}^{\theta}(t) = \bigg(\frac{I_{5}(t)H_{2}(t)}{I_{6}(t)H_{5}(t)} - \frac{I_{2}(t)}{I_{6}(t)}\bigg)\bigg(1-\frac{I_{5}(t)H_{6}(t)}{I_{6}(t)H_{5}(t)}\bigg)^{-1}, \\
    &\tilde{i}^{Y}(t) = \bigg(\frac{I_{5}(t)H_{3}(t)}{I_{6}(t)H_{5}(t)} - \frac{I_{3}(t)}{I_{6}(t)}\bigg)\bigg(1-\frac{I_{5}(t)H_{6}(t)}{I_{6}(t)H_{5}(t)}\bigg)^{-1}, \\
    &\tilde{i}^{q}(t) = \bigg(\frac{I_{5}(t)H_{4}(t)}{I_{6}(t)H_{5}(t)} - \frac{I_{4}(t)}{I_{6}(t)}\bigg)\bigg(1-\frac{I_{5}(t)H_{6}(t)}{I_{6}(t)H_{5}(t)}\bigg)^{-1}, \\
    &\tilde{i}^{R}(t) = \bigg(\frac{I_{5}(t)H_{7}(t)}{I_{6}(t)H_{5}(t)} - \frac{I_{7}(t)}{I_{6}(t)}\bigg)\bigg(1-\frac{I_{5}(t)H_{6}(t)}{I_{6}(t)H_{5}(t)}\bigg)^{-1}, \\
    &\tilde{i}^{P}(t) = \bigg(\frac{I_{5}(t)H_{8}(t)}{I_{6}(t)H_{5}(t)} - \frac{I_{8}(t)}{I_{6}(t)}\bigg)\bigg(1-\frac{I_{5}(t)H_{6}(t)}{I_{6}(t)H_{5}(t)}\bigg)^{-1}, \\
    &\tilde{i}^{A}(s,t) = \bigg(\frac{I_{5}(t)}{I_{6}(t)}\bigg(c_{s}\frac{H_{2}(s)}{H_{5}(t)}-\frac{1}{\epsilon}\frac{H_{4}(s)}{H_{5}(t)}+\frac{H_{8}(s)}{H_{5}(t)}\bigg) \\
    &\qquad\qquad\qquad\qquad-c_{s}\frac{I_{2}(s)}{I_{6}(t)}+ \frac{1}{\epsilon}\frac{I_{4}(s)}{I_{6}(t)} - \frac{I_{8}(s)}{I_{6}(t)}\bigg)\bigg(1-\frac{I_{5}(t)H_{6}(t)}{I_{6}(t)H_{5}(t)}\bigg)^{-1}.
\end{split}
\end{equation}
Secondly, the deterministic functions $\tilde{h}^{X}, \tilde{h}^{\theta}, \tilde{h}^{Y}, \tilde{h}^{q}, \tilde{h}^{R}, \tilde{h}^{P}, \tilde{h}^{A}$ are given by 
\begin{equation} \label{def:h tilde}
\begin{split}
    \tilde{h}^{X}(t) &= \bigg(-\frac{H_{6}(t)}{H_{5}(t)}\tilde{i}^{X}(t) - \frac{H_{1}(t)}{H_{5}(t)}\bigg), \\
    \tilde{h}^{\theta}(t) &= \bigg(-\frac{H_{6}(t)}{H_{5}(t)}\tilde{i}^{\theta}(t) - \frac{H_{2}(t)}{H_{5}(t)}\bigg), \\
    \tilde{h}^{Y}(t) &= \bigg(-\frac{H_{6}(t)}{H_{5}(t)}\tilde{i}^{Y}(t) - \frac{H_{3}(t)}{H_{5}(t)}\bigg), \\
    \tilde{h}^{q}(t) &= \bigg(-\frac{H_{6}(t)}{H_{5}(t)}\tilde{i}^{q}(t) - \frac{H_{4}(t)}{H_{5}(t)}\bigg), \\
    \tilde{h}^{R}(t) &= \bigg(-\frac{H_{6}(t)}{H_{5}(t)}\tilde{i}^{R}(t) - \frac{H_{7}(t)}{H_{5}(t)}\bigg), \\
    \tilde{h}^{P}(t) &= \bigg(-\frac{H_{6}(t)}{H_{5}(t)}\tilde{i}^{P}(t) - \frac{H_{8}(t)}{H_{5}(t)}\bigg), \\
    \tilde{h}^{A}(s, t) &= \bigg(-\frac{H_{6}(t)}{H_{5}(t)}\tilde{i}^{A}(s,t) -c_{s}\frac{H_{2}(s)}{H_{5}(t)}+ \frac{1}{\epsilon}\frac{H_{4}(s)}{H_{5}(t)} - \frac{H_{8}(s)}{H_{5}(t)}\bigg)
\end{split}
\end{equation}
Thirdly, introduce the function
\begin{equation}
    \tilde{j}(t) = \bigg(1 + \frac{J_{5}(t)}{J_{7}(t)}\tilde{h}^{R}(t) + \frac{J_{6}(t)}{J_{7}(t)}\tilde{i}^{R}(t)\bigg)^{-1}
\end{equation}
and we can in turn present the deterministic functions $j^{X}, j^{\theta}, j^{Y}, j^{q}, j^{P}, j^{A}$ as
\begin{equation} \label{def:j}
\begin{split}
    j^{X}(t) &= -\bigg(\frac{J_{5}(t)}{J_{7}(t)}\tilde{h}^{X}(t) + \frac{J_{6}(t)}{J_{7}(t)}\tilde{i}^{X}(t) + \frac{J_{1}(t)}{J_{7}(t)}\bigg)\tilde{j}(t), \\
    j^{\theta}(t) &= -\bigg(\frac{J_{5}(t)}{J_{7}(t)}\tilde{h}^{\theta}(t) + \frac{J_{6}(t)}{J_{7}(t)}\tilde{i}^{\theta}(t) + \frac{J_{2}(t)}{J_{7}(t)}\bigg)\tilde{j}(t), \\
    j^{Y}(t) &= -\bigg(\frac{J_{5}(t)}{J_{7}(t)}\tilde{h}^{Y}(t) + \frac{J_{6}(t)}{J_{7}(t)}\tilde{i}^{Y}(t) + \frac{J_{3}(t)}{J_{7}(t)}\bigg)\tilde{j}(t), \\
    j^{q}(t) &= -\bigg(\frac{J_{5}(t)}{J_{7}(t)}\tilde{h}^{q}(t) + \frac{J_{6}(t)}{J_{7}(t)}\tilde{i}^{q}(t) + \frac{J_{4}(t)}{J_{7}(t)}\bigg)\tilde{j}(t), \\
    j^{P}(t) &= -\bigg(\frac{J_{5}(t)}{J_{7}(t)}\tilde{h}^{P}(t) + \frac{J_{6}(t)}{J_{7}(t)}\tilde{i}^{P}(t) + \frac{J_{8}(t)}{J_{7}(t)}\bigg)\tilde{j}(t), \\
    j^{A}(s,t) &= -\bigg(c_{s}\frac{J_{2}(s)}{J_{7}(t)}-\frac{1}{\epsilon}\frac{J_{4}(s)}{J_{7}(t)} + \frac{J_{8}(s)}{J_{7}(t)} + \frac{J_{5}(t)}{J_{7}(t)}\tilde{h}^{A}(s,t) + \frac{J_{6}(t)}{J_{7}(t)}\tilde{i}^{A}(s,t)\bigg)\tilde{j}(t)
\end{split}
\end{equation}
Fourthly, the deterministic functions  $i^{X}, i^{\theta}, i^{Y}, i^{q}, i^{P}, i^{A}$ are given by
\begin{equation} \label{def:i}
\begin{split}
    i^{X}(t) &= \tilde{i}^{X}(t) + \tilde{i}^{R}(t)j^{X}(t), \\
    i^{\theta}(t) &= \tilde{i}^{\theta}(t) + \tilde{i}^{R}(t)j^{\theta}(t), \\
    i^{Y}(t) &= \tilde{i}^{Y}(t) + \tilde{i}^{R}(t)j^{Y}(t), \\
    i^{q}(t) &= \tilde{i}^{q}(t) + \tilde{i}^{R}(t)j^{q}(t), \\ 
    i^{P}(t) &= \tilde{i}^{P}(t) + \tilde{i}^{R}(t)j^{P}(t), \\
    i^{A}(s,t) &= \tilde{i}^{A}(s,t) + \tilde{i}^{R}(t)j^{A}(s,t). \\
\end{split}
\end{equation}
Fifthly, the deterministic functions $h^{X}, h^{\theta}, h^{Y}, h^{q}, h^{P}, h^{A}$ are given by
\begin{equation} \label{def:h}
\begin{split}
    h^{X}(t) &= \tilde{h}^{X}(t) + \tilde{h}^{R}(t)j^{X}(t), \\
    h^{\theta}(t) &= \tilde{h}^{\theta}(t) + \tilde{h}^{R}(t)j^{\theta}(t), \\
    h^{Y}(t) &= \tilde{h}^{Y}(t) + \tilde{h}^{R}(t)j^{Y}(t), \\
    h^{q}(t) &= \tilde{h}^{q}(t) + \tilde{h}^{R}(t)j^{q}(t), \\ 
    h^{P}(t) &= \tilde{h}^{P}(t) + \tilde{h}^{R}(t)j^{P}(t), \\
    h^{A}(s,t) &= \tilde{h}^{A}(s,t) + \tilde{h}^{R}(t)j^{A}(s,t) \\
\end{split}
\end{equation}
Finally, to give the functions $g^{X}, g^{\theta}, g^{Y}, g^{P}, g^{A}$ we first introduce
\begin{equation}
    \tilde{g}(t) = \bigg(1+\frac{G_{5}(t)}{G_{4}(t)}h^{q}(t) + \frac{G_{6}(t)}{G_{4}(t)}i^{q}(t) +\frac{G_{7}(t)}{G_{4}(t)}j^{q}(t)\bigg)^{-1}
\end{equation}
and then these functions, appearing in \eqref{eq:optimal q}, are
\begin{equation} \label{def:g}
\begin{split}
    &g^{X}(t) = -\bigg(\frac{G_{5}(t)}{G_{4}(t)}h^{X}(t) + \frac{G_{6}(t)}{G_{4}(t)}i^{X}(t) + \frac{G_{7}(t)}{G_{4}(t)}j^{X}(t) + \frac{G_{1}(t)}{G_{4}(t)}\bigg)\tilde{g}(t), \\
    &g^{\theta}(t) = -\bigg(\frac{G_{5}(t)}{G_{4}(t)}h^{\theta}(t) + \frac{G_{6}(t)}{G_{4}(t)}i^{\theta}(t) + \frac{G_{7}(t)}{G_{4}(t)}j^{\theta}(t) + \frac{G_{2}(t)}{G_{4}(t)}\bigg)\tilde{g}(t), \\
    &g^{Y}(t) = -\bigg(\frac{G_{5}(t)}{G_{4}(t)}h^{Y}(t) + \frac{G_{6}(t)}{G_{4}(t)}i^{Y}(t) + \frac{G_{7}(t)}{G_{4}(t)}j^{Y}(t) + \frac{G_{3}(t)}{G_{4}(t)}\bigg)\tilde{g}(t), \\
    &g^{P}(t) = -\bigg(\frac{G_{5}(t)}{G_{4}(t)}h^{P}(t) + \frac{G_{6}(t)}{G_{4}(t)}i^{P}(t) + \frac{G_{7}(t)}{G_{4}(t)}j^{P}(t) + \frac{G_{8}(t)}{G_{4}(t)}\bigg)\tilde{g}(t), \\
    &g^{A}(s,t) = -\bigg(\frac{G_{5}(t)}{G_{4}(t)}h^{A}(s, t) + \frac{G_{6}(t)}{G_4(t)}i^{A}(s, t) + \frac{G_{7}(t)}{G_{4}(t)}j^{A}(s,t) \\
    &\qquad\qquad\qquad\quad+c_{s}\frac{G_{2}(s)}{G_{4}(t)} - \frac{1}{\epsilon}\frac{G_{4}(s)}{G_{4}(t)} + \frac{G_{8}(s)}{G_{4}(t)}\bigg)\tilde{g}(t)
\end{split}
\end{equation}

\section{Proof of Theorem \ref{thm:optimal control full information}} \label{sec:pf-full}

The optimal unwind strategy $q$ in the full information case also satisfies an FBSDE.

\begin{lemma} \label{lemma:FBSDE full information}
A control $q\in\mathcal{A}$, where $\mathcal{A}$ is given in \eqref{def:admissset full info}, solves the optimisation problem \eqref{def:objective} if and only if $(X^{q}, \theta^{q}, Y^{q}, q, \Gamma^{q}, \Psi^{q})$ is the solution to the following linear forward backward SDE system
\begin{equation} \label{eq:FBSDE full info}
    \begin{dcases}
        &dX^{q}_{t} =  \left(\theta^{q}_{t} + q_{t}\right)dt + \sigma dW^{Z}_{t}, \\
        &d\theta^{q}_{t} = \left(a_{t}\theta^{q}_{t} + b_{t}q_{t}\right)dt+c_{t}dA_{t} + d_{t}dW^{\theta}_{t}, \\
        &dY^{q}_{t} = -\beta Y^{q}_{t}dt + \lambda q_{t} dt, \\
        &dq_{t} = -\frac{1}{\epsilon}\left(dA_{t}+d\overline{M}_{t}-\beta Y^{q}_{t}dt + \beta \Gamma^{q}_{t}dt  - a_{t}\Psi^{q}_{t}dt + dR_{t} + r_{t}\widetilde{R}_{t}dt\right), \\
        &d\Gamma^{q}_{t} = \beta \Gamma^{q}_{t}dt - \lambda q_{t} dt + dN_{t}, \\
        &d\Psi^{q}_{t} = -a_{t}\Psi^{q}_{t}dt + dM_{t} + r_{t}\widetilde{R}_{t}dt,
    \end{dcases}
\end{equation}
with the initial and terminal conditions: 
$$(X_0^{q}, \theta^{q}_0, Y^{q}_0)=(x,\theta_0,0), \qquad  (q_T,\Gamma_T^q,\Psi_T^q) = \left( -\frac{1}{\epsilon}(Y^{q}_{T}+2\alpha X^{q}_{T}),0,0 \right).$$
\end{lemma}
\begin{proof}
    From  \eqref{def:theta}, we have 
\begin{equation}
\begin{split}
    \theta_{t}^{q+\delta\eta} &
    = \theta_{0}\exp\left(\int_{0}^{t}a_{r}dr\right) +\int_{0}^{t}b_{s}\exp\left(\int_{s}^{t}a_{r}dr\right)\left(q_{s}+\delta\eta_{s}\right)ds \\
    &\quad +\int_{0}^{t}\exp\left(\int_{s}^{t}a_{r}dr\right)c_{s}dA_{s} + \int_{0}^{t}\exp\left(\int_{s}^{t}a_{r}dr\right)d_{s}dW^{\theta}_{s}  \\
    &= \theta^{q}_{t} + \delta \int_{0}^{t} b_{s}\exp\left(\int_{s}^{t}a_{r}dr\right) \eta_{s} ds
\end{split}
\end{equation}
which is of the same form as in expression for $\hat{\theta}^{q+\delta\eta}_{t}$ found in \eqref{eq:thetahat iota eta}. From this, we use \eqref{def:X} to see that
\begin{equation}
\begin{split}
    X^{q+\delta\eta}_{t} 
    &= x + \int_{0}^{t} \left(\theta^{q+\delta\eta}_{s}+q_{s} + \delta\eta_{s}\right)ds + \sigma W^{Z}_{t}\\
    &= x + \int_{0}^{t}  \left(\theta_{s}^{q} + \delta  \int_{0}^{s} b_{r}\exp\left(\int_{r}^{s}a_{u}du\right) \eta_{r} dr  + q_{s} + \delta \eta_{s}\right)ds + \sigma W^{Z}_{t}\\
    &= X^{q}_{t} + \delta \int_{0}^{t} \left(\int_{0}^{s} b_{r}\exp\left(\int_{r}^{s}a_{u}du\right) \eta_{r} dr + \eta_{s} \right) ds
\end{split}
\end{equation}
which is of the same form as found in the expression for $\hat{X}^{q+\delta \eta}_{t}$ in \eqref{eq:Xhat iota eta}. Additionally, it can be seen using \eqref{def:Y} that
\begin{equation}
    Y^{q+\delta\eta}_{t} = Y^{q}_{t} + \delta\int_{0}^{t}e^{-\beta(t-s)}\lambda\eta_{s}ds.
\end{equation}
The rest of the proof follows the same steps as the proof for Lemma \ref{lemma:FBSDE} hence it is omitted. 
\end{proof}

\begin{proof}[Proof of Theorem \ref{thm:optimal control full information}] 
    The proof follows exactly the same lines as that of Theorem \ref{thm:optimal control} in Appendix  \ref{sec:control proof}, except that we change the definitions of $\mathbf{X}^{q}$ and $\mathbf{M}$ to the following
\begin{equation}
    \mathbf{X}^{q}_{t} = 
    \begin{pmatrix}
        X^{q}_{t} \\
        \theta^{q}_{t} \\
        Y^{q}_{t} \\ 
        q_{t} \\
        \Gamma^{q}_{t} \\
        \Psi^{q}_{t} \\
        \widetilde{R}_{t} \\
        P_{t} 
    \end{pmatrix}
    \qquad \text{and} \qquad
    \mathbf{M}_{t} = 
    \begin{pmatrix}
        \sigma W^{Z}_{t} \\
        \int_{0}^{t}d_{s}dW^{\theta}_{s} + \int_{0}^{t}c_{s}dA_{s} \\
        0 \\ 
        -\frac{1}{\epsilon}\left(A_{t} + \overline{M}_{t}+R_{t}\right) \\
        N_{t} \\
        M_{t} \\
        \widetilde{R}_{t} \\
        A_{t} + \overline{M}_{t}
    \end{pmatrix},
    \qquad (0\leq t \leq T).
\end{equation}
Since all integrals with respect to terms in $\mathbf{M}$, except for the signal $A$, disappear under conditional expectation, the difference in noise between the two definitions of $\mathbf{M}$ makes no difference to the coefficients in the optimal control. 
\end{proof}

\end{appendices}

\end{document}